\documentclass[aps,preprint,floatfix,nofootinbib,showpacs]{revtex4-1}
\pdfoutput=1
\usepackage{dcolumn}
\usepackage{bm}
\usepackage{graphicx}
\usepackage{amssymb,amsmath}
\usepackage{multirow}
\usepackage{color,url}
\usepackage{tabu}
\usepackage{array}
\usepackage[colorlinks=true,urlcolor=blue,anchorcolor=blue
,citecolor=blue,filecolor=blue,linkcolor=blue,menucolor=blue
,linktocpage=true,pdfproducer=medialab,pdfa=true]{hyperref}
\usepackage{colordvi}
\usepackage{subcaption}
\def\ga{\mathrel{\raise.3ex\hbox{$>$\kern-.75em\lower1ex\hbox{$\sim$}}}}
\def\la{\mathrel{\raise.3ex\hbox{$<$\kern-.75em\lower1ex\hbox{$\sim$}}}}
\def\beqa{\begin{eqnarray}}
\def\eeqa{\end{eqnarray}}
\begin{document}

\title{
  Sensitivity Reach on the Heavy Neutral Leptons and $\tau$-Neutrino Mixing
  $|U_{\tau N}|^2 $ at the HL-LHC }
\def\slash#1{#1\!\!/}

\renewcommand{\thefootnote}{\arabic{footnote}}
\renewcommand{\thefootnote}{\arabic{footnote}}

\author{
  Kingman Cheung$^{1,2,3}$, Yi-Lun Chung$^2$, Hiroyuki Ishida$^4$,
  Chih-Ting Lu$^5$ }
\affiliation{
 $^1$ Physics Division, National Center for Theoretical Sciences, Hsinchu, Taiwan 300 \\
 $^2$ Department of Physics, National Tsing Hua University,
Hsinchu 300, Taiwan \\
 $^3$ Division of Quantum Phases and Devices, School of Physics, 
Konkuk University, Seoul 143-701, Republic of Korea \\
 $^4$ KEK Theory Center, Tsukuba, Ibaraki 305-0801, Japan \\
 $^5$ School of Physics, KIAS, Seoul 130-722, Republic of Korea
}
\date{\today}

\begin{abstract}
The model of heavy neutral leptons (HNLs) is one of the well-motivated
models beyond the standard model from both theoretical and
phenomenological point of views. It is an indispensable ingredient to
explain the puzzle of tiny neutrino masses and the origin of the
matter-antimatter asymmetry in our Universe, based on the models in which
the simplest Type-I seesaw mechanism can be embedded.  The HNL with
a mass up to the electroweak scale is an attractive scenario
which can be readily tested in present or near-future experiments
including the LHC.  In this work, we study the decay rates of HNLs and
find the sensitive parameter space of the mixing angles between
the active neutrinos
and HNLs.  
Since there are fewer collider studies of the mixing between $ \nu_{\tau} $ and HNL in literature compared with those of $\nu_e$
and $\nu_{\mu}$ for the HNL of mass in the electroweak scale,
we focus on the
channel $ pp\rightarrow W^{\pm(\ast)} + X\rightarrow \tau^{\pm} N + X$
to search for HNLs at the LHC 14 TeV. The targeted signature consists of
three prompt charged leptons, which include at least two tau leptons.
After the signal-background analysis, we further set sensitivity
bounds on the mixing $ |U_{\tau N}|^2 $ with $ M_N $ at High-Luminosity
LHC (HL-LHC).
We predict the testable bounds from HL-LHC can be stronger 
than the previous LEP constraints and Electroweak Precision Data (EWPD), 
especially for $ M_N \lesssim $ 50 GeV can reach
down to $ |U_{\tau N}|^2 \approx 5\times 10^{-6} $.
\end{abstract}

\maketitle

\section{Introduction}

Neutrino oscillation is one of the definite evidences of physics beyond the standard model, 
which implies that at least two of three active neutrinos are massive. 
However, there is no clear answer for the origin of neutrino-mass generation. 
Further, the matter-antimatter asymmetry in our Universe is another
mystery that the SM cannot explain.
To address these problems
the conventional Type-I seesaw mechanism
\cite{Minkowski:1977sc,Yanagida:1979as,Yanagida:1980xy,GellMann:1980vs,Ramond:1979,Glashow:1979,Mohapatra:1979ia}
with at least two superheavy right-handed neutrinos 
is one of the the simplest possibilities and widely discussed so far. 
Thanks to the existence of heavy Majorana neutrinos, 
the observed tiny neutrino masses are naturally explained 
and their decays can be the source of the baryon asymmetry of the
Universe (BAU) through a well-known mechanism called thermal
leptogenesis~\cite{Fukugita:1986hr}. 

Hence, if heavy Majorana neutrinos are discovered,
it would be a clear signal of new physics without any doubts.
Unfortunately, since the thermal leptogenesis requires 
the scale of the Majorana neutrinos to be superheavy, 
say more than $10^9~{\rm GeV}$~\cite{Davidson:2002qv}, 
and the conventional Type-I seesaw can be perturbatively applied up to around the GUT scale, $10^{15}~{\rm GeV}$, 
we cannot directly produce and test such heavy particles in
near-future terrestrial experiments. 
However, this is not the end of the story 
because the allowed mass range for the heavy Majorana neutrinos can be
very wide below the GUT scale. 
On the other hand, once the mass of the heavy Majorana neutrinos, 
which contribute to the seesaw mechanism, 
becomes below the pion mass in the minimal model, 
it would conflict with the constraints from the Big Bang Nucleosynthesis,
since its lifetime becomes longer than 1 sec~\cite{Ruchayskiy:2012si}. 
Therefore, the Type-I seesaw mechanism itself can be valid for the mass range of right-handed neutrinos 
between $\sim \mathcal{O}(100~{\rm MeV})$ and the GUT scale.

Among a bunch of possibilities, the one with heavy Majorana neutrinos
below the electroweak scale is an attractive scenario 
which can be readily tested in present or near future experiments.
A model called the Neutrino Minimal Standard Model ($\nu$MSM)~\cite{Asaka:2005an,Asaka:2005pn}, 
in which the SM is extended only by introducing three heavy Majorana neutrinos,
possesses two such neutrinos around the electroweak scale 
and one in the keV scale which also serves as a dark matter candidate. 
Since the neutrino Yukawa coupling of the keV-scale Majorana neutrino is 
so tiny compared with the other two that 
we can completely separate its physics from the others 
and simply focus on the dynamics of the other two heavier Majorana neutrinos, 
namely, the contribution from the keV-scale Majorana neutrino to
the seesaw neutrino mass is small enough 
and the lightest active neutrino mass is suppressed enough 
compared with the solar neutrino mass scale. 
The other two Majorana neutrinos, which have the mass above the pion mass
and below the EW scale,
are responsible for the explanations of the observed atmospheric and
solar neutrino mass scales 
and baryogenesis via neutrino oscillation~\cite{Akhmedov:1998qx,Asaka:2005pn}. 

Generically, the mass eigenstates of the heavy Majorana neutrinos 
are called heavy neutral leptons (HNLs) and labeled as $N$. 
The HNLs can be searched for at terrestrial experiments 
and, especially the testability at beam dump experiments 
where bunches of kaon and B mesons are produced 
when HNLs are lighter than the parent mesons as firstly proposed by \cite{Shrock:1980vy,Shrock:1980ct,Shrock:1981wq}  
(see e.g. \cite{Atre:2009rg,Asaka:2011pb,Asaka:2016rwd,Abada:2019bac,Chun:2019nwi,Bryman:2019bjg} for recent relevant works).
Furthermore, the HNLs can also be searched for at colliders like the
LHC as well and searchable range of HNL mass becomes wider than
the beam damp experiments 
(see e.q. 
\cite{Kersten:2007vk,Atre:2009rg,Blondel:2014bra,Deppisch:2015qwa,Drewes:2016jae,Cai:2017mow,Helo:2018qej,Liu:2019qfa} and references therein).  
Actually, the lepton-number-violating (LNV) channels are the most
specular signals 
and the definite discriminator of the models 
because the HNLs uniquely break lepton number which the SM always preserves. 
Not only for that but the lepton-number-conserving (LNC) channels 
can also provide strong hints for searching for the HNLs.

Although the mixing between $\nu_\tau$ and HNL is more challenging to be probed compared with those of $\nu_e$ and $\nu_\mu$, there already exist some studies for $M_N \sim \mathcal{O}(1 \mathchar`- 5)$ GeV in Ref.~\cite{Bondarenko:2018ptm,Cvetic:2019shl}, $M_N \sim \mathcal{O}(1 \mathchar`- 20)$ GeV in Refs.~\cite{Abada:2018sfh,Cottin:2018nms,Hernandez:2018cgc,Drewes:2019vjy}, and $M_N > 150$ GeV in Ref.~\cite{Andres:2017daw,Pascoli:2018rsg}. However, for $25 < M_N < 150$ GeV, the detectability of the mixing between $\nu_\tau$ and HNL is not well-studied at the LHC. In this work, we focus on the channel $pp\rightarrow W^{\pm(\ast)} + X\rightarrow \tau^{\pm} N + X $ to search for HNLs with $25 < M_N < 150$ GeV at the High-Luminosity LHC (HL-LHC).   
In this work, we focus on the channel $pp\rightarrow W^{\pm(\ast)} + X\rightarrow \tau^{\pm} N + X $~%
\footnote{
Actually, the HNL production in $e^+ e^-$ collider has a long
history~\cite{Gronau:1984ct,Perl:1984yp,Gilman:1985tr,Gilman:1986mz,Hagiwara:1987ub,Dittmar:1989yg,Ma:1989jpa,Dicus:1991wj}. Instead of the charged current interaction in hadron colliders, the neutral current interaction is used to search for HNLs in $e^+ e^-$ colliders.
}  
to search for HNLs with $25 < M_N < 150$ GeV at the High-Luminosity LHC (HL-LHC). 
Our characteristic signature consists of three prompt charged leptons, 
where at least two tau leptons are included. 
With a detailed signal-background analysis
we can set sensitivity bounds on the mixing angle $|U_{\tau N}|^2$ with
$M_N $ at the HL-LHC. 
Especially, it can be improved by a factor of five over the previous analyses when $M_N \lesssim 50~{\rm GeV}$. 
This is a significant improvement over previous studies.

The organization of the paper is as follows.  We highlight some
details of the model that are relevant to our study and calculate the
decay rates of HNLs in Sec.~\ref{Sec:Model}.  In Sec.~\ref{Sec:Conct},
we survey the valid parameter space for the mixing of the active
neutrinos with HNLs in various HNL mass ranges up to the electroweak
scale.  In Sec.~\ref{Sec:Strategy}, we give details about the search
for HNL with $\tau$ leptons at the HL-LHC. In Sec.~\ref{Sec:Analyses},
we present the signal-background analysis and the results, and obtain
the sensitivity bounds on the mixing $|U_{\tau N}|^2$.  We conclude in
Sec.~\ref{Sec:Conclusions}.

\section{The Neutrino Minimal Standard Model}\label{Sec:Model}

\subsection{The model}

In this section, we highlight some details of the
Neutrino Minimal Standard Model ($\nu$MSM) which are relevant to our study.
After introducing three gauge-singlet right-handed neutrino fields into the SM, 
the total Lagrangian can be written as 
\begin{align}
\mathcal{L} = 
\mathcal{L}_{\rm SM} 
  + i \, \overline{\nu_{RI}} \gamma^\mu \partial_\mu \nu_{RI}
  -
  \left(
  F_{\alpha I} \, \overline{\ell_{\alpha}} \, \Phi \, \nu_{RI}
  +
  \frac{M_I}{2} \, \overline{\nu_{RI}^c} \, \nu_{RI}
  +
  h.c.
  \right) \,,\label{Eq:Lag}
\end{align}
where $\mathcal{L}_{\rm SM}$ is the SM Lagrangian 
based on $SU(3)_c \times SU(2)_L \times U(1)_Y$ gauge symmetry, 
the index $\alpha$ denotes the active flavors running for $e, \mu,$ and $\tau$,
and $I$ is the HNL-flavor index running from $1$ to $3$.
The fields $\ell$, $\Phi$, and $\nu_{R}$ are the lepton doublet, the
Higgs doublet, and the right-handed neutrino singlet, respectively. 
$F_{\alpha I}$'s are the neutrino Yukawa coupling constants 
and $M_I$'s are the Majorana masses for the right-handed neutrinos. 

After the Higgs field acquires the vacuum expectation value, 
there are two kinds of neutrino masses, namely, the Dirac neutrino masses 
defined as $(M_D)_{\alpha I} \equiv F_{\alpha I} \langle \Phi \rangle$ 
and the Majorana neutrino masses, $M_I$. 
In the mass basis of neutrinos, the tiny active neutrino masses can be explained 
by the hierarchical ratio between Dirac and Majorana masses 
as $M_D^2/M_I$ realized by the seesaw mechanism. 
In the mass basis, 
the HNLs are composed of mostly right-handed neutrinos but also small portion of left-handed neutrinos, 
thus, HNLs can have gauge interactions through the mixing 
denoted as $U_{\alpha I} \equiv (M_D)_{\alpha I}/M_I$. 
Therefore, HNLs can be searched for at terrestrial experiments. 

As discussed in a number works in literature
~(see e.g. \cite{Chun:2017spz} and references therein and also related papers)
a certain amount of mass degeneracy between two HNLs is necessary for
the success of baryogenesis. 
Then, we can simply rewrite the Majorana masses as
$M_{2,3} = M_N \pm \Delta M/2$ 
where $M_N$ is the common mass and $\Delta M$ denotes the slight mass difference.
We do not stick ourselves to the valid parameter space for
baryogenesis in the following studies, though. 
Between these two mass parameters, 
the common mass scale is more important than their slight 
difference for the purpose of HNLs searches since $\Delta M / M \ll 1$. 
Therefore, we can safely neglect the correction of $\Delta M$ 
and simply multiply a factor of 2 when we want to estimate 
physical observables, such as cross sections, for HNLs in the $\nu$MSM. 
In the following analyses and discussion, however, 
we focus on the case with one HNL just for simplicity 
and denote the mixing angle as $U_{\alpha N}$.

\subsection{Decay rates of the Heavy Neutral Leptons}

Based on the mass range of HNLs, we can calculate its decay rate in three mass ranges: 
(1) low mass region ($ M_N\ll m_{W,Z} $), 
(2) medium mass region ($ M_N\lesssim m_{t} $) 
and (3) high mass region ($M_N\gg m_{W,Z} $). 
Here we only focus on the low and medium mass ranges in this study.\footnote{As complementary studies including heavier mass region, please see e.g.\cite{Alva:2014gxa,Pascoli:2018rsg,Pascoli:2018heg}. Actually, the reason why we focus on such a low mass region is motivated from the model, 
so that higher mass region is beyond our scope. 
Indeed,the $N-\nu_{\tau}$ mixing for $m_N > 150$ GeV was also covered in Ref.~\cite{Pascoli:2018rsg,Pascoli:2018heg}.
}

\begin{figure}[t!]
\centering
\includegraphics[width=5in]{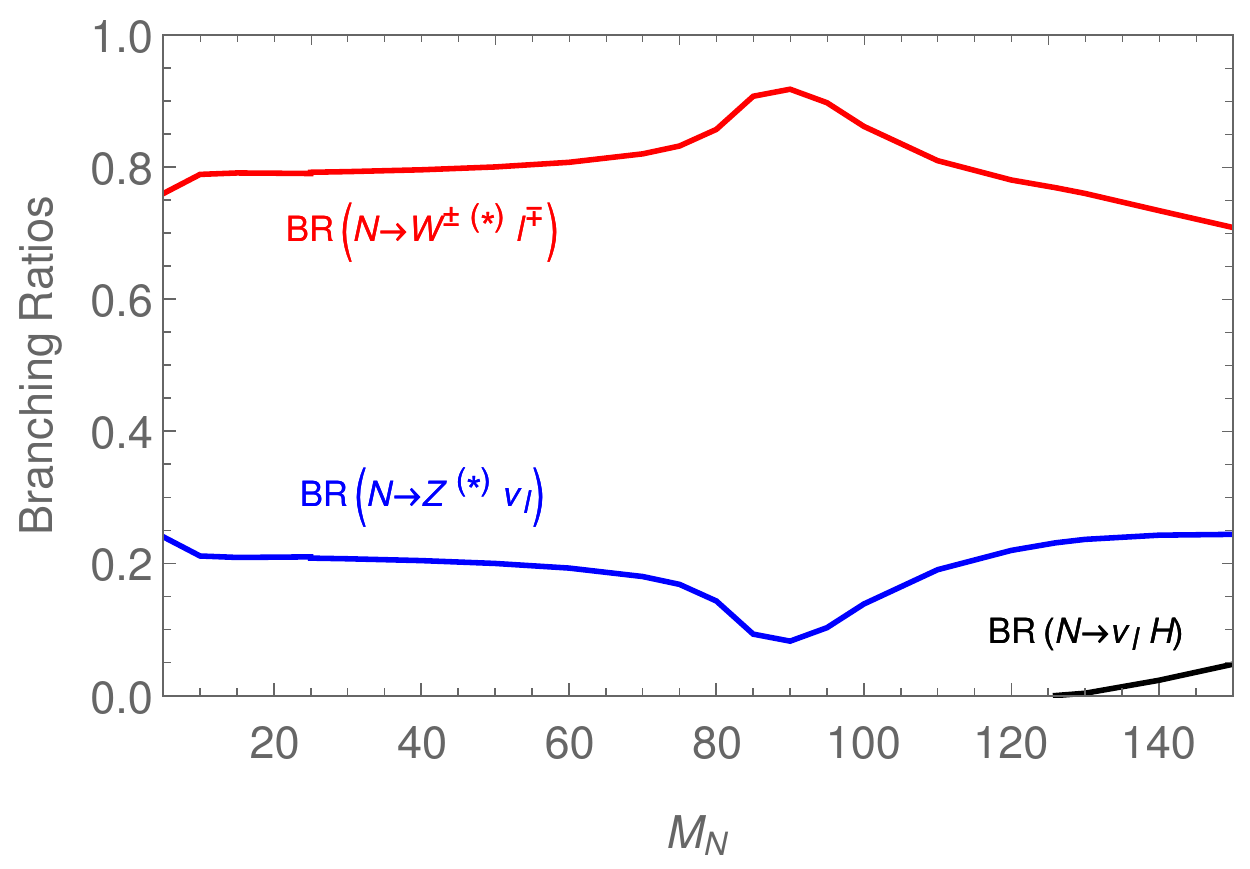}
\caption{
  The branching ratios of the HNL
  with the assumption $|U_{e N}|^2 = |U_{\mu N}|^2 = |U_{\tau N}|^2$ 
  for the decay modes $N \rightarrow W^{\pm(\ast)} l^{\mp}_\alpha$, 
  $N\rightarrow Z^{(\ast)} \nu_\alpha$ and $N \rightarrow \nu_\alpha H $ 
  of HNL in the low and medium mass regions.
}\label{fig:N_BR}
\end{figure}

In the low and medium mass ranges of HNL, the major decay modes are
$N \rightarrow W^{\pm(\ast)} l^{\mp}_\alpha$ and $N \rightarrow Z^{(\ast)} \nu_\alpha$,
where $W, Z$ bosons can be either on-shell or off-shell depending on $M_N $. 
Once HNL is heavier than the Higgs boson, 
the $N \rightarrow \nu_\alpha H $ decay mode is also open.
\footnote{The partial decay width $ \Gamma (N \rightarrow \nu_\alpha H^{\ast})$ 
  is much smaller than the other two partial decay widths via the propagators of $W$ or $Z$ boson 
  when $M_N < m_H$, so we can safely ignore this small contribution in our calculation.}
All detailed formulas for these partial decay widths
are collected in Appendix~\ref{app:decay_width}. 
The branching ratios with the assumption $|U_{e N}|^2 = |U_{\mu N}|^2 = |U_{\tau  N}|^2$ 
for the above decay modes of HNL in the above mass ranges
are shown in Fig.~\ref{fig:N_BR}.%
\footnote{Numerically, we take $M_N \leq 25$ GeV for the low mass range 
  and $25 < M_N \leq 150$ GeV for the medium mass range.}
Since $BR(N \rightarrow W^{\pm(\ast)} l^{\mp}_\alpha)$ is dominant for the
whole mass range, 
we focus on $ N \rightarrow W^{\pm(\ast)} l^{\mp}_\alpha$ in the following study.

\begin{figure}[t!]
\centering
\includegraphics[width=3in]{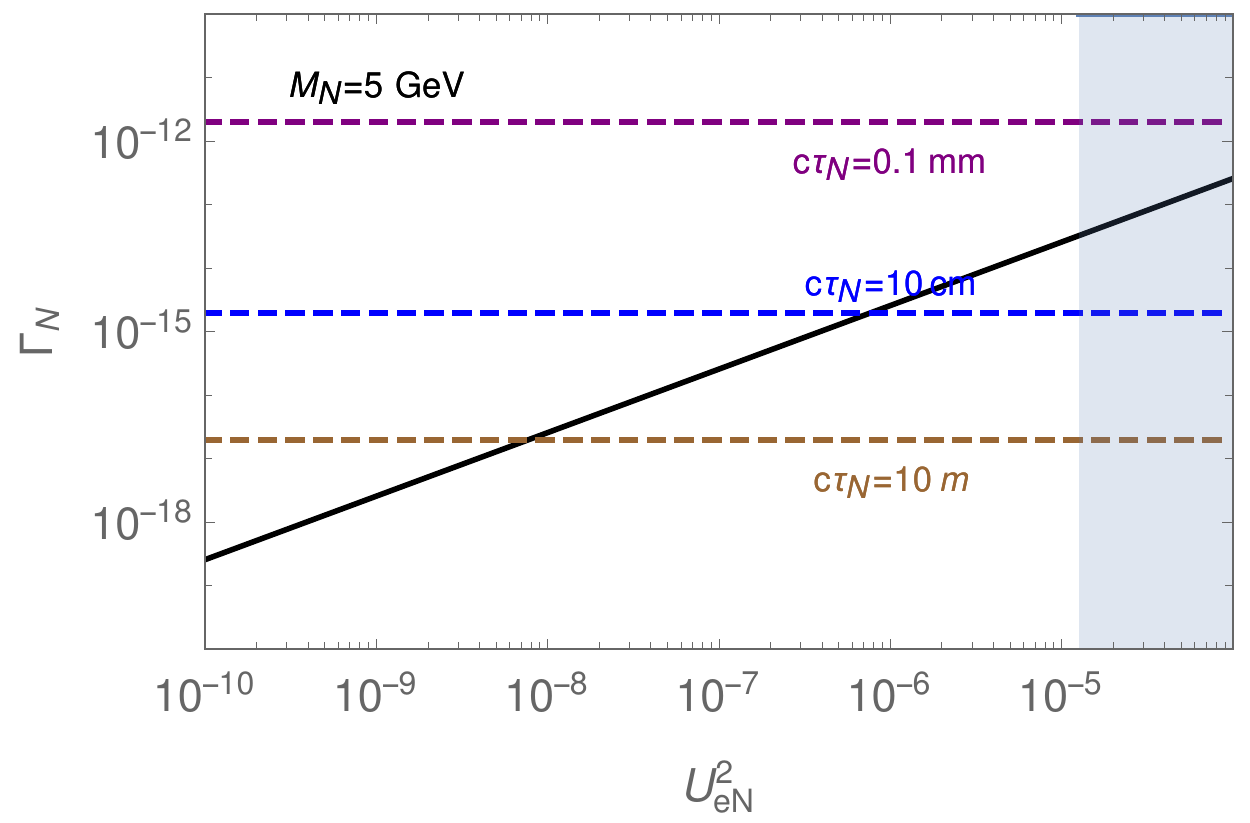}
\includegraphics[width=3in]{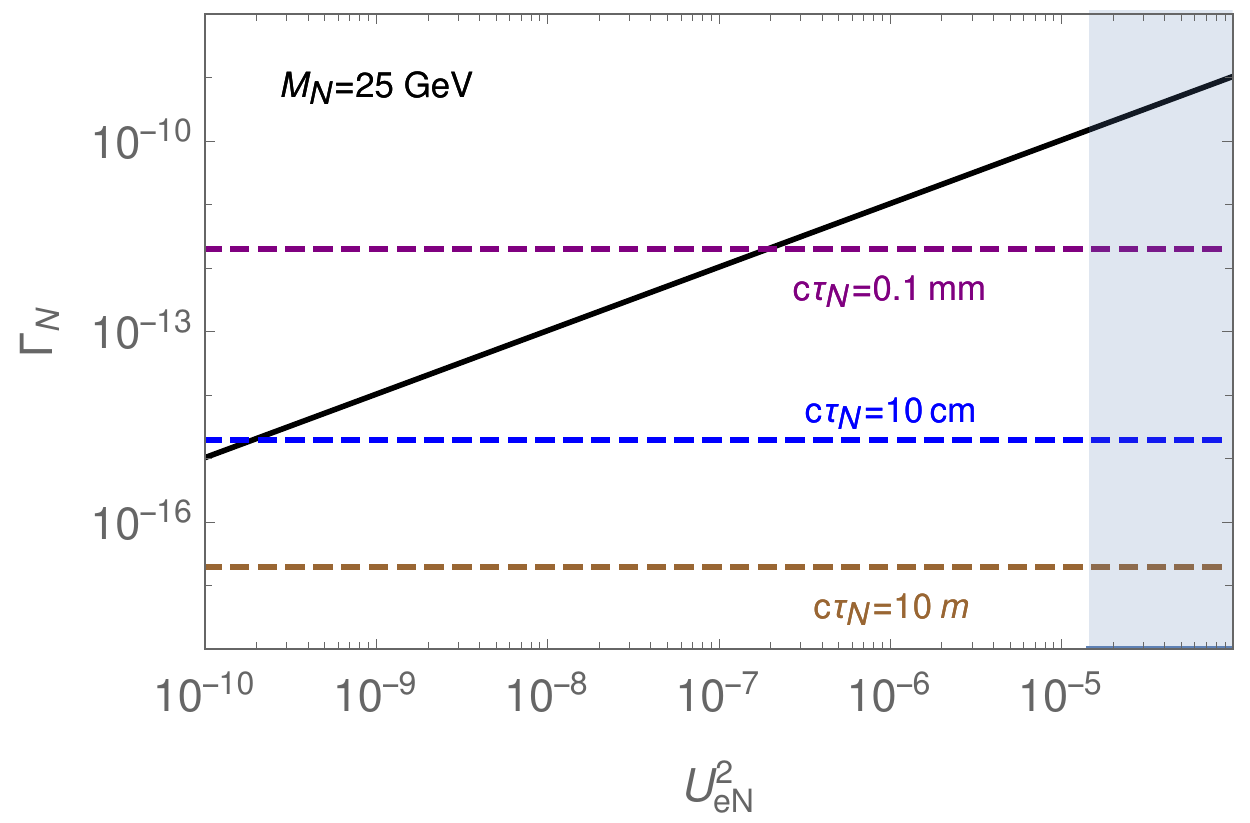}
\includegraphics[width=3in]{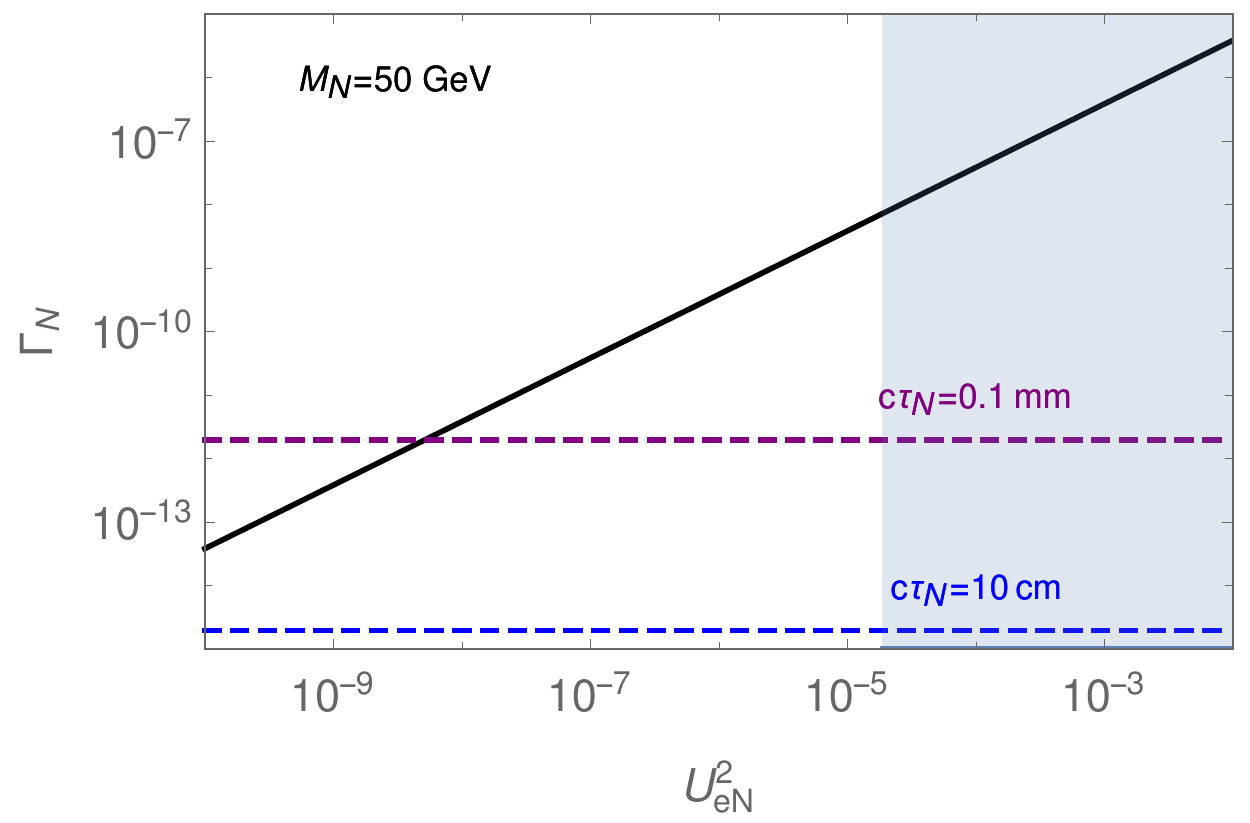}
\includegraphics[width=3in]{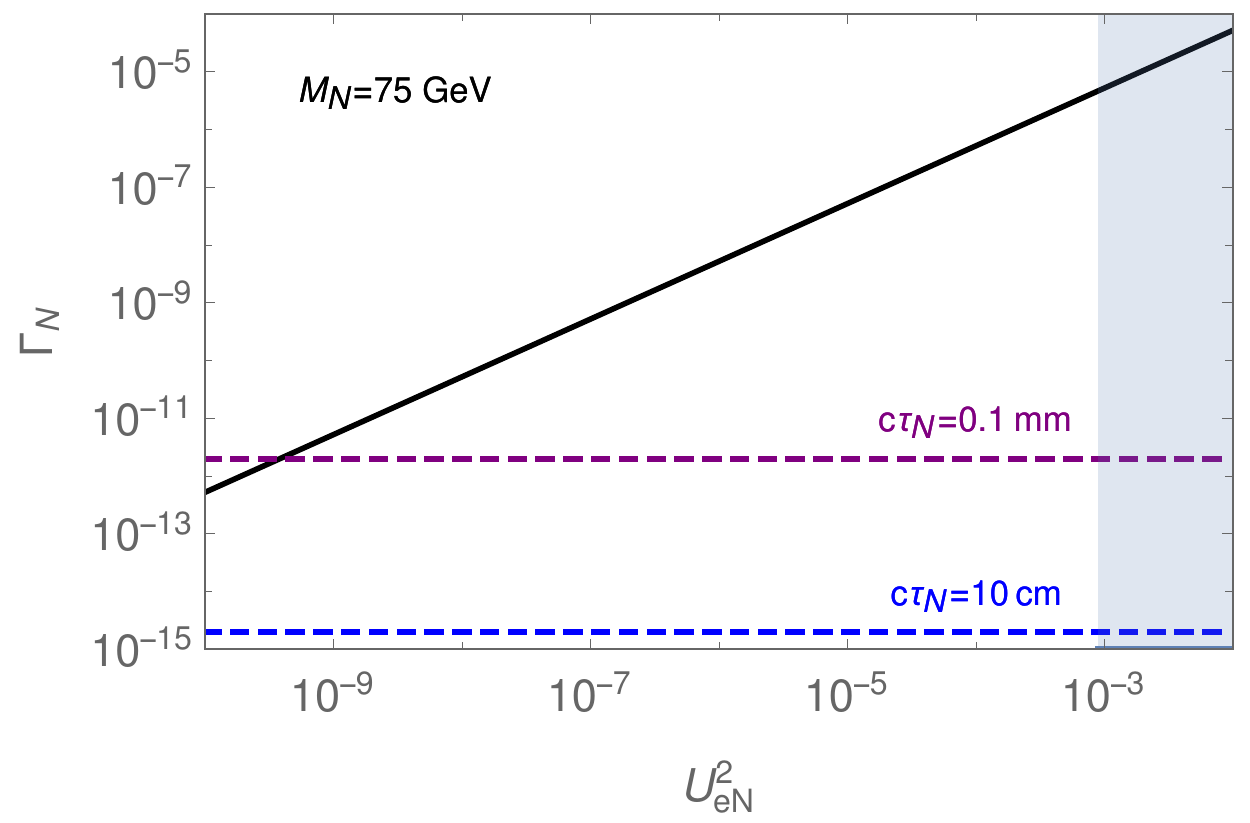}
\caption{
  The decay width $\Gamma_N$ versus the mixing parameter $U^2_{eN}$ (solid line)
  in the parameter space of ($U^2_{eN} $, $ \Gamma_N$) with
  $M_N$ = $5$ GeV (upper-left),   $25$ GeV (upper-right),
  $50$ GeV (lower-left) and $75$ GeV (lower-right).
  The shaded regions come from various constraints shown in
  Fig.~\ref{fig:e_cons}. The three dashed lines indicate the benchmark 
  decay lengths of $c \tau_N$ = $0.1$ mm (purple),
  $10$ cm (blue) and $10$ m (brown).
\label{fig:e_mix}
}
\end{figure}

\begin{figure}[t!]
\centering
\includegraphics[width=3in]{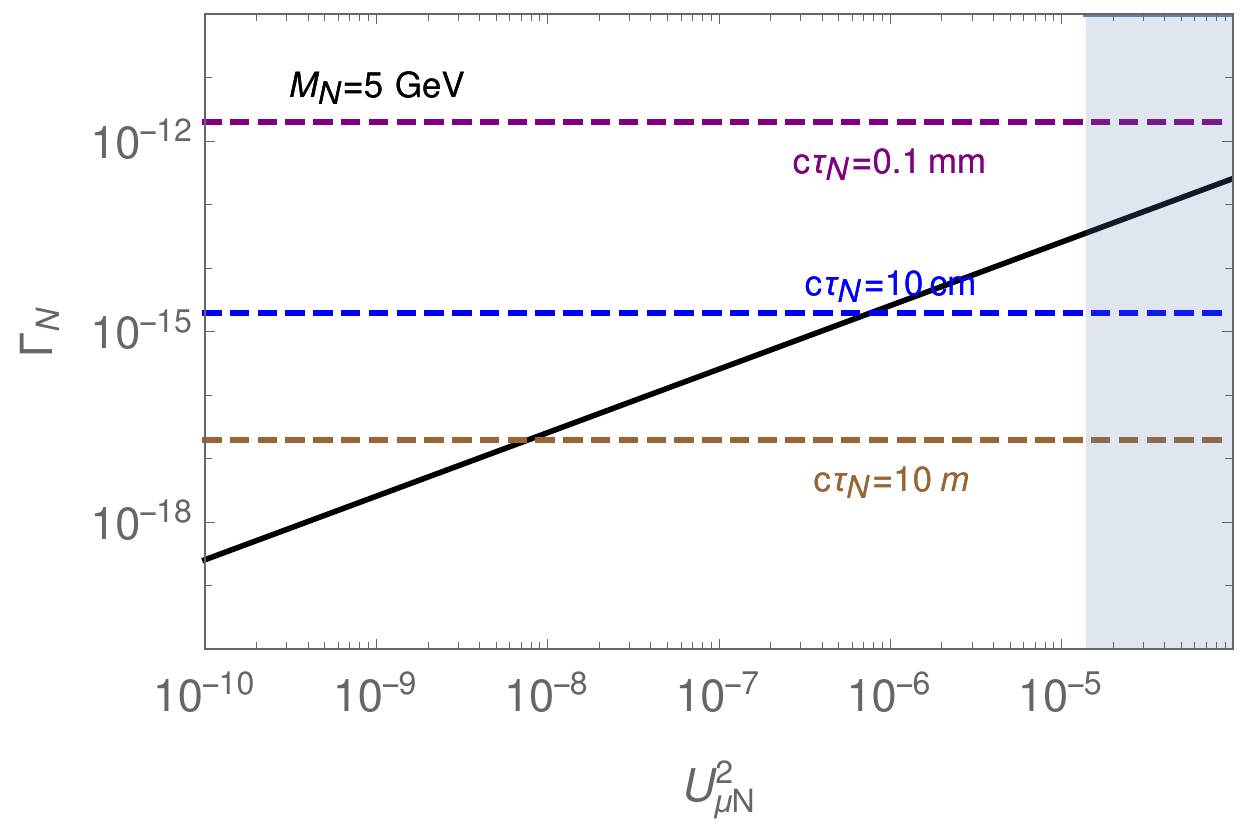}
\includegraphics[width=3in]{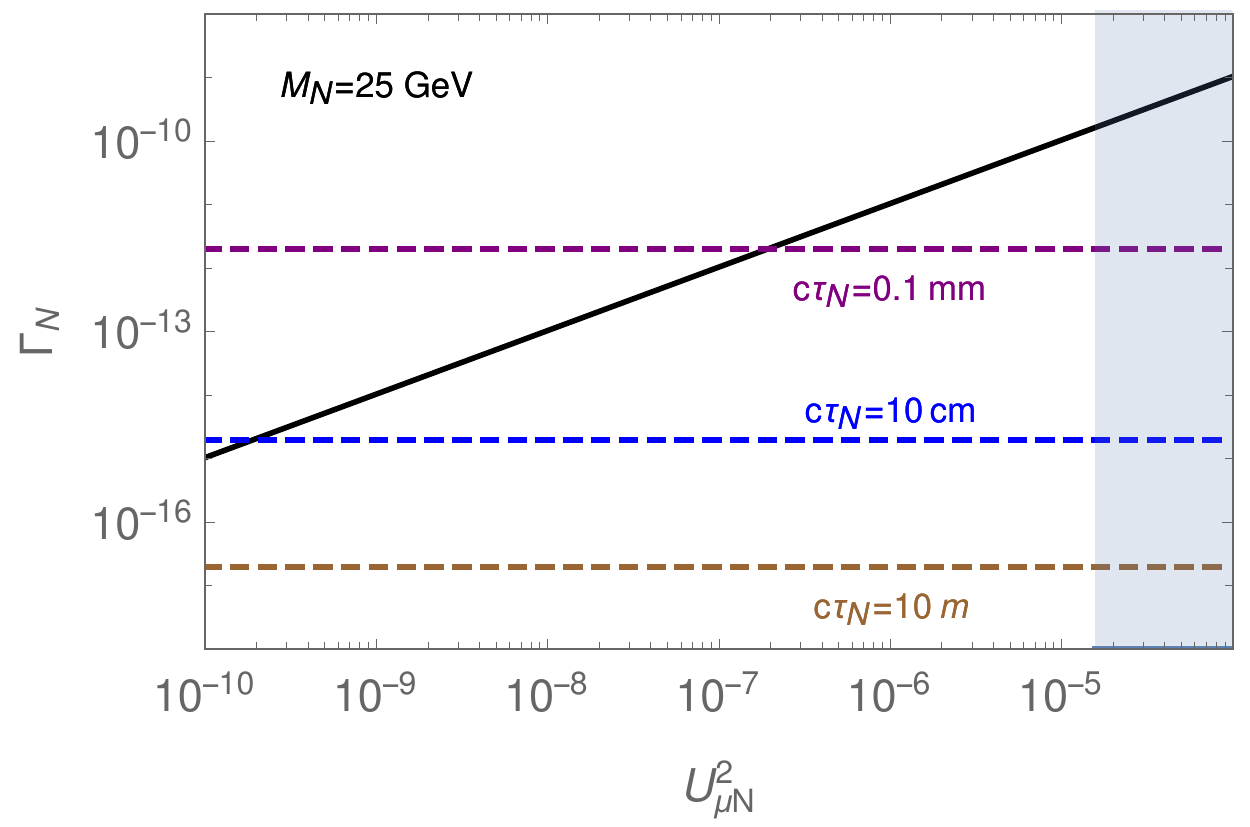}
\includegraphics[width=3in]{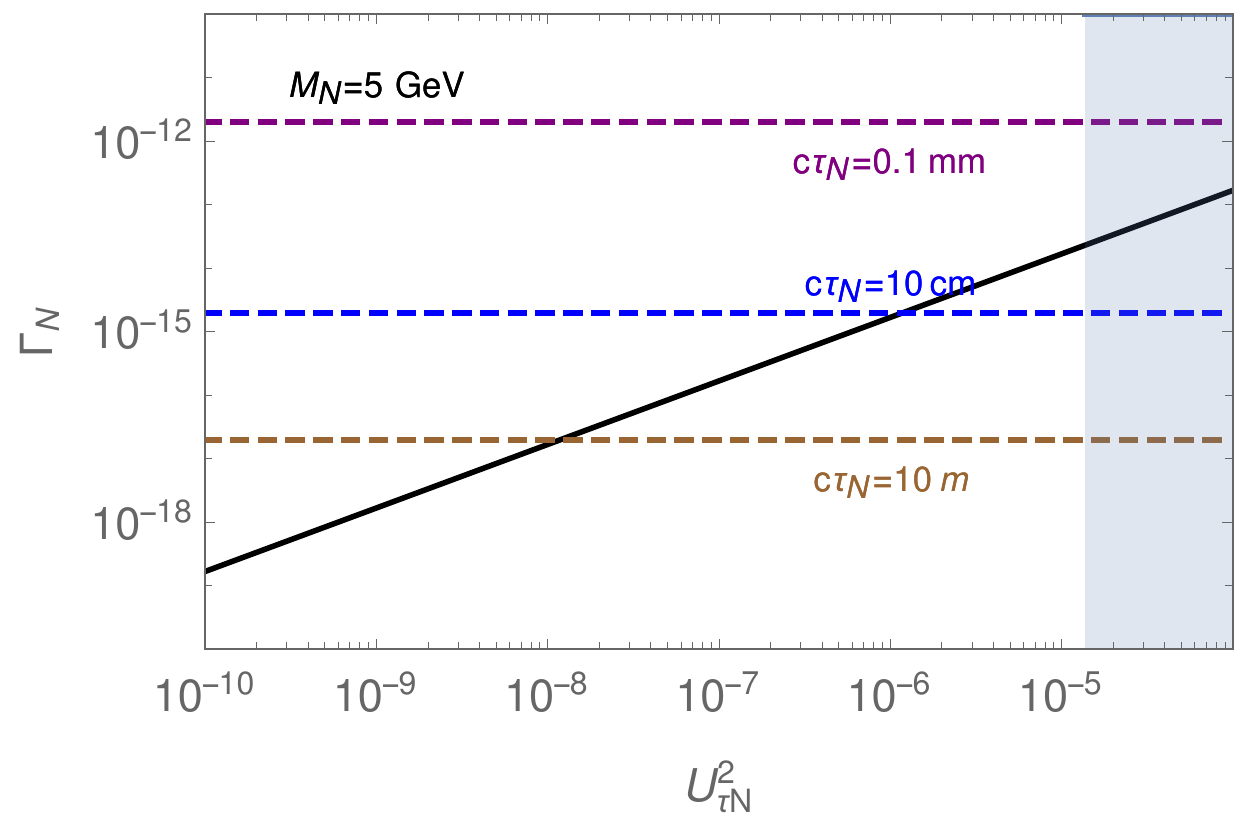}
\includegraphics[width=3in]{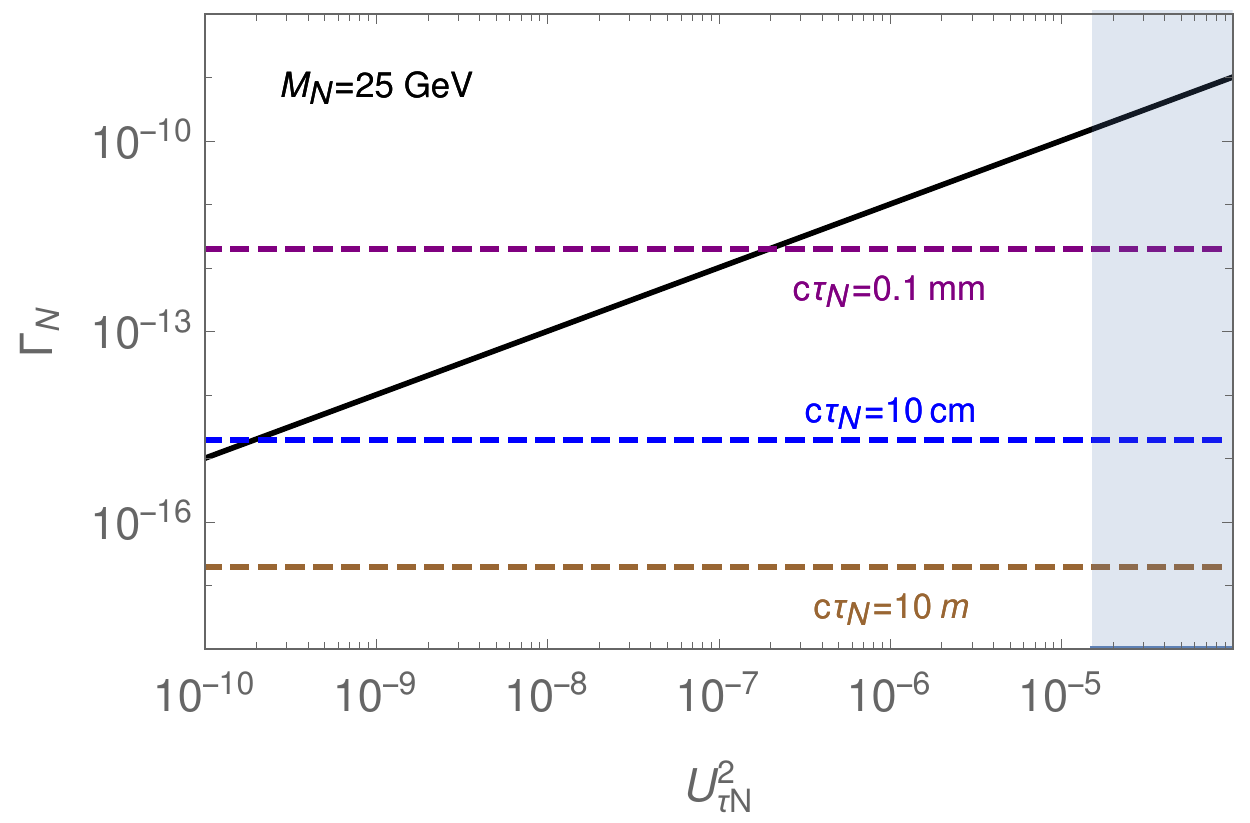}
\caption{
  The decay width $\Gamma_N$ versus the mixing parameter $U^2_{\mu N}$ (upper
  panels) or $U^2_{\tau N}$ (lower panels) in the parameter space of 
  ($U^2_{\mu N}$, $\Gamma_N$) with $ M_N $ = $5$ GeV (upper-left) and
  $25$ GeV (upper-right), and of  
  ($U^2_{\tau N}$, $\Gamma_N$) with $ M_N $ = $5$ GeV (lower-left)
  and $25$ GeV (lower-right). 
  The shaded regions come from various constraints shown in 
  Fig.~\ref{fig:mu_cons} and Fig.~\ref{fig:tau_cons}, respectively. 
  The three dashed lines indicate the benchmark decay 
  lengths of $c \tau_N $ = $0.1$ mm (purple),
  $10$ cm (blue) and $10$ m (brown).
\label{fig:mu_mix}
}
\end{figure}

The dependence of the total decay rate $\Gamma_N$ 
on the square of mixing parameter $U^2_{\alpha N}$ ($\alpha = e, \mu, \tau$) 
is numerically studied below.  We first show $\Gamma_{N}$ 
verse $U^2_{eN}$ with $ M_N $ = $5$, $25$, $50$ and $75$ GeV in Fig.~\ref{fig:e_mix}. 
Since we ignore the fermion mass in the final state 
for the medium mass range in our numerical calculations, 
there is no difference among the lepton flavors in this mass range. 
We show $\Gamma_{N}$ verse $U^2_{\mu N}$ and 
$\Gamma_{N}$ verse $U^2_{\tau  N}$ with only $M_N $ = $5$ and $25$ GeV in Fig.~\ref{fig:mu_mix}. 
The shaded regions come from the constraints
shown in Figs.~\ref{fig:e_cons} to \ref{fig:tau_cons} in the next section. 
Three dashed lines indicate the benchmark decay lengths of 
$c\tau_N $ = $0.1$ mm (purple), $10$ cm (blue) and $10$ m (brown).  
We observe that once $M_N \gtrsim$ $50$ GeV and $U^2_{\alpha N} \gtrsim 10^{-8}$, 
the decay length of HNL is quite small such that we can simply take 
the decay of HNL as prompt in most of the parameter space for each lepton 
flavor. In contrast, the low mass HNL with tiny $U^2_{\alpha N}$ can 
easily generate the displaced vertex signature after it 
has been produced at colliders~\cite{Alekhin:2015byh,Kling:2018wct,Curtin:2018mvb,Lee:2018pag,Abada:2018sfh,Dercks:2018wum,Alimena:2019zri,Aielli:2019ivi,Hirsch:2020klk}, 
which is of immense interest in the upcoming LHC run.

\section{Constraints for Heavy Neutral Leptons}\label{Sec:Conct}

\begin{figure}[t]
\centering
\includegraphics[width=5in]{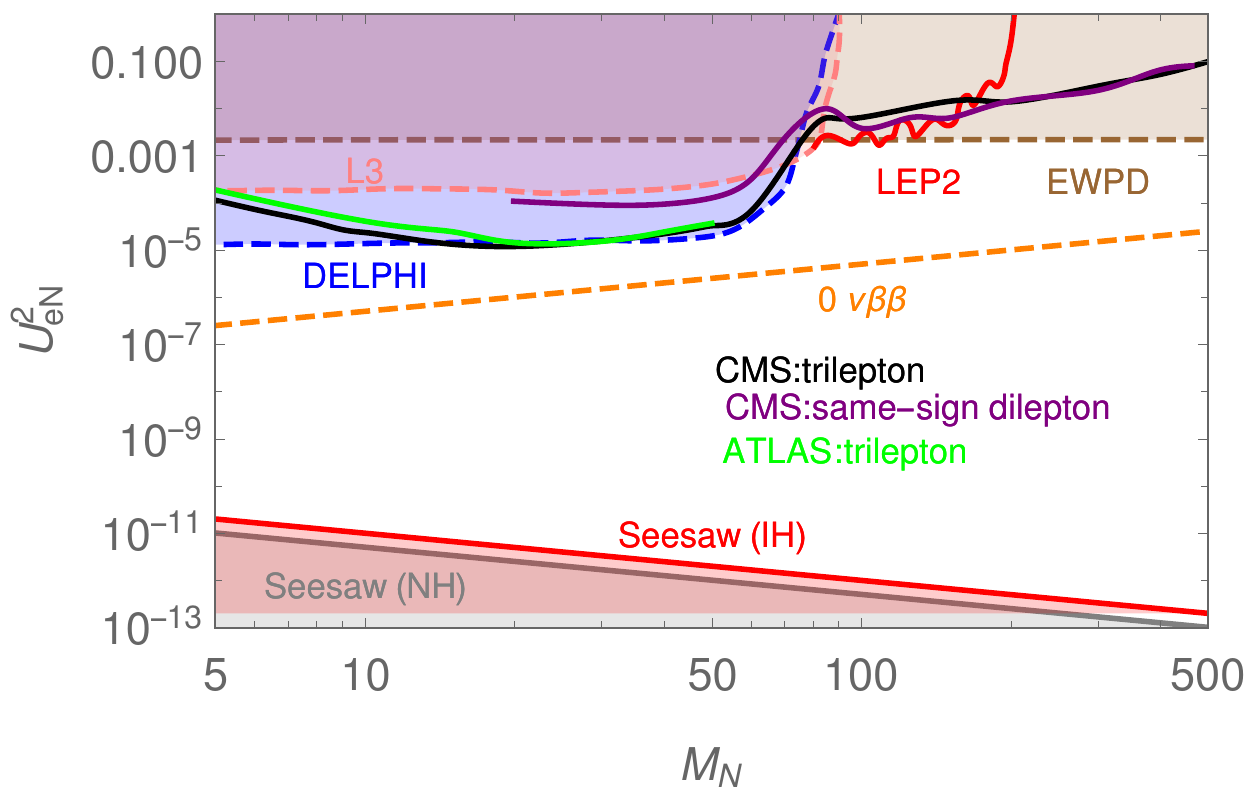}
\caption{
  The allowed parameter space of ($M_N $, $ |U_{eN}|^2$).
  We display the main constraints from 
  EWPD~\cite{delAguila:2008pw,Akhmedov:2013hec,Basso:2013jka,deBlas:2013gla,Antusch:2015mia} (brown dashed line), 
  L3~\cite{Adriani:1992pq,Acciarri:1999qj,Achard:2001qv} (pink dashed line), 
  DELPHI~\cite{Abreu:1996pa} (blue dashed line), 
  LEP2~\cite{Adriani:1992pq,Acciarri:1999qj,Achard:2001qv} (red solid line), 
  CMS-13TeV trilepton~\cite{Sirunyan:2018mtv} (black solid line), 
  CMS-13TeV same-sign dilepton~\cite{Sirunyan:2018xiv}( purple solid line), 
  ATLAS-13TeV trilepton~\cite{Aad:2019kiz} (green solid line), 
  $0 \nu \beta \beta$ (orange dashed line) 
  and Seesaw (NH) (Seesaw(IH)) (gray solid line (red solid line)) on the plane.
}\label{fig:e_cons} 
\end{figure}

\begin{figure}[t]
\centering
\includegraphics[width=5in]{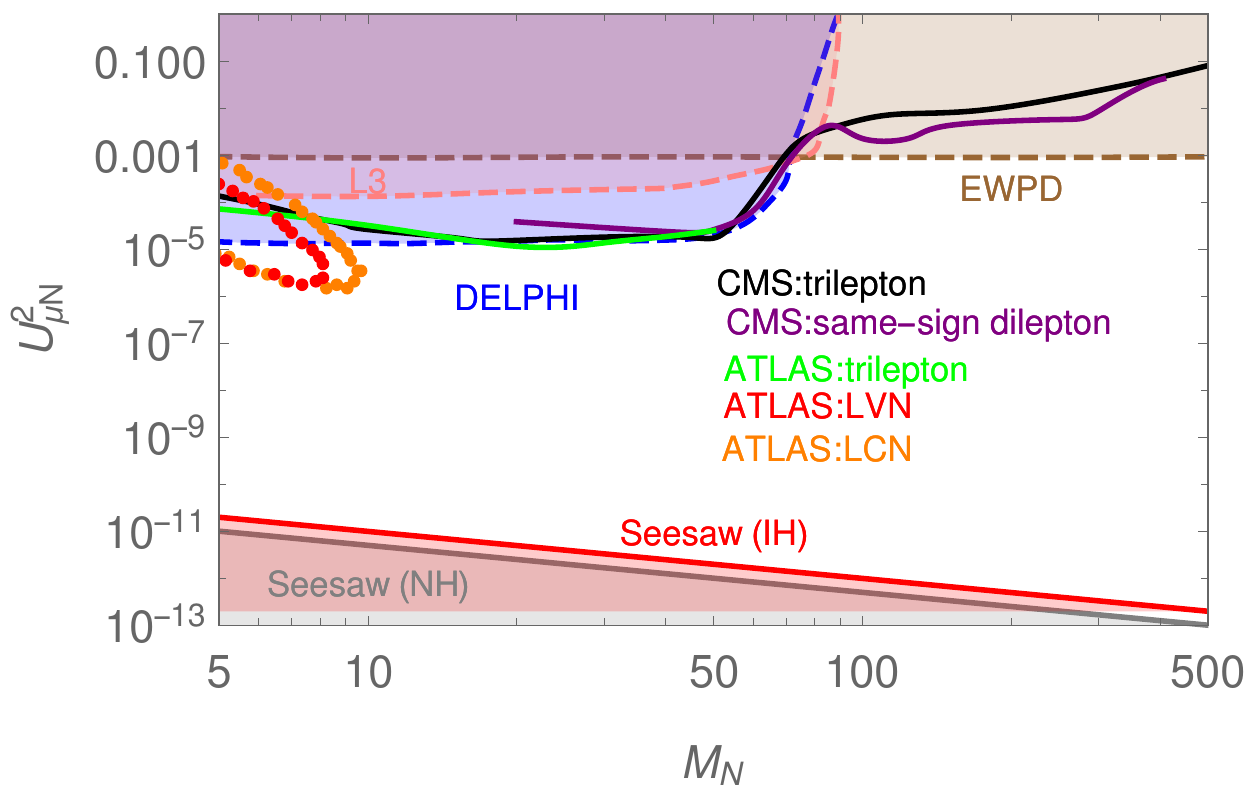}
\caption{
  The allowed parameter space of ($ M_N $, $ |U_{\mu N}|^2 $).
  We display the main constraints from 
  EWPD~\cite{delAguila:2008pw,Akhmedov:2013hec,Basso:2013jka,deBlas:2013gla,Antusch:2015mia} (brown dashed line),
  L3~\cite{Adriani:1992pq,Acciarri:1999qj,Achard:2001qv} (pink dashed line), 
  DELPHI~\cite{Abreu:1996pa} (blue dashed line), 
  CMS-13TeV trilepton~\cite{Sirunyan:2018mtv} (black solid line), 
  CMS-13TeV same-sign dilepton~\cite{Sirunyan:2018xiv} (purple solid line),
  ATLAS-13TeV trilepton, LVN and LCN~\cite{Aad:2019kiz} (green solid line,
  red dotted line and orange dotted line) 
  and Seesaw (NH) (Seesaw(IH)) (gray solid line (red solid line)) on the plane.
}\label{fig:mu_cons}
\end{figure}

\begin{figure}
\centering
\includegraphics[width=5in]{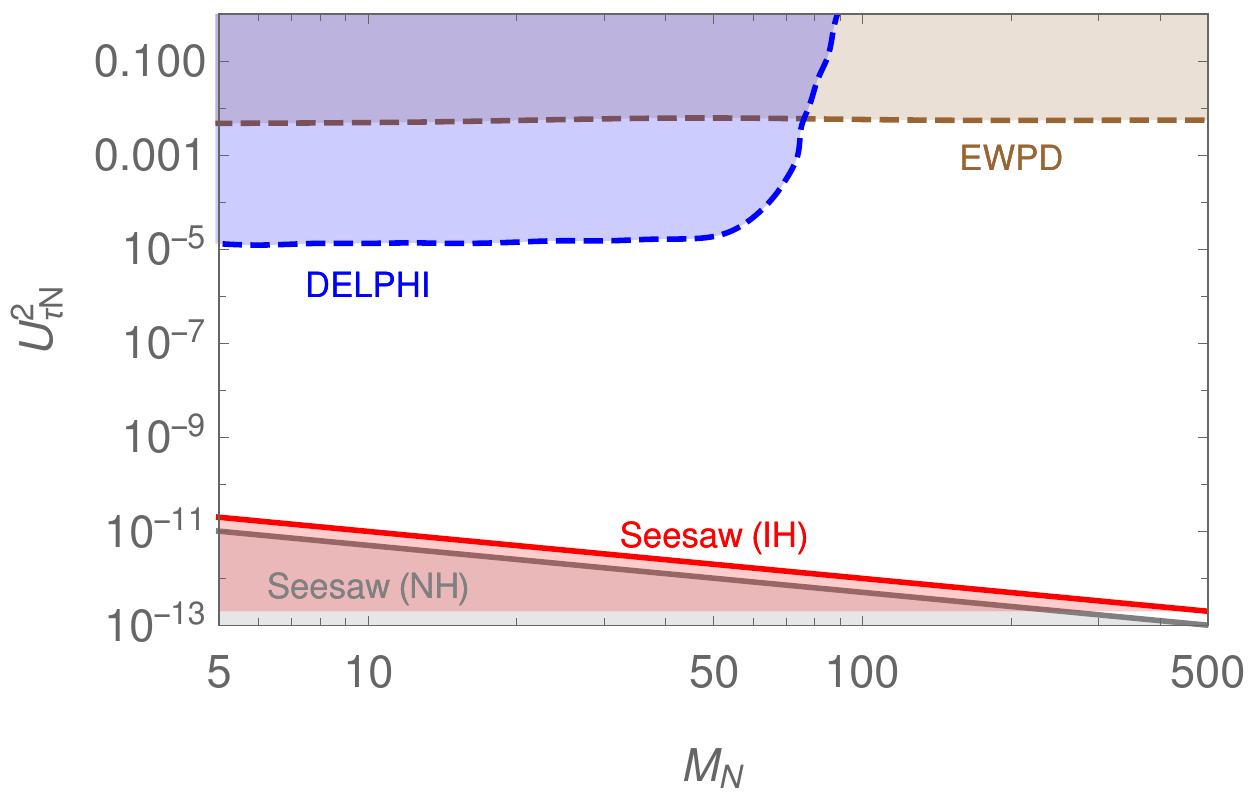}
\caption{
  The allowed parameter space of ($ M_N $, $ |U_{\tau N}|^2 $). 
  We display the main constraints from 
  EWPD~\cite{delAguila:2008pw,Akhmedov:2013hec,Basso:2013jka,deBlas:2013gla,Antusch:2015mia} (brown dashed line), 
  DELPHI~\cite{Abreu:1996pa} (blue dashed line) 
  and Seesaw (NH) (Seesaw(IH)) (gray solid line (red solid line)) on the plane.
}\label{fig:tau_cons}
\end{figure}


In this section, we summarize various constraints on the mixing
$|U_{\alpha N}|^2$ ($\alpha = e, \mu, \tau $) 
in the mass range of $ M_N $ from $5$ to $500$ GeV. 
We categorize these constraints as follows.
\begin{enumerate}
\item Electroweak Precision Data (EWPD)~\cite{delAguila:2008pw,Akhmedov:2013hec,Basso:2013jka,deBlas:2013gla,Antusch:2015mia},
\item Large Electron–Positron (LEP) Collider experiments,
  including L3~\cite{Adriani:1992pq,Acciarri:1999qj,Achard:2001qv},
  DELPHI~\cite{Abreu:1996pa}, and
  LEP2~\cite{Adriani:1992pq,Acciarri:1999qj,Achard:2001qv},
\item Large Hadron Collider (LHC) experiments, including CMS-13TeV
  trilepton~\cite{Sirunyan:2018mtv}, CMS-13TeV same-sign
  dilepton~\cite{Sirunyan:2018xiv} and ATLAS-13TeV trilepton~\cite{Aad:2019kiz},
\item Neutrinoless double beta ($0 \nu \beta \beta$) decay, and
\item Theoretical lower bound of the seesaw mechanism. 
\end{enumerate}
We first show the valid parameter space of ($M_N $, $ |U_{eN}|^2$) in Fig.~\ref{fig:e_cons}.  
Generally, $ |U_{eN}|^2\lesssim 2\times 10^{-5} $ for $ M_N \lesssim 50$ GeV. 
The main constraints for this mass region come from DELPHI~\cite{Abreu:1996pa}, 
CMS-13TeV trilepton~\cite{Sirunyan:2018mtv} 
and ATLAS-13TeV trilepton searches~\cite{Aad:2019kiz}. 
On the other hand, 
$|U_{eN}|^2 \lesssim 2.2 \times 10^{-3}$ for $M_N \gtrsim 100$ GeV 
from constraints of 
LEP2~\cite{Adriani:1992pq,Acciarri:1999qj,Achard:2001qv} and
EWPD~\cite{delAguila:2008pw,Akhmedov:2013hec,Basso:2013jka,deBlas:2013gla,Antusch:2015mia}. 
The jump of the $|U_{eN}|^2$ constraints from $M_N \approx 50$ to $100$ GeV 
comes from the threshold of gauge boson masses $m_{W,Z}$. 
In addition, we follow Eq.~(2.18) in Ref.~\cite{Atre:2009rg} for the 
constraint of $0 \nu \beta \beta$ decay which is the strongest in
Fig.~\ref{fig:e_cons}. 

Similarly, the valid parameter space of ($M_N $, $ |U_{\mu N}|^2$) is shown in Fig.~\ref{fig:mu_cons}. 
Again, $|U_{\mu N}|^2 \lesssim 2 \times 10^{-5}$ for $M_N \lesssim 50$ GeV, 
but $|U_{\mu N}|^2 \lesssim 9 \times 10^{-4}$ for $M_N \gtrsim 100$ GeV. 
Interestingly, the search for displaced-vertex signature of muons from HNL 
in the case of lepton-number violation (LNV) 
and lepton-number conservation (LNC) was published in Ref.~\cite{Aad:2019kiz} from the ATLAS Collaboration.  
The above searches set a stronger constraint for $M_N < 10$ GeV.  

Finally, we show the valid parameter space of ($ M_N$, $|U_{\tau N}|^2$) in Fig.~\ref{fig:tau_cons}.
The main constraints only come from
EWPD~\cite{delAguila:2008pw,Akhmedov:2013hec,Basso:2013jka,deBlas:2013gla,Antusch:2015mia}
and DELPHI~\cite{Abreu:1996pa} with $|U_{\tau N}|^2\lesssim 5.5\times 10^{-3}$ for $M_N \gtrsim 100$ GeV. 
 
We observe that the constraints on the mixing between $\nu _{\tau}$ 
and HNLs are relatively weaker than both $\nu _e $ and $ \nu _{\mu}$ in the electroweak scale HNLs.  
On the other hand, we approximately apply the $m_3$ value of Normal Hierarchical (NH) case 
and $m_2, m_1$ values of Inverted Hierarchical (IH) case 
from PDG 2018~\cite{Tanabashi:2018oca}, respectively, 
to set the theoretical lower bound of the seesaw mechanism 
for the mixing angles ($ M_N $, $ |U_{\alpha N}|^2$) in Figs.~\ref{fig:e_cons} to \ref{fig:tau_cons}.

\section{Search for the HNL with $ \tau $ lepton at HL-LHC}\label{Sec:Strategy}

To our knowledge there have not been any concrete analyses for
the sensitivity reach of $ U^2_{\tau N} $ 
for HNLs around the EW scale at the LHC.
Here we propose to search for HNLs with the signatures consisting of three
prompt charged leptons in the final state, of 
which at least two are tau leptons.
We first study the kinematical behavior of the HNL in the production channel, 
$ pp\rightarrow W^{\pm(\ast)} + X\rightarrow \tau^{\pm} N + X $, 
and then discuss the signatures for various final states from the HNL decays 
and discuss possible SM backgrounds. 
Finally, the details of simulations and event selections for both signals and SM backgrounds are displayed. 

\subsection{Kinematical behavior of the HNL in the production channel
  $ pp\rightarrow W^{\pm(\ast)} + X\rightarrow \tau^{\pm} N + X $}

\begin{figure}
\centering
\includegraphics[width=3.2in]{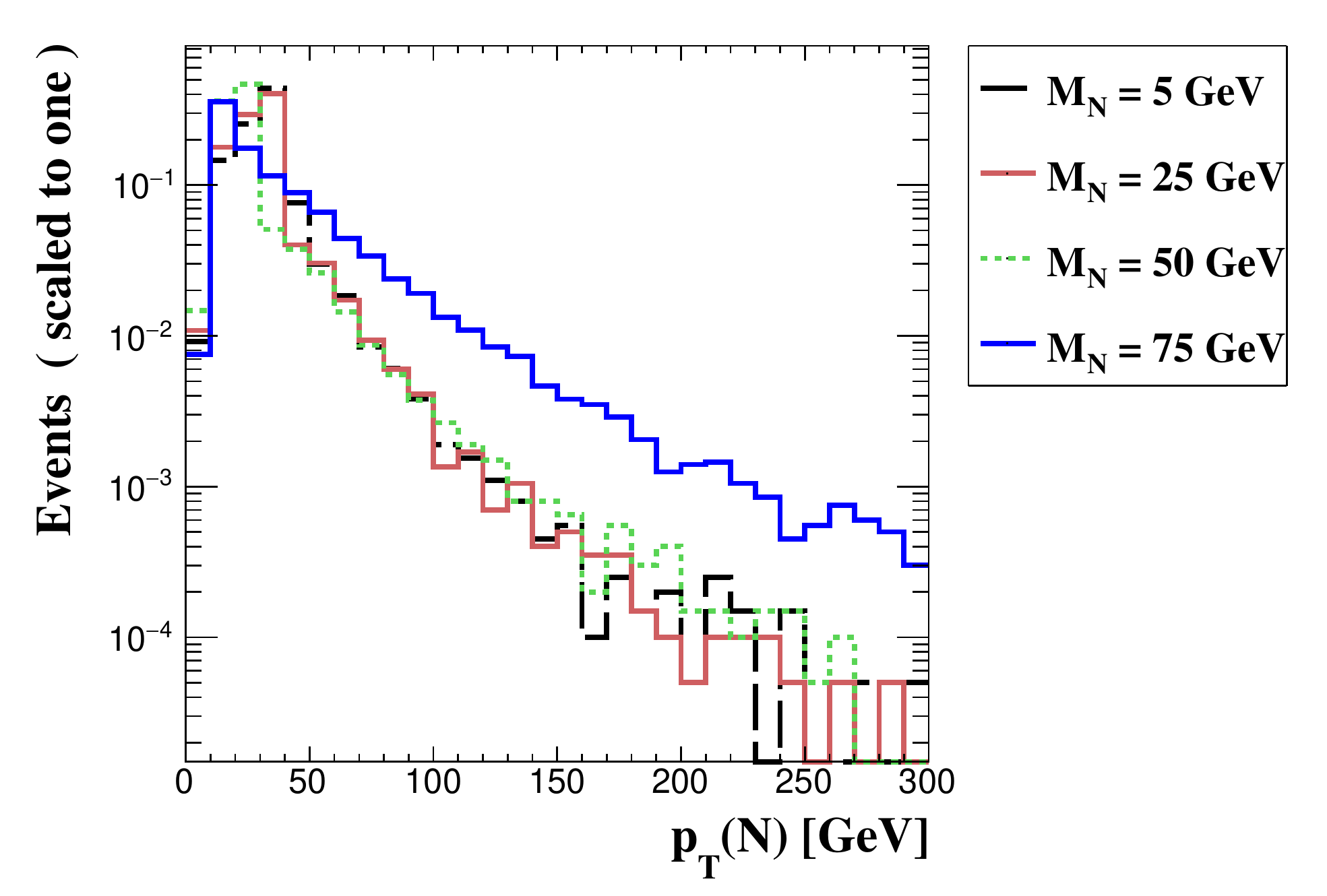}
\includegraphics[width=3.2in]{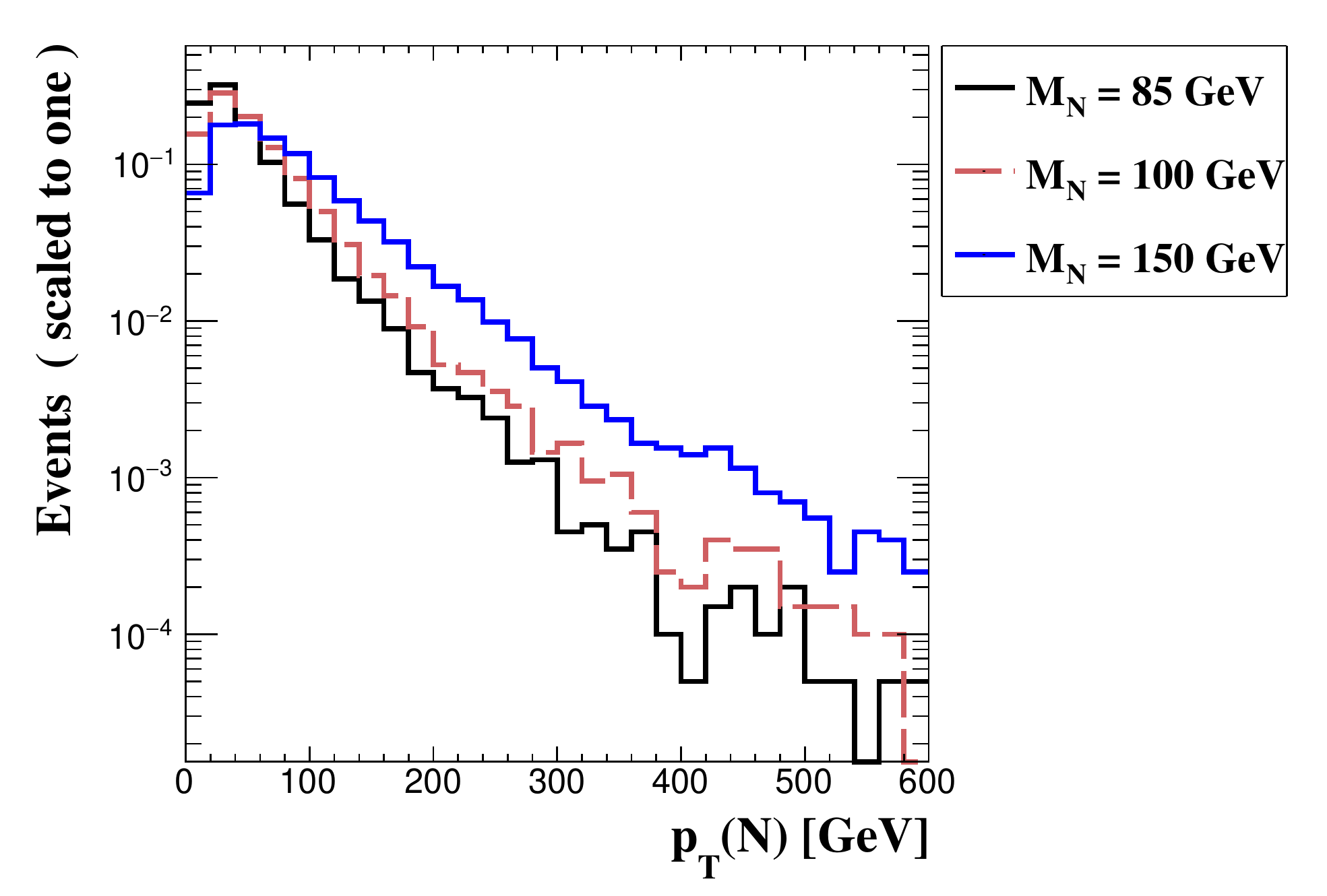}
\caption{
  The transverse momentum $p_T(N)$ distribution of the HNL
  in the process $pp \rightarrow W^{\pm(\ast)} \rightarrow \tau^{\pm} N + X$
  at $ \sqrt{s} = 14 $ TeV for some benchmark points with
  $M_N < m_W$ (left) and $M_N > m_W$ (right) at parton level.
}\label{fig:N_pT}
\end{figure}

\begin{figure}
\centering
\includegraphics[width=3.2in]{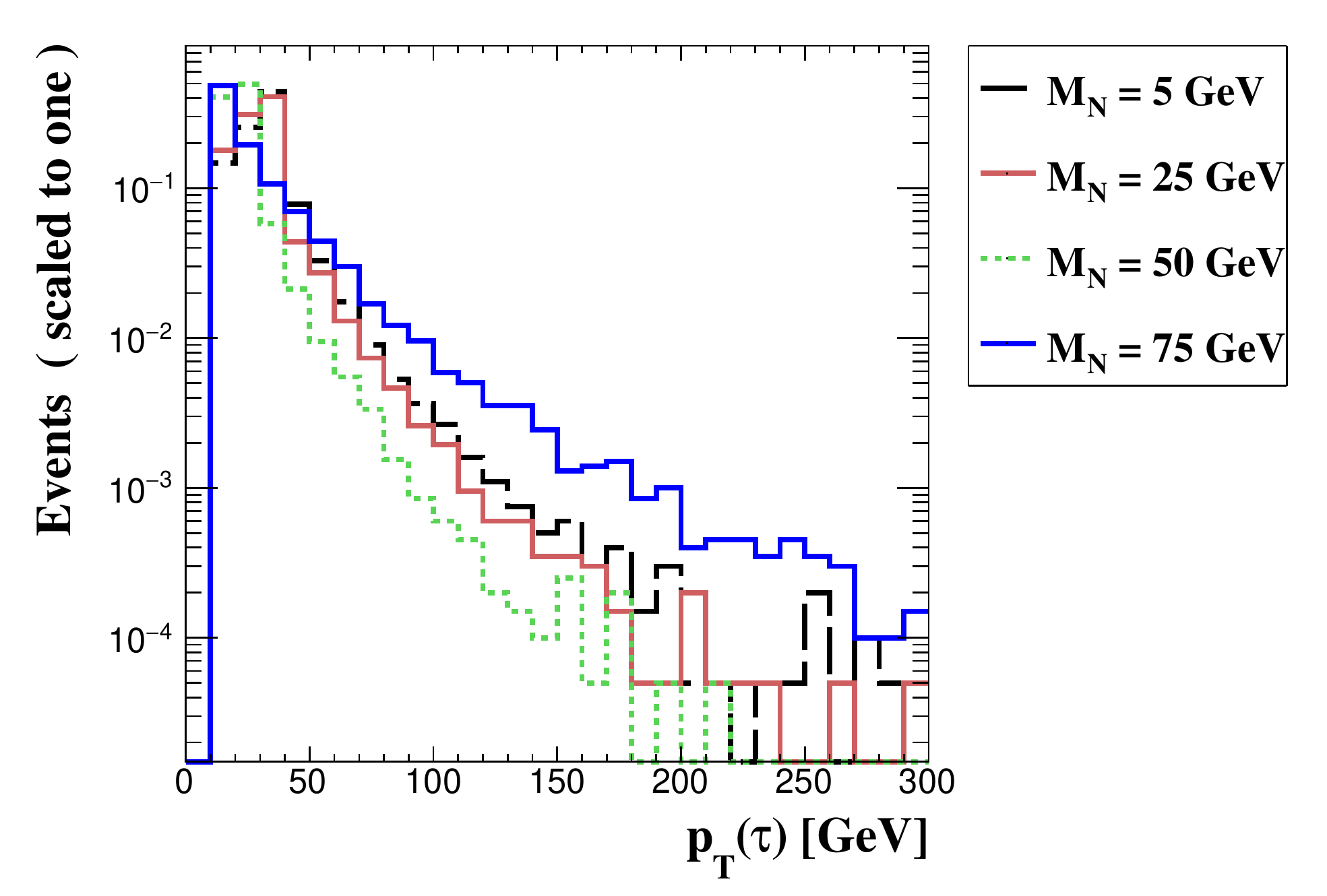}
\includegraphics[width=3.2in]{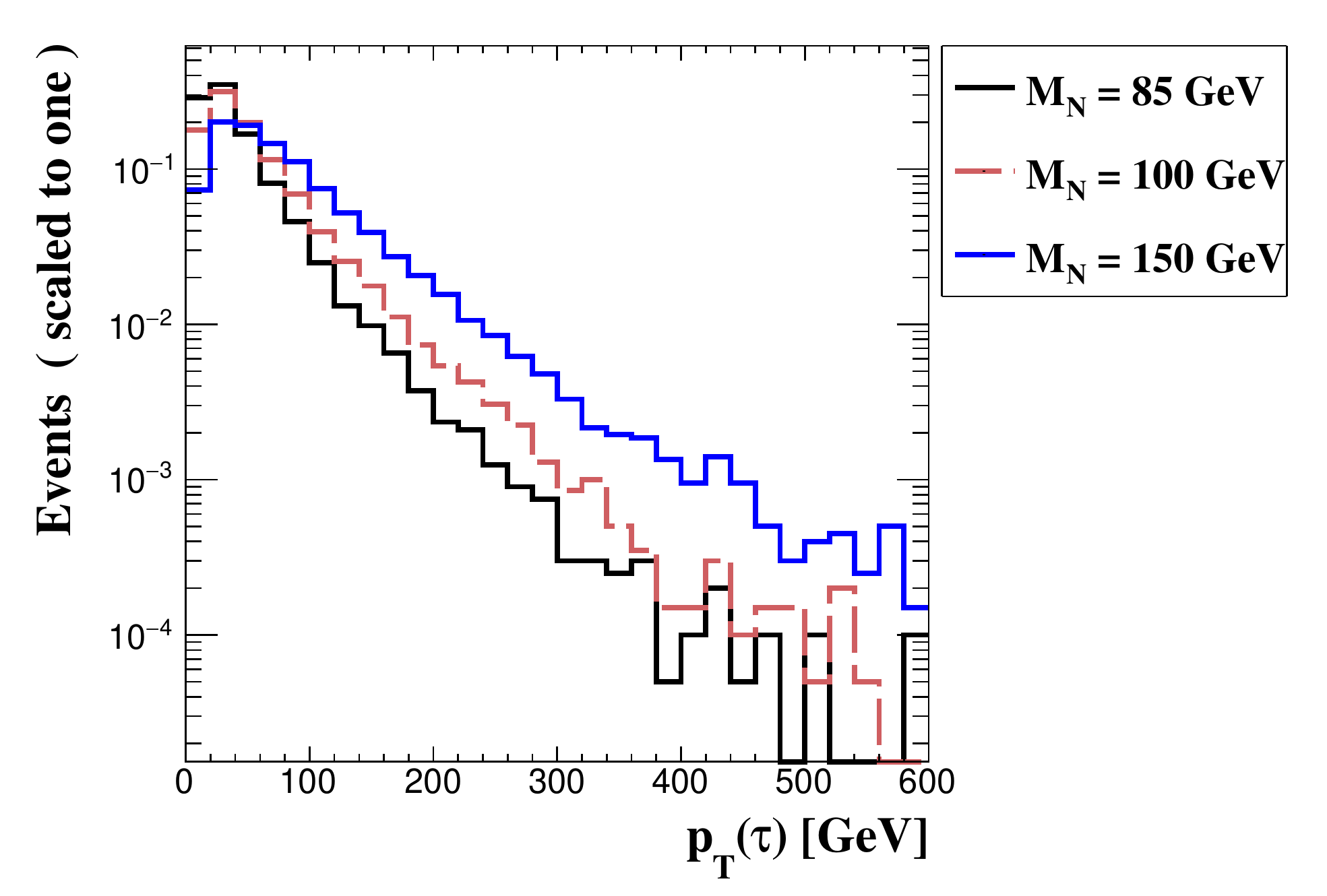}
\caption{
  The transverse momentum $p_T(\tau )$ distribution of the $\tau$ lepton
  in the process $pp \rightarrow W^{\pm(\ast)} \rightarrow \tau^{\pm} N + X$
  at $ \sqrt{s} = 14 $ TeV for the same benchmark points as
  Fig.~\ref{fig:N_pT}.
}\label{fig:tau_pT}
\end{figure}

Based on the fact that the constraints on the mixing  between
$\nu _{\tau}$ and HNLs  
are relatively weaker than those of $\nu _e$ and $\nu _{\mu}$ in
various HNL mass ranges, 
we study the channel
$pp\rightarrow W^{\pm(\ast)} + X \rightarrow \tau^{\pm} N + X$ 
at the LHC 14 TeV to search for HNLs in this work.
We first set $U^2_{eN} = U^2_{\mu N} = 0$ 
and only focus on the $U^2_{\tau N}$ dependence in the above production channel. 
The $W$ boson propagator can be either on-shell or off-shell depending on the mass of HNLs. 
We apply the Heavy Neutrino model file~\cite{Degrande:2016aje} 
from the model database of \textbf{FeynRules}~\cite{Alloul:2013bka} 
and use \textbf{Madgraph5 aMC@NLO}~\cite{Alwall:2014hca,Frederix:2018nkq} 
to simulate this production channel at tree level and include the emission of up to two additional partons. 
The $p_T (N)$ and $p_T (\tau)$ distributions for some benchmark points 
with $M_N < m_W$ ($M_N > m_W$) at parton level are shown 
in the left (right) panel of Figs.~\ref{fig:N_pT}
and~\ref{fig:tau_pT}, respectively. 
Because of the mass thresholds of the $W$ boson and HNLs,
the $p_T (N)$ and $p_T(\tau)$ 
are relatively soft for $M_N < m_W$, especially for the case of $M_N = 75$ GeV. 
To identify and detect these soft final states are the main issue of this study. 
On the other hand, a detail study for the situation of $M_N \sim m_W$ is
needed, and we leave this part in future.

The decay length $L_N$ of the HNLs can be simply estimated by 
$L_N = \gamma c \tau_N$ where $\tau_N = 1/\Gamma_N$ 
and the Lorentz boost factor $\gamma$ can be approximated as $p_T(N)/M_N$. 
We expect that HNL is not very boosted in this production channel 
except for $M_N  = 5$ GeV in Fig.~\ref{fig:N_pT}. 
Combined with the information from Figs.~\ref{fig:e_mix} and~\ref{fig:mu_mix}, 
there is still large allowed parameter space for prompt decays of
HNLs in this production channel. 
Therefore, we focus on the case with prompt decays of HNLs first 
and leave the displaced vertex of HNLs aside in this paper.

\subsection{Signature of the signals and possible SM backgrounds}

We first divide the signal region to two parts: 
(1) on-shell $W$ boson production region and 
(2) off-shell $W$ boson production region. 
Different analysis strategies will be applied to each signal 
and SM backgrounds in these two regions. 
We focus on those final states with two $\tau$ leptons 
and one additional charged lepton in this work, 
and will explore the signature of two same-sign $\tau$ leptons with two jets 
as Ref.~\cite{Keung:1983uu,Das:2017gke,Das:2018usr} in the future. 
As we known, the $\tau^{\pm} \tau^{\pm} j j$ search channel would suffer from the severe QCD backgrounds such that signal events may be easily submerged.
Conversely, the signature of two $\tau$ leptons 
with one additional charged lepton can effectively reduce those huge QCD backgrounds, 
but we need to carefully exploit kinematic properties of the final states 
to discriminate between the signal and SM backgrounds.

Consider the following signal process
\begin{equation}
\begin{aligned}
pp 
\rightarrow 
W^{\pm(\ast)} \rightarrow \tau^{\pm} N 
&\rightarrow 
\tau^{\pm} \tau^{\pm} l^{\mp}_\alpha \overline{\nu_{\alpha}} (\nu_{\alpha}) \\
&\hookrightarrow 
\tau^{\pm} \tau^{\mp} l^{\pm}_\alpha \nu_{\alpha} (\overline{\nu_{\alpha}})
\end{aligned}
\end{equation}
where $\alpha = e, \mu, \tau $. 
We can further classify the final states in the following three categories: 
(1) Two same-sign $\tau$s, $e/ \mu$ 
and $mET$ ($\tau^{\pm} \tau^{\pm} e^{\mp} (\mu^{\mp}) \overline{\nu_{e,\mu}} (\nu_{e,\mu})$), 
(2) Two opposite-sign $\tau$s, $e/ \mu$ 
and $mET$ ($\tau^{+} \tau^{-} e^{\pm} (\mu^{\pm}) \nu_{e,\mu} (\overline{\nu_{e,\mu}}$) and  
(3) Three $\tau$s and $mET$ ($\tau^{\pm} \tau^{\pm} \tau^{\mp} \overline{\nu_{\tau}} (\nu_{\tau})$). 
We will ignore the analysis of three $\tau$s and $mET$ final state, as 
we cannot distinguish the Majorana or Dirac nature of the HNL via
the three $\tau$ leptons and $mET$ final state, in contrast to
the first two categories.

As shown before, there are still some possibilities 
to search for displaced $\tau$ leptons events from the low $M_N$
region with small mixing angles.
This kind of signature has been studied in Ref.~\cite{Cottin:2018nms}. 
Therefore, we mainly focus on the prompt $\tau$s in this work.   
On the other hand, $\tau$ leptons have both hadronic and leptonic decay modes. 
We choose hadronic $\tau$ lepton decays for all $\tau$ leptons in our study with the following two main reasons. 
First, hadronic $\tau$ lepton decays account for approximately
$65\%$ of all possible $ \tau $ lepton decay modes. 
Therefore, we can save more $\tau$ lepton decay events from hadronic decay modes than leptonic decay modes. 
Second, leptonic $\tau$ lepton decays can mimic the signals of only $e$'s 
and $\mu$'s in the final state which cannot be distinguished at the LHC.

There are some irreducible and reducible SM backgrounds for the above
three categories of signatures.
We first consider the signal signature with two same-sign 
$\tau$s, $e/ \mu$ and $mET$, the backgrounds of which include
\begin{enumerate}
\item Irreducible SM backgrounds: \\
$W^{\pm}W^{\pm}W^{\mp}$.
\item Reducible SM backgrounds: \\
(1) EW processes : $ W^+W^-Z/H/\gamma^{\ast} $. \\
(2) $ t\overline{t} $ associated processes: 
$ t\overline{t}W^{\pm}/Z/H/\gamma^{\ast} $ and $ t\overline{t} + nj $ $(n = 0 \mathchar`- 2)$. \\
(3) QCD multijets.
\end{enumerate}  
Then we consider the signal signature with two opposite-sign
$ \tau $s, $ e/ \mu $ and $ mET $, the backgrounds of  which include
\begin{enumerate}
\item Irreducible SM backgrounds: \\
$ W^{\pm}Z/H/\gamma^{\ast} $, and $ W^{\pm}W^{\pm}W^{\mp} $.
\item Reducible SM backgrounds: \\
(1) EW processes : $ ZZ/H/\gamma^{\ast} $ and $ W^+W^-Z/H/\gamma^{\ast} $. \\
(2) $ t\overline{t} $ associated processes: 
$ t\overline{t}W^{\pm}/Z/H/\gamma^{\ast} $ and $ t\overline{t} + nj $ $(n = 0 \mathchar`- 2)$. \\
(3) $ \tau^+\tau^- + nj $ $(n = 0 \mathchar`- 2)$. \\
(4) QCD multijets.
\end{enumerate}
Finally, the sources of SM backgrounds for the signal signature 
with three prompt $\tau$s and $mET$ are similar to
those of two opposite-sign $\tau$s, $e/ \mu$ and $mET$. 
We will not repeatedly list them again.

\subsection{Simulations and event selections}

We use \textbf{Madgraph5 aMC@NLO}~\cite{Alwall:2014hca,Frederix:2018nkq} 
to calculate the signal and background processes at leading order (LO) and generate MC events, 
perform parton showering and hadronization by \textbf{Pythia8}~\cite{Sjostrand:2007gs}, 
and employ the detection simulations by \textbf{Delphes3}~\cite{deFavereau:2013fsa} with the ATLAS template. 
The NNPDF2.3LO PDF set was used and ME-PS matching 
with MLM prescription~\cite{Mangano:2006rw,Alwall:2007fs} was applied for all the signal and major SM backgrounds.
We include the emission of up to two additional partons for the signals 
with a matching scale set to be $ 30 $ GeV for $M_N \lesssim 120$ GeV 
and about one quarter of the $M_N$ for $M_N > 120$ GeV. 
On the other hand, the matching scales for $t \overline{t} + n j$ and $\tau^+\tau^- + n j$ 
$(n = 0 \mathchar`- 2)$ are set to be $20$ GeV and $30$ GeV, respectively. 
All jets are reconstructed using the the anti-$k_T$ algorithm~\cite{Cacciari:2008gp} 
in \textbf{FastJets}~\cite{Cacciari:2011ma} with a radius parameter of $R=0.6$. 
The procedures of hadronic tau lepton decay and reconstruction are as follows. We first set tau leptons to automatically decay through Pythia8, and use the ATLAS template in Delphes3 for the tau tagging algorithm according to the efficiencies shown in Ref.~\cite{ATLAS:2019uhp} to reconstruct hadronic tau lepton decay using
the visible final states. Notice the tau lepton cannot be fully reconstructed because the part from neutrino becomes missing energy and 
is ignored from the hadronic tau reconstruction. 
Furthermore, the electron, muon efficiencies in Delphes3 are modified 
to include the low $ P_T $ regions inspired from the Ref.~\cite{Aad:2019qnd}.
In order to study the Majorana nature of HNLs at the LHC, 
we classify our simulations and event selections in 
(1) two same-sign $\tau$s, $e/ \mu$ and $mET$ 
and (2) two opposite-sign $\tau$s, $\mu$ and $mET$.%
\footnote{
In order to suppress the SM background contributions from both $\tau^+\tau^- + n j$ 
and $t \overline{t} + n j $ $(n = 0 \mathchar`- 2)$ with non-negligible jet fake to electron rate, 
we don't take into account of the signature with two opposite-sign $\tau$s, $e$ and $ mET $ in this study.}

\subsubsection{Two same-sign $\tau$s, $e/ \mu$ and $mET$}

In this scenario, we require two same-sign $\tau$ leptons with an
additional $e/ \mu$ 
in the final state with the following cut flow.
\begin{enumerate}
\item For $M_N < m_W$, we specifically take two soft same-sign $\tau$ leptons 
and an extra soft $e/ \mu$ as the selection of signals in our events with the following conditions,\footnote{
Note that the $p_T$ cuts on the hadronic tau-leptons are slightly below
the recomended values in the public trigger menu~\cite{menu}. Nevertheless,
it would only lead to marginal decrease in projected sensitivities. 
On the other hand,
except for the known public trigger thresholds for lepton pairs or single tau-lepton inclusive processes as shown in Ref.~\cite{menu},
we envision a trilepton trigger that includes hadronic tau-lepton candidates.
}  
\begin{align}
N(\tau^{\pm},l^{\mp}) \geqslant 2,1,\quad 5 < P^{l}_T < 40\;{\rm  GeV},
\quad |\eta^{l}| < 2.5,\quad
20 < P^{\tau_1 (\tau_2)}_T < 50 (40)\; {\rm GeV},\quad |\eta^{\tau}|<2.5\,,
\label{eq:trigger-ss1}
\end{align}
where $ l = e, \mu $.
Since $ \tau $ leptons and $ e/ \mu $ are relatively soft in this case compared with SM backgrounds, 
we reject those high $ P_T $ regions to reduce background contributions 
inspired from the Ref.~\cite{Florez:2016lwi,Aboubrahim:2017aen,Sirunyan:2018iwl,Sirunyan:2019mlu,Aad:2019qnd}. 
We follow Ref.~\cite{ATLAS:2019uhp} with $p^{min}_T > 10$ GeV for jet-seeding of visible hadronic tau to start with and only visible hadronic tau candidates with $p^{min}_T > 20$ GeV are used. However, we think it is still worthwhile to tell the readers about the situation with $p^{min}_T > 15$ GeV as in Ref.~\cite{Florez:2016lwi,Aboubrahim:2017aen}, so we place the cut flow tables for this case into Appendix~\ref{app:tables_figures}.
On the other hand, for $ M_N > m_W $, we only choose the following conditions for them: 
\begin{align}
N(\tau^{\pm},l^{\mp})\geqslant 2,1,\quad P^{l}_T > 10\;{\rm  GeV},
\quad |\eta^{l}| < 2.5,\quad
P^{\tau}_T > 20\; {\rm GeV},\quad |\eta^{\tau}|<2.5\,,
\label{eq:trigger-ss2}
\end{align}
where $ l = e, \mu $.
Besides, the two same-sign $ \tau $ candidates must be angularly
separated enough  by requiring $\Delta R_{\tau^{\pm}\tau^{\pm}} > 0.6$
in order to avoid overlapping.
Other isolation criteria among $e, \mu, \tau$ and jets are the same as the default settings of Delphes3. 

\item In order to reduce the $\tau$ lepton pair from the
  Drell-Yan process, we veto any opposite-sign $\tau$ lepton for both
  the signal and backgrounds with 
\begin{align}
N(\tau^{\mp})= 0\quad {\rm with} \quad
P^{\tau}_T > 20\; {\rm GeV},\quad |\eta^{\tau}|<2.5\,.
\label{eq:veto-os-tau}
\end{align}

\item To suppress the contributions from backgrounds of
  $t \overline{t}$ associated processes, 
  we reject the high missing transverse momentum $P^{miss}_T$ events by
  requiring
\begin{align}
P^{miss}_T < 40\, (M_N / 2)\; {\rm GeV}\,,
\label{eq:ptmiss}
\end{align}
for $ M_N < m_W $ ($ M_N > m_W $).

\item To further reduce the contributions from backgrounds of $ t\overline{t} $ associated processes, 
we apply the $b$-veto for both the signal and backgrounds with 
\begin{align}
N(b)=0\quad {\rm with} \quad
P^b_T > 20\; {\rm GeV},\quad |\eta^b|<2.5\,.
\label{eq:b-veto}
\end{align}
Moreover, for $ M_N > m_W $, we further reduce background contributions 
by requiring the inclusive scalar sum of jet $E_T$, $H_T$~\cite{Pascoli:2018heg,Liu:2019ayx}, to satisfy 
\begin{align}
H_T < 200\; {\rm GeV}.
\end{align}
The inclusive $H_T$ distributions for both signals and backgrounds are shown in Appendix~\ref{app:tables_figures}.

\item We require the minimum invariant mass for one of $ \tau $ leptons and an extra $ e/ \mu $ to satisfy
\begin{align}
M_{\tau^{\pm}\ell_1^\mp} < M_N\,.
\label{eq:mNcut}
\end{align}
This $ \tau $ lepton is most likely to be the second energetic one for small $ M_N $, 
but it becomes hard to be distinguished as $ M_N $ increases.  Here we
use the transverse mass distribution for $MT_{\tau^{\pm}\ell_1^\mp
  P_T^{miss}}$ to find the correct $ \tau $ lepton from the HNL
decay. We plot both $MT_{\tau_1^\pm \ell_1^\mp P_T^{miss}}$ and
$MT_{\tau_2^\pm \ell_1^\mp P_T^{miss}}$ distributions, and choose the one that
closely indicates the mass of the HNL. The same $\tau$ lepton is used
to form the invariant mass $M_{\tau^\pm \ell_1^\mp}$.


\item Finally, if $M_N < m_W$, the invariant mass of two same-sign $\tau$ leptons 
and an extra $e/ \mu$ system is required to have 
\begin{align}
M_{\tau_1^{\pm}\tau_2^{\pm}\ell_1^\mp} < m_W\,.
\label{eq:mWcut}
\end{align}
\end{enumerate}

\subsubsection{Two opposite-sign $\tau$s, $\mu$ and $mET$}

In this scenario, we require two opposite-sign $\tau$ leptons 
and an extra $\mu$ in the final state with the following cut flow.
\begin{enumerate}
\item For $M_N < m_W$, we specifically take two soft opposite-sign $\tau$ leptons 
and an extra soft $\mu$ as the selection of signals in our events with the following conditions,
\begin{align}
N(\tau,\mu)\geqslant 2,1,\quad 5 < P^{\mu}_T < 40\;{\rm  GeV},
\quad |\eta^{\mu}| < 2.5,\quad
20 < P^{\tau_1 (\tau_2)}_T < 50 (40)\; {\rm GeV},\quad |\eta^{\tau}|<2.5\,,
\label{eq:trigger-os1}
\end{align} 
On the other hand, for $M_N > m_W$, we choose instead
the following conditions for them: 
\begin{align}
N(\tau,\mu)\geqslant 2,1,\quad P^{\mu}_T > 15\;{\rm  GeV},
\quad |\eta^{\mu}| < 2.5,\quad
P^{\tau}_T > 20\; {\rm GeV},\quad |\eta^{\tau}|<2.5\,,
\label{eq:trigger-os2}
\end{align}
Compared with Eq.~(\ref{eq:trigger-ss2}), 
we require a stronger $P^{\mu}_T$ cut to further suppress soft radiation muons 
from $\tau\tau +nj$ and $t\bar{t} + n j$ processes. 
Again, $\Delta R_{\tau^+\tau^-} > 0.6$ and other isolation criteria are set to avoid overlaps. 

\item In order to reduce SM backgrounds with more than three $ \tau $ leptons, 
we veto any same-sign $\tau$ lepton for both signal and backgrounds 
with the same conditions as Eq.~(\ref{eq:veto-os-tau}).


\item To further reduce the contributions from backgrounds of $t \overline{t}$ associated processes, 
we apply the following cuts for both signal and backgrounds: 
high $P^{miss}_T$ rejection as Eq.~(\ref{eq:ptmiss}), $ b$-veto as Eq.~(\ref{eq:b-veto}). 
In addition, the cut $H_T < 200\; {\rm GeV}$ is applied for $M_N > m_W$. 

\item We require the minimum invariant mass for the $\tau$ leptons 
and an extra $\mu$ with opposite charges to satisfy Eq.~(\ref{eq:mNcut}).
Compared with the case of same-sign $\tau$s, 
it becomes more precise to pick up the correct $\tau$ lepton from the HNL
decay. 

\item Finally, if $M_N < m_W$, the invariant mass of two opposite-sign $\tau$ leptons 
and an extra $\mu$ system is required to have 
\begin{align}
M_{\tau^+\tau^-\mu} < m_W\,.
\label{eq:mWcut2}
\end{align}
 \end{enumerate}

\section{Analysis and results at HL-LHC}\label{Sec:Analyses}

\subsection{Same-sign tau leptons plus a charged lepton}

\begin{table}[h!]
\scriptsize
\begin{center}
\begin{tabular}{c|ccccc }
\hline\hline
\multicolumn{6}{c}{\textbf{Two Same-Sign $ \tau $s Selection Flow Table}}\\
\hline\hline
\multirow{2}{*}{\textbf{Process}}&\textbf{$\sigma$}&\textbf{Preselection}&\textbf{$P_T^{miss}<$ 40 GeV}&\textbf{b veto}&\textbf{Invariant Mass Selection}\\
&(fb)&\textbf{$A\epsilon$ (\%)}&\textbf{$A\epsilon$ (\%)}&\textbf{$A\epsilon$ (\%)}&\textbf{$A\epsilon$ (\%)}\\
\hline
$M_N$ = 25 GeV   &$2.851$&$9.834\times10^{-1}$&$9.071\times10^{-1}$&$8.860\times10^{-1}$&$5.663\times10^{-1}$\\
$W^{\pm}W^{\pm}W^{\mp}$ &$1.828\times10^{-1}$&$1.029$	&$5.110\times10^{-1}$&$5.030\times10^{-1}$	&$1.890\times10^{-2}$\\
$W^+W^-Z/H/\gamma$ &$1.065\times10^{-1}$ &$6.755\times10^{-1}$ &$3.047\times10^{-1}$ &$2.990\times10^{-1}$ &$1.440\times10^{-2}$\\
$t\bar{t}+nj$&$2.357\times10^4$ &$6.415\times10^{-2}$ &$1.282\times10^{-2}$ &$	1.864\times10^{-3}$ &$	2.530\times10^{-5}$\\
\hline\hline

\multirow{2}{*}{\textbf{Process}}&\textbf{$\sigma$}&\textbf{Preselection}&\textbf{$P_T^{miss}<$ 40 GeV}&\textbf{b veto}&\textbf{Invariant Mass Selection}\\
&(fb)&\textbf{$A\epsilon$ (\%)}&\textbf{$A\epsilon$ (\%)}&\textbf{$A\epsilon$ (\%)}&\textbf{$A\epsilon$ (\%)}\\
\hline
$M_N$ = 50 GeV   &$2.068$&$1.023$&$9.465\times10^{-1}$&$9.255\times10^{-1}$&$7.584\times10^{-1}$\\
$W^{\pm}W^{\pm}W^{\mp}$ &$1.828\times10^{-1}$&$1.029$	&$5.110\times10^{-1}$&$5.030\times10^{-1}$	&$6.114\times10^{-2}$\\
$W^+W^-Z/H/\gamma$ &$1.065\times10^{-1}$ &$6.755\times10^{-1}$ &$3.047\times10^{-1}$ &$2.990\times10^{-1}$ &$4.565\times10^{-2}$\\
$t\bar{t}+nj$&$2.357\times10^4$ &$6.415\times10^{-2}$ &$1.282\times10^{-2}$ &$	1.864\times10^{-3}$ &$	1.602\times10^{-4}$\\
\hline\hline

\multirow{2}{*}{\textbf{Process}}&\textbf{$\sigma$}&\textbf{Preselection}&\textbf{$P_T^{miss}<$ 40 GeV}&\textbf{b veto}&\textbf{Invariant Mass Selection}\\
&(fb)&\textbf{$A\epsilon$ (\%)}&\textbf{$A\epsilon$ (\%)}&\textbf{$A\epsilon$ (\%)}&\textbf{$A\epsilon$ (\%)}\\
\hline
$M_N$ = 75 GeV   &$8.935\times10^{-2}$&$7.486\times10^{-1}$&$6.723\times10^{-1}$&$6.512\times10^{-1}$&$4.861\times10^{-1}$\\
$W^{\pm}W^{\pm}W^{\mp}$ &$1.828\times10^{-1}$&$1.029$	&$5.110\times10^{-1}$&$5.030\times10^{-1}$	&$6.752\times10^{-2}$\\
$W^+W^-Z/H/\gamma$ &$1.065\times10^{-1}$ &$6.755\times10^{-1}$ &$3.047\times10^{-1}$ &$2.990\times10^{-1}$ &$4.955\times10^{-2}$\\
$t\bar{t}+nj$&$2.357\times10^4$ &$6.415\times10^{-2}$ &$1.282\times10^{-2}$ &$	1.864\times10^{-3}$ &$	1.771\times10^{-4}$\\
\hline\hline
\end{tabular}
\end{center}
\caption{
The two same-sign $\tau$s selection flow table for HNLs 
with benchmark points of $M_N$ = $25$, $50$ and $75$ GeV with $U^2_{\tau N}=10^{-5}$. 
The preselection and invariant mass selection are written in the main text. 
The $A \epsilon$ for each selection is the total accepted efficiency in each step.}
\label{ss_cut_soft}
\end{table}

\begin{table}[t!]
\tiny
\begin{center}
\begin{tabular}{c|cccccc }
\hline\hline
\multicolumn{7}{c}{\textbf{Two Same-Sign $ \tau $s Selection Flow Table}}\\
\hline\hline
\multirow{2}{*}{\textbf{Process}}&\textbf{$\sigma$}&\textbf{Preselection}&\textbf{$P_T^{miss}<$ 85/2 GeV}&\textbf{b veto}&\textbf{$H_T < 200$ GeV}&\textbf{Invariant Mass Selection}\\
&(fb)&\textbf{$A\epsilon$ (\%)}&\textbf{$A\epsilon$ (\%)}&\textbf{$A\epsilon$ (\%)}&\textbf{$A\epsilon$ (\%)}&\textbf{$A\epsilon$ (\%)}\\
\hline
$M_N$ = 85 GeV   &$1.102\times10^{-2}$&$2.488	$&$1.525$&$1.484$&$1.321$&$1.124$\\
$W^{\pm}W^{\pm}W^{\mp}$ &$1.713\times10^{-1}$&$5.454$&$1.577$&$1.547$&$	1.374$&$7.939\times10^{-1}$\\
$W^+W^-Z/H/\gamma$ &$5.824\times10^{-2}$&8.036&	1.937&1.892&1.395&$7.277\times10^{-1}$\\
$t\bar{t}+nj$&$2.240\times10^4$&$6.030\times10^{-1}$&$1.218\times10^{-1}$&$1.801\times10^{-2}$&$4.654\times10^{-3}$&$2.428\times10^{-3}$\\
\hline\hline

\multirow{2}{*}{\textbf{Process}}&\textbf{$\sigma$}&\textbf{Preselection}&\textbf{$P_T^{miss}<$ 100/2 GeV}&\textbf{b veto}&\textbf{$H_T < 200$ GeV}&\textbf{Invariant Mass Selection}\\
&(fb)&\textbf{$A\epsilon$ (\%)}&\textbf{$A\epsilon$ (\%)}&\textbf{$A\epsilon$ (\%)}&\textbf{$A\epsilon$ (\%)}&\textbf{$A\epsilon$ (\%)}\\
\hline
$M_N$ = 100 GeV   &$8.461\times10^{-3}$&$2.834$&$1.702$&$1.656$&$1.447$&$1.144$\\
$W^{\pm}W^{\pm}W^{\mp}$ &$1.713\times10^{-1}$&$5.455$&$2.019$&$1.980$&$	1.751$&$1.170$\\
$W^+W^-Z/H/\gamma$&$5.824\times10^{-2}$&8.036&	2.496&2.438&1.779&$1.085$\\
$t\bar{t}+nj$&$2.240\times10^4$&$6.031\times10^{-1}$&$1.607\times10^{-1}$&$2.383\times10^{-2}$&$5.969\times10^{-3}$&$4.199\times10^{-3}$\\
\hline\hline

\multirow{2}{*}{\textbf{Process}}&\textbf{$\sigma$}&\textbf{Preselection}&\textbf{$P_T^{miss}<$ 125/2 GeV}&\textbf{b veto}&\textbf{$H_T < 200$ GeV}&\textbf{Invariant Mass Selection}\\
&(fb)&\textbf{$A\epsilon$ (\%)}&\textbf{$A\epsilon$ (\%)}&\textbf{$A\epsilon$ (\%)}&\textbf{$A\epsilon$ (\%)}&\textbf{$A\epsilon$ (\%)}\\
\hline
$M_N$ = 125 GeV   &$3.486\times10^{-3}$&$8.227$&$5.871$&$5.708$&$4.683$&$4.087$\\
$W^{\pm}W^{\pm}W^{\mp}$ &$1.713\times10^{-1}$&$5.455$&$2.709$&$2.656$&$2.322$&$1.801$\\
$W^+W^-Z/H/\gamma$ &$5.824\times10^{-2}$&8.036&	3.403&3.324&2.378&$1.7103$\\
$t\bar{t}+nj$&$2.240\times10^4$&$6.031\times10^{-1}$&$2.280\times10^{-1}$&$3.420\times10^{-2}$&$8.044\times10^{-3}$&$6.526\times10^{-3}$\\
\hline\hline

\multirow{2}{*}{\textbf{Process}}&\textbf{$\sigma$}&\textbf{Preselection}&\textbf{$P_T^{miss}<$ 150/2 GeV}&\textbf{b veto}&\textbf{$H_T < 200$ GeV}&\textbf{Invariant Mass Selection}\\
&(fb)&\textbf{$A\epsilon$ (\%)}&\textbf{$A\epsilon$ (\%)}&\textbf{$A\epsilon$ (\%)}&\textbf{$A\epsilon$ (\%)}&\textbf{$A\epsilon$ (\%)}\\
\hline
$M_N$ = 150 GeV   &$1.758\times10^{-3}$&$1.201\times10^{1}$&$8.953$&$8.703$&$6.637$&$6.047$\\
$W^{\pm}W^{\pm}W^{\mp}$ &$1.713\times10^{-1}$&$5.455$&$3.297$&$3.231$&$	2.788$&$2.340$\\
$W^+W^-Z/H/\gamma$ &$5.824\times10^{-2}$&8.036&	4.200&4.099&2.864&$2.264$\\
$t\bar{t}+nj$&$2.240\times10^4$&$6.031\times10^{-1}$&$2.951\times10^{-1}$&$4.361\times10^{-2}$&$9.966\times10^{-3}$&$8.853\times10^{-3}$\\
\hline\hline
\end{tabular}
\end{center}
\caption{
The same as Table~\ref{ss_cut_soft}, 
but for HNLs with benchmark points of $M_N$ = $85$, $100$, $125$ and $150$ GeV 
with $U^2_{\tau N}=10^{-5}$.}
\label{ss_cut_hard}
\end{table}

In this section, we display our results based on the simulation and analysis strategies in the previous section. 
First, we explain our results for the channel of two same-sign $\tau$s, $e/ \mu$ and $mET$. 
The cut flow tables for $M_N < m_W$ ($M_N = 25, 50, 75$ GeV) 
and $ M_N > m_W $ ($M_N = 85, 100, 125, 150$ GeV) are shown 
in the Table~\ref{ss_cut_soft} and Table~\ref{ss_cut_hard}, respectively. 
Here we set $U^2_{\tau N}=10^{-5}$ for all benchmark points. 
We list three major SM backgrounds in these two tables: 
$W^{\pm} W^{\pm} W^{\mp}$, $W^+ W^- Z/H/\gamma$ and $t \bar{t} + n j$. 
The $t \bar{t} + n j$ is the dominant one among them before applying the
selection cuts. 
On the other hand, the notation of \textbf{Preselection} includes 
Eqs.~(\ref{eq:trigger-ss1}), (\ref{eq:trigger-ss2}) and (\ref{eq:veto-os-tau}) 
and \textbf{Invariant Mass Selection} includes Eqs.~(\ref{eq:mNcut}) and (\ref{eq:mWcut}) (when $M_N < m_W$).

\begin{figure}[h]
\centering
	\begin{subfigure}{0.45\textwidth}
         \centering
        \includegraphics[width=0.8\linewidth]{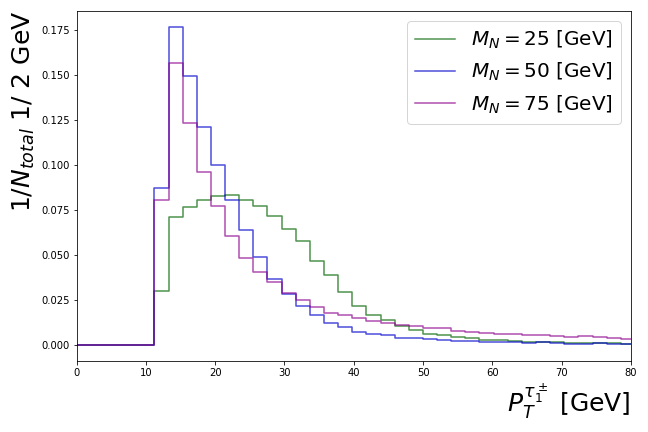}
         \caption{The $p_T$ distribution for leading $\tau^\pm$}
     \end{subfigure}
     \begin{subfigure}{0.45\textwidth}
         \centering
        \includegraphics[width=0.8\linewidth]{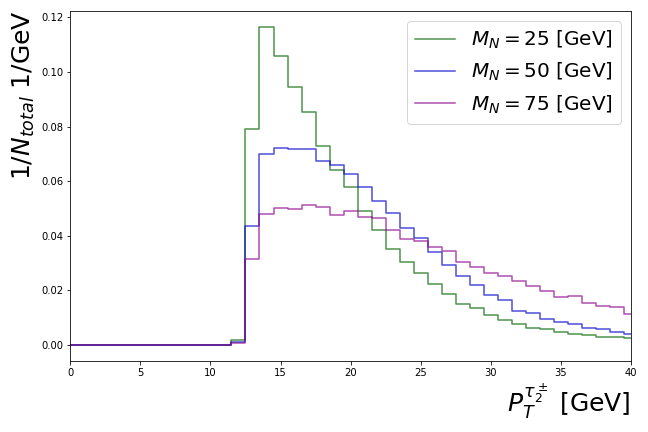}
         \caption{The $p_T$ distribution for subleading $\tau^\pm$}
     \end{subfigure}
     \begin{subfigure}{0.45\textwidth}
         \centering
        \includegraphics[width=0.8\linewidth]{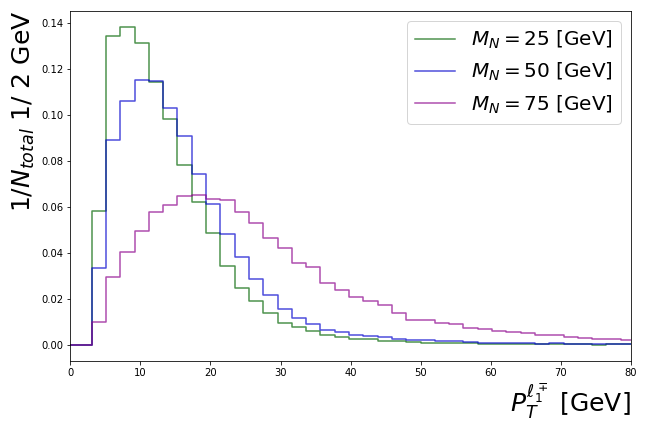}
         \caption{The $p_T$ distribution for leading $\ell^\mp$}
     \end{subfigure}
	 \begin{subfigure}{0.45\textwidth}
         \centering
        \includegraphics[width=0.8\linewidth]{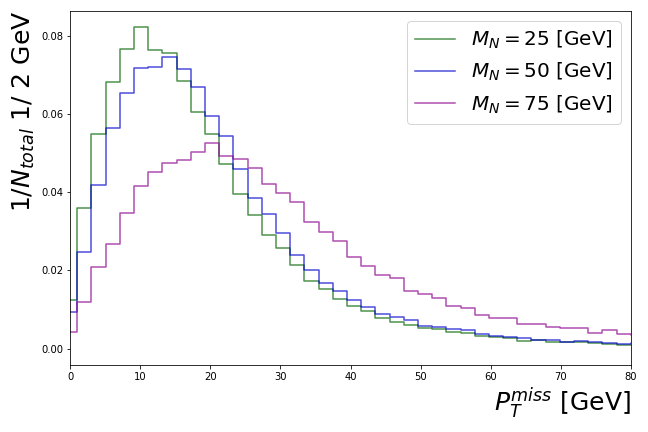}
         \caption{The $p_T$ distribution for missing energy}
     \end{subfigure}
     \begin{subfigure}{0.45\textwidth}
         \centering
        \includegraphics[width=0.8\linewidth]{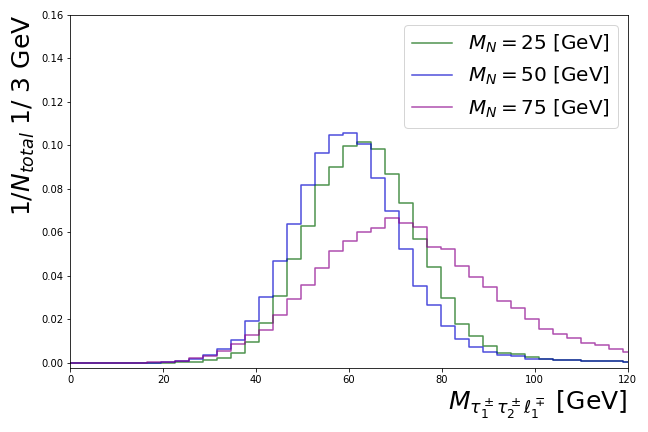}
         \caption{The distribution for $M_{\tau_1^\pm\tau_2^\pm\ell_1^\mp}$  }
     \end{subfigure}
	 \begin{subfigure}{0.45\textwidth}
         \centering
        \includegraphics[width=0.8\linewidth]{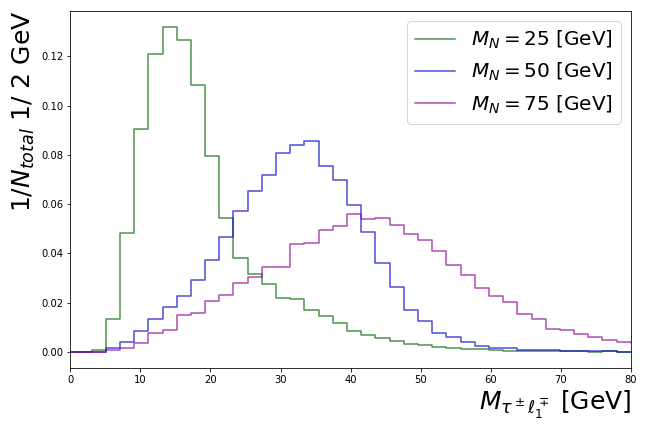}
         \caption{The distribution for $M_{\tau^\pm\ell_1^\mp}$  }
     \end{subfigure}
     \begin{subfigure}{0.45\textwidth}
         \centering
        \includegraphics[width=0.8\linewidth]{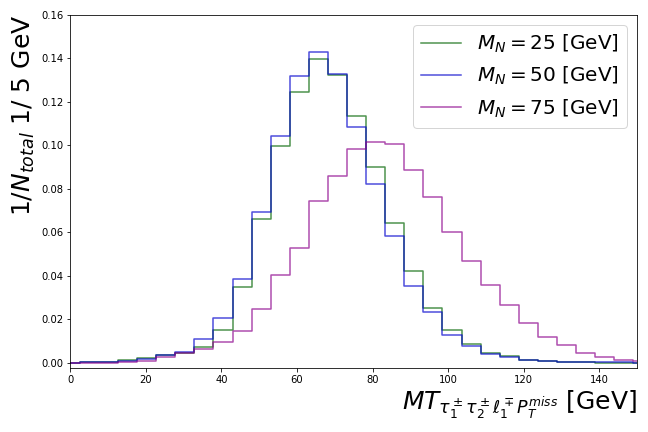}
         \caption{The distribution for $MT_{\tau_1^\pm\tau_2^\pm\ell_1^\mp P_T^{miss}}$  }
     \end{subfigure}
	 \begin{subfigure}{0.45\textwidth}
         \centering
        \includegraphics[width=0.8\linewidth]{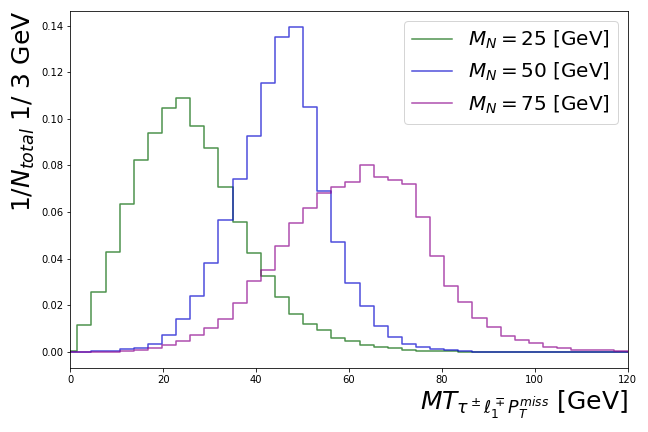}
         \caption{The distribution for $MT_{\tau^\pm\ell_1^\mp P_T^{miss}}$  }
     \end{subfigure}
\caption{
\textbf{Two Same-Sign $\tau$s:} 
Various kinematical distributions for the signal with
the benchmark points of $M_N$ = $25$, $50$ and $75$ GeV. 
Notice the distributions in (e), (f), (g) and (h) passed the
preselection criteria. 
}\label{fig:M_N_ss_Soft}
\end{figure}

\begin{figure}[h]
\centering
	\begin{subfigure}{0.45\textwidth}
         \centering
        \includegraphics[width=0.8\linewidth]{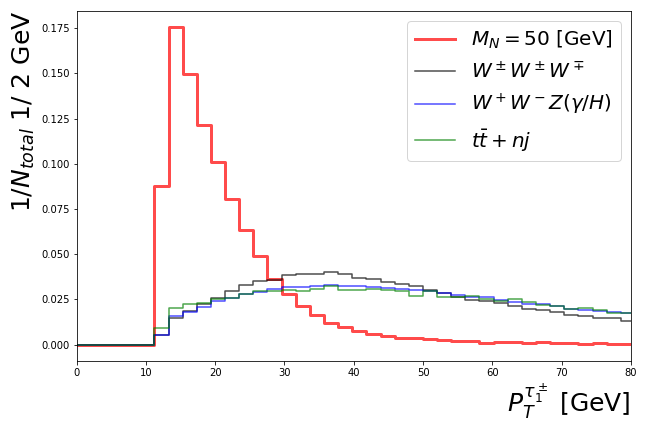}
         \caption{The $p_T$ distribution for leading $\tau^\pm$}
     \end{subfigure}
     \begin{subfigure}{0.45\textwidth}
         \centering
        \includegraphics[width=0.8\linewidth]{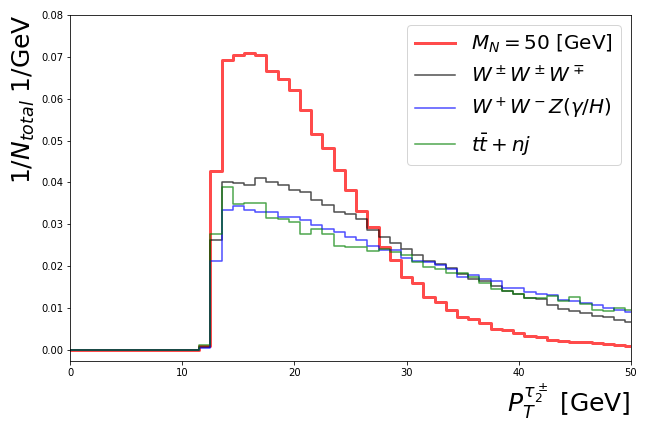}
         \caption{The $p_T$ distribution for subleading $\tau^\pm$}
     \end{subfigure}
     \begin{subfigure}{0.45\textwidth}
         \centering
        \includegraphics[width=0.8\linewidth]{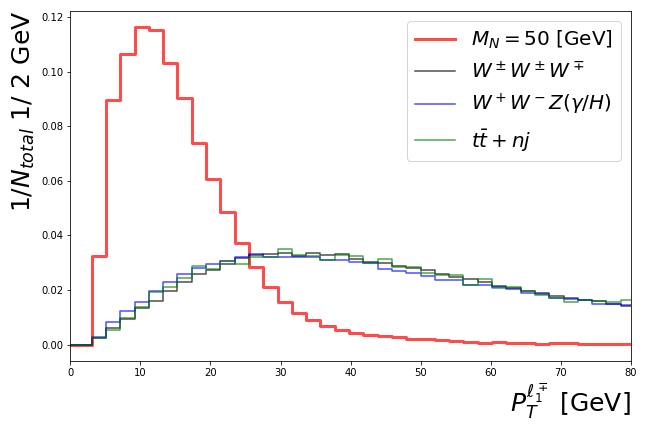}
         \caption{The $p_T$ distribution for leading $\ell^\mp$}
     \end{subfigure}
	 \begin{subfigure}{0.45\textwidth}
         \centering
        \includegraphics[width=0.8\linewidth]{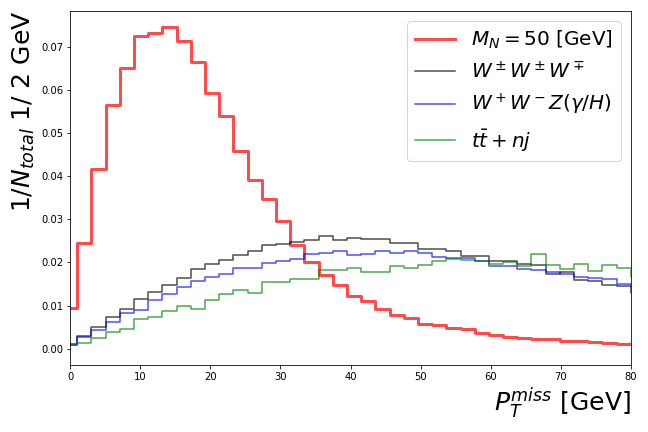}
         \caption{The $p_T$ distribution for missing energy}
     \end{subfigure}
     \begin{subfigure}{0.45\textwidth}
         \centering
        \includegraphics[width=0.8\linewidth]{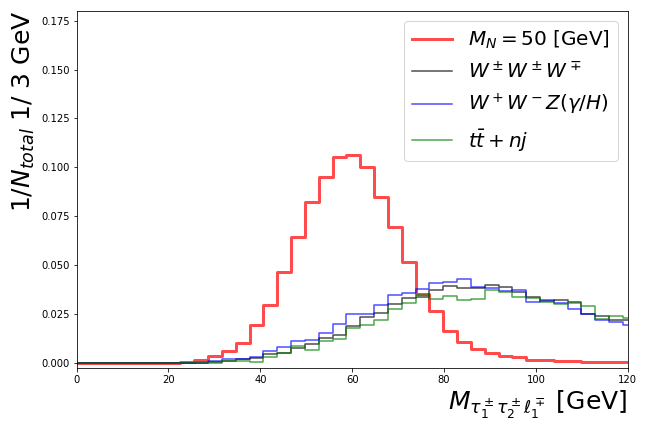}
         \caption{The distribution for $M_{\tau_1^\pm\tau_2^\pm\ell_1^\mp}$  }
     \end{subfigure}
	 \begin{subfigure}{0.45\textwidth}
         \centering
        \includegraphics[width=0.8\linewidth]{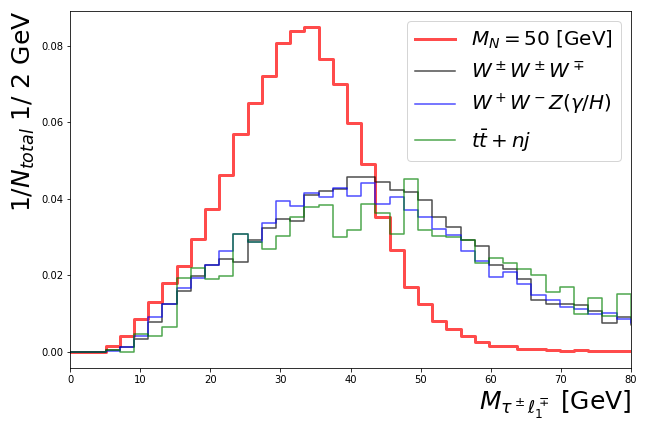}
         \caption{The distribution for $M_{\tau^\pm\ell_1^\mp}$  }
     \end{subfigure}
      \begin{subfigure}{0.45\textwidth}
         \centering
        \includegraphics[width=0.8\linewidth]{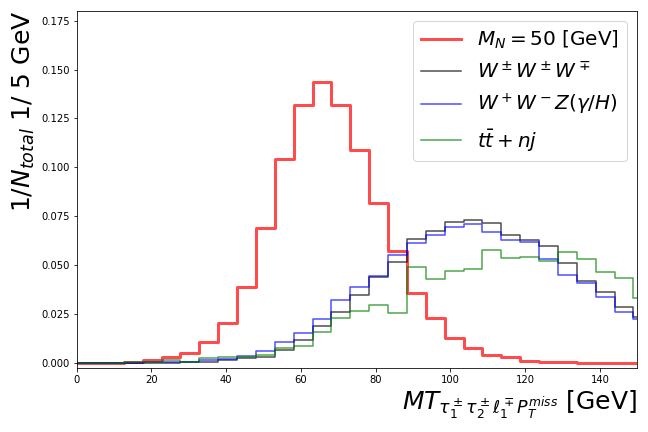}
         \caption{The distribution for $MT_{\tau_1^\pm\tau_2^\pm\ell_1^\mp  P_T^{miss}}$  }
     \end{subfigure}
	 \begin{subfigure}{0.45\textwidth}
         \centering
        \includegraphics[width=0.8\linewidth]{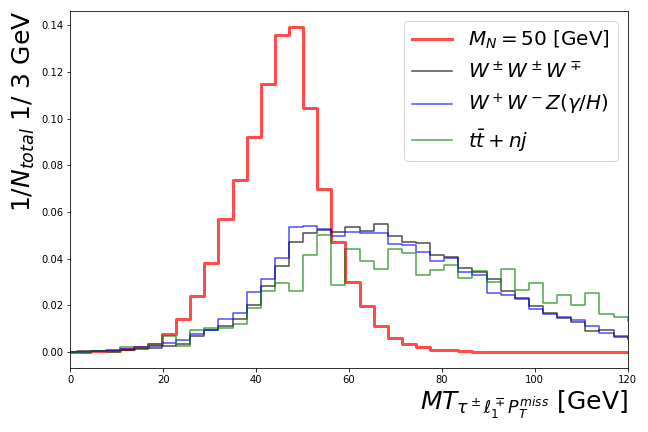}
         \caption{The distribution for $MT_{\tau^\pm\ell_1^\mp P_T^{miss}}$  }
     \end{subfigure}
\caption{
The same as Fig.~\ref{fig:M_N_ss_Soft}, 
but for the signal with the benchmark point of $M_N = 50$ GeV and
major SM backgrounds.
}\label{fig:M_N_ss_N50}
\end{figure}

For $ M_N < m_W $, after passing all selection cuts, 
we can find the signal efficiencies around $0.49 \mathchar`- 0.57\%$, 
the efficiencies of $W^{\pm} W^{\pm} W^{\mp}$ and $W^+W^- Z/H/\gamma$ are
less than $6.8 \times 10^{-2}\%$ and $5.0 \times 10^{-2}\%$, 
and that of $t \bar{t} + n j$ is even smaller, less than $2.5 \times 10^{-5}\%$.
\footnote{
  The tiny efficiency of $t \bar{t} + n j$ also causes unavoidable
  large statistical fluctuations, 
  even we already generated more than $1.2 \times 10^7$ Monte Carlo events.}
Some kinematical distributions for the signal with $M_N = 25$, $50$ and 75 GeV 
are shown in Fig.~\ref{fig:M_N_ss_Soft}. 
Notice that the distributions in (e), (f), (g) and (h) pass
the preselection criteria. 
All $\tau_1$, $\tau_2$ and $\ell_1$ are relatively soft as shown in (a), (b)
and (c) on Fig~\ref{fig:M_N_ss_Soft}. 
In order to pick out these soft objects, we focus on low $P_T$ regions as
in Eq.~(\ref{eq:trigger-ss1}). 
Similar to the soft charged leptons, the $ P^{miss}_T $ is also soft as
shown in (d) in Fig~\ref{fig:M_N_ss_Soft}, 
so we further reject the high $P^{miss}_T$ regions as in Eq.~(\ref{eq:ptmiss}). 
Finally, Eqs.~(\ref{eq:mNcut}) and (\ref{eq:mWcut}) can help us to
select the major parts of the signal 
as shown in (e) and (f) in Fig.~\ref{fig:M_N_ss_Soft}. 
On the other hand, the transverse mass distribution
for $M_T(P^{\tau^{\pm}_1}_T,P^{\tau^{\pm}_2}_T,P^{\ell^{\mp}_1}_T,P^{miss}_T)$ and
$M_T(P^{\tau^{\pm}}_T,P^{\ell^{\mp}_1}_T,P^{miss}_T)$
in (g) and (h) in Fig.~\ref{fig:M_N_ss_Soft} clearly show the
resonance structure of both $m_W$ and $M_N$, respectively.
In Fig.~\ref{fig:M_N_ss_N50}, 
we also display these kinematical distributions for the signal $M_N = 50$ GeV
and three major SM backgrounds. 
We can clearly see that these analysis strategies for this scenario
in the previous section 
can successfully distinguish most parts of the signal from the SM backgrounds.

\begin{figure}[h]
\centering
	\begin{subfigure}{0.45\textwidth}
         \centering
        \includegraphics[width=0.8\linewidth]{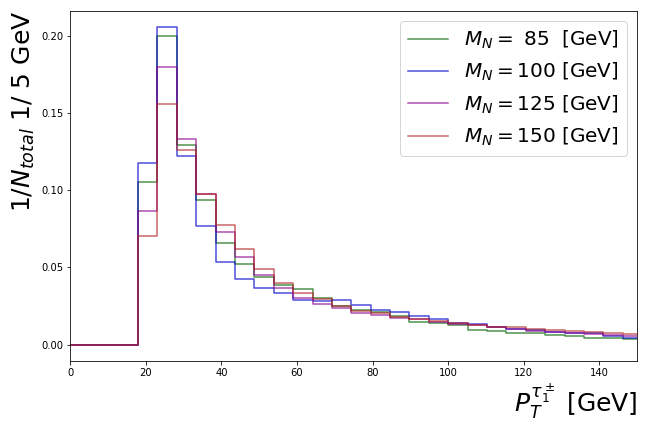}
         \caption{The $p_T$ distribution for leading $\tau^\pm$}
     \end{subfigure}
     \begin{subfigure}{0.45\textwidth}
         \centering
        \includegraphics[width=0.8\linewidth]{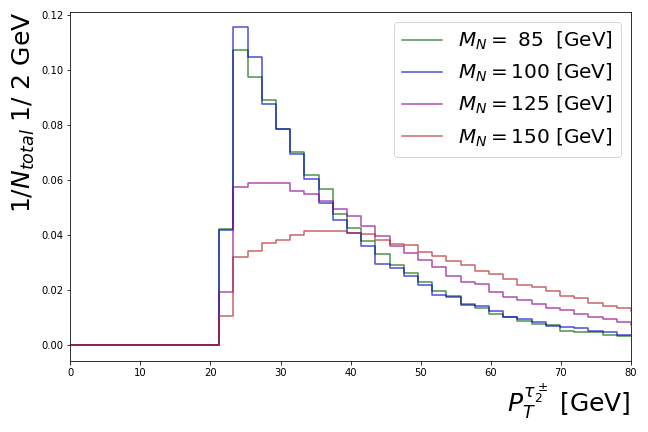}
         \caption{The $p_T$ distribution for subleading $\tau^\pm$}
     \end{subfigure}
     \begin{subfigure}{0.45\textwidth}
         \centering
        \includegraphics[width=0.8\linewidth]{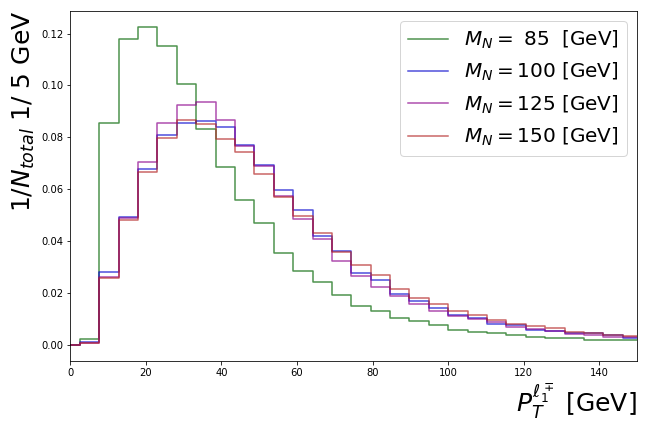}
         \caption{The $p_T$ distribution for leading $\ell^\mp$}
     \end{subfigure}
	 \begin{subfigure}{0.45\textwidth}
         \centering
        \includegraphics[width=0.8\linewidth]{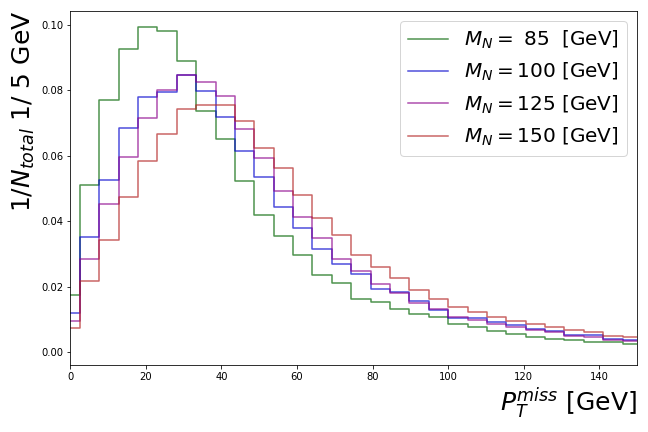}
         \caption{The $p_T$ distribution for missing energy}
     \end{subfigure}
	 \begin{subfigure}{0.45\textwidth}
         \centering
        \includegraphics[width=0.8\linewidth]{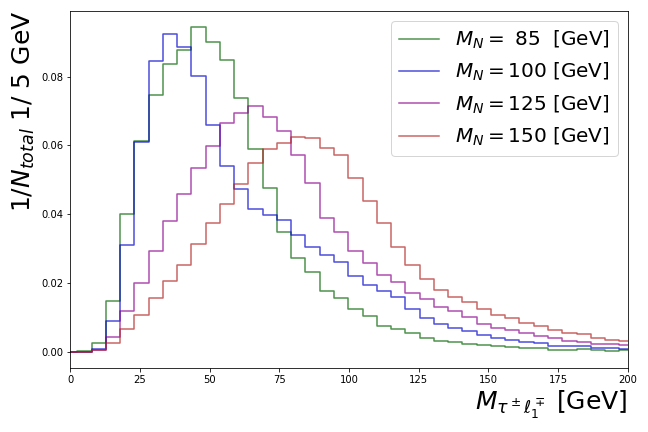}
         \caption{The distribution for $M_{\tau^\pm\ell_1^\mp}$  }
     \end{subfigure}
       \begin{subfigure}{0.45\textwidth}
         \centering
        \includegraphics[width=0.8\linewidth]{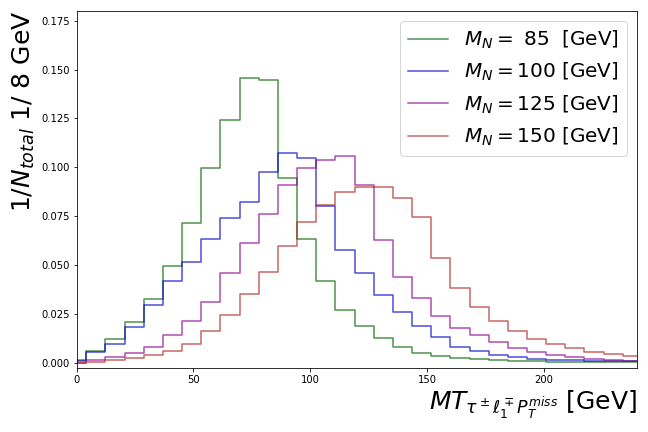}
         \caption{The distribution for $MT_{\tau^\pm\ell_1^\mp P_T^{miss}}$ }
     \end{subfigure}
\caption{
\textbf{Two Same-Sign $\tau$s:} 
various kinematical distributions for the signal with the
benchmark points of $M_N = 85$, $100$, $125$ and $150$ GeV. 
Notice the distributions in (e) and (f) passed the preselection criteria. 
}\label{fig:M_N_ss_Hard}
\end{figure}

\begin{figure}[h]
\centering
	\begin{subfigure}{0.45\textwidth}
         \centering
        \includegraphics[width=0.8\linewidth]{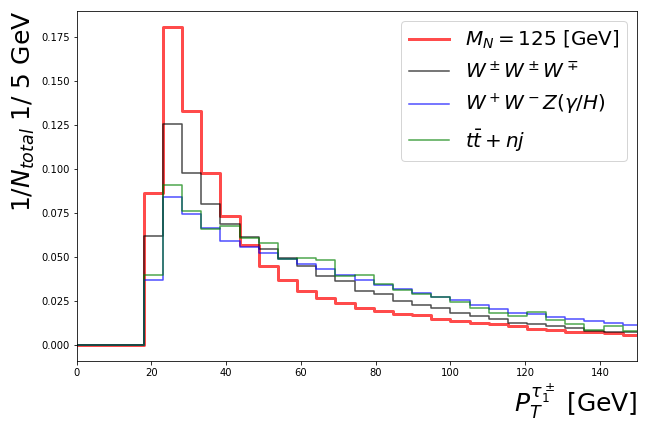}
         \caption{The $p_T$ distribution for leading $\tau^\pm$}
     \end{subfigure}
     \begin{subfigure}{0.45\textwidth}
         \centering
        \includegraphics[width=0.8\linewidth]{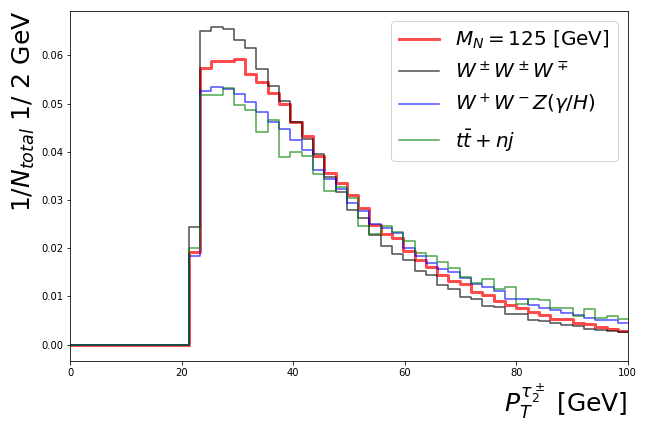}
         \caption{The $p_T$ distribution for subleading $\tau^\pm$}
     \end{subfigure}
     \begin{subfigure}{0.45\textwidth}
         \centering
        \includegraphics[width=0.8\linewidth]{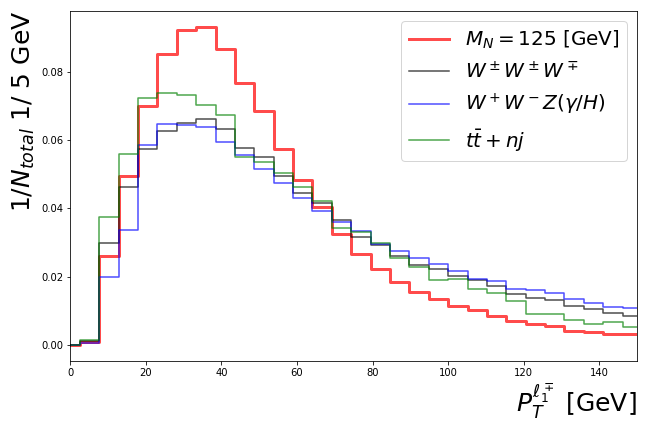}
         \caption{The $p_T$ distribution for leading $\ell^\mp$}
     \end{subfigure}
	 \begin{subfigure}{0.45\textwidth}
         \centering
        \includegraphics[width=0.8\linewidth]{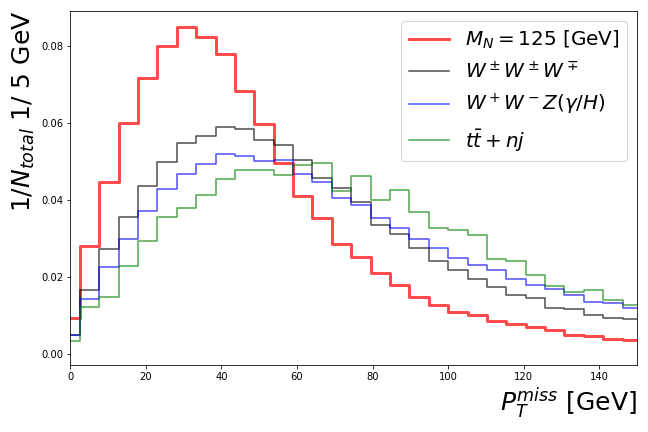}
         \caption{The $p_T$ distribution for missing energy}
     \end{subfigure}
	 \begin{subfigure}{0.45\textwidth}
         \centering
        \includegraphics[width=0.8\linewidth]{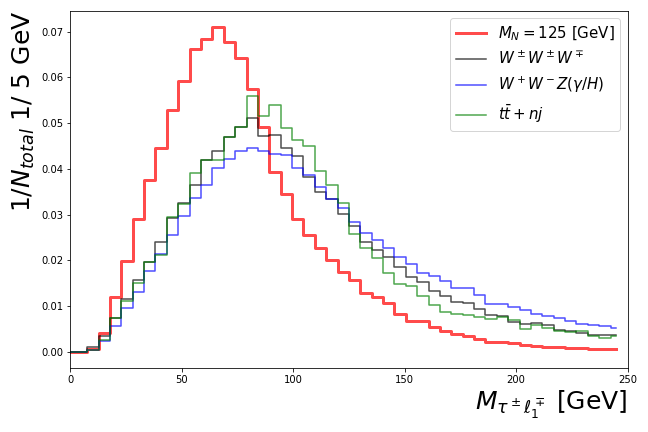}
         \caption{The distribution for $M_{\tau^\pm\ell_1^\mp}$  }
     \end{subfigure}
     \begin{subfigure}{0.45\textwidth}
         \centering
        \includegraphics[width=0.8\linewidth]{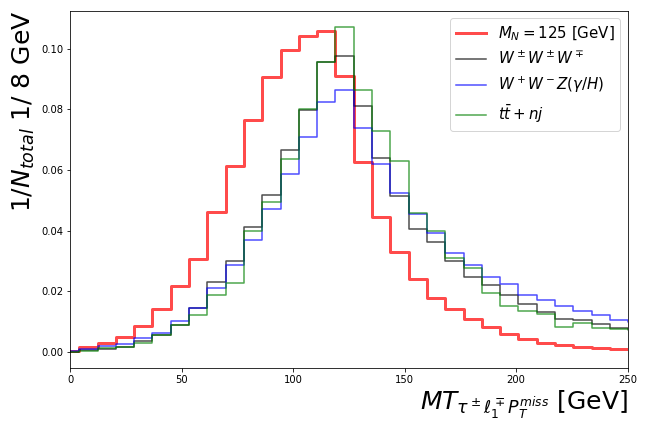}
         \caption{The distribution for $MT_{\tau^\pm\ell_1^\mp P_T^{miss}}$ }
     \end{subfigure}
\caption{
The same as Fig.~\ref{fig:M_N_ss_Hard}, 
but for the signal with the benchmark point of $M_N = 125$ GeV and
major SM backgrounds.
}\label{fig:M_N_ss_125}
\end{figure}

For $M_N > m_W$, 
after passing all selection cuts, we can find the signal efficiencies
around $1.1 \mathchar`- 6.0\%$, 
the efficiencies of $W^{\pm} W^{\pm} W^{\mp}$ and $W^+W^- Z/H/\gamma$ are
less than $2.3\%$, 
and the efficiencies of $t \bar{t} + n j$ is even smaller,
less than $8.9 \times 10^{-3}\%$. 
Some kinematical distributions for the signal with
$M_N = 85$, $100$, $125$ and $150$ GeV are shown 
in Fig.~\ref{fig:M_N_ss_Hard}. 
Notice that the distributions in (e) and (f) pass the preselection criteria. 
In contrast to the case $M_N < m_W$, 
as shown in (a), (b) and (c) in Fig.~\ref{fig:M_N_ss_Hard}, 
$\tau$ leptons and $e/ \mu$ can have long tail $P_T$ distributions
with the increase in the mass of HNLs. 
We can also find most of $P^{miss}_T$ distributions in this scenario 
are less than $M_N /2$ as shown in (d) in Fig.~\ref{fig:M_N_ss_Hard}. 
For the benchmark points of $ M_N > 85$ GeV,
$N$ decays into an on-shell $W$ boson and a relatively soft $\tau$ because
of the mass threshold.
Thus, the subleading $\tau$ lepton shows a soft $P_T$ spectrum
especially for the low mass shown in panel (b) of Fig.~\ref{fig:M_N_ss_Hard}.
Both the invariant mass $M_{\tau^\pm \ell_1^\mp}$ (panel (e))
and the transverse mass (panel (f)) distributions clearly
correlate with the mass of the HNL.
In Fig.~\ref{fig:M_N_ss_125}, 
we also display these kinematical distributions for the signal benchmark
$M_N = 125$ GeV and three major SM backgrounds.
All the major backgrounds show relatively harder spectra in
$P_T^{\ell_1^\mp}$, $P_T^{miss}$, $M_{\tau^\pm \ell_1^\mp}$, and
$M_{T_{\tau^\pm \ell_1^\mp P_T^{miss}}}$. One can make use of these features to
discriminate the signal from the backgrounds.

\begin{table}[h!]
\scriptsize
\begin{center}
\begin{tabular}{c|cccccc }

\hline\hline
\multicolumn{6}{c}{\textbf{Two Opposite-Sign $\tau$s Selection Flow Table}}\\
\hline\hline
\multirow{2}{*}{\textbf{Process}}&\textbf{$\sigma$}&\textbf{Preselection}&\textbf{$P_T^{miss}<$ 40 GeV}&\textbf{b veto}&\textbf{Invariant Mass Selection}\\
&(fb)&\textbf{$A\epsilon$ (\%)}&\textbf{$A\epsilon$ (\%)}&\textbf{$A\epsilon$ (\%)}&\textbf{$A\epsilon$ (\%)}\\
\hline
$M_N$ = 25 GeV   &$2.291$&$8.487\times10{-1}$&$7.629\times10{-1}$&$7.478\times10{-1}$&$6.132\times10{-1}$\\
$W^{\pm}Z/H/\gamma$ &$1.599\times10^{2}$&$7.696\times10^{-1}$&$5.737\times10^{-1}$&$5.652\times10^{-1}$&$1.700\times10^{-2}$\\
$ZZ/\gamma$ &$2.400\times10^{1}$&$7.990\times10^{-1}$&$7.066\times10^{-1}$&$6.967\times10^{-1}$&$3.970\times10^{-2}$\\
$\tau\tau+nj$ &$9.559\times10^{5}$&$3.680\times10^{-4}$&	$3.476\times10^{-4}$&$3.476\times10^{-4}$&$0$\\
$t\bar{t}+nj$&$2.987\times10^4$&$2.164\times10^{-2}$&$4.353\times10^{-3}$&$5.804\times10^{-4}$&$5.128\times10^{-5}$\\
\hline\hline

\multirow{2}{*}{\textbf{Process}}&\textbf{$\sigma$}&\textbf{Preselection}&\textbf{$P_T^{miss}<$ 40 GeV}&\textbf{b veto}&\textbf{Invariant Mass Selection}\\
&(fb)&\textbf{$A\epsilon$ (\%)}&\textbf{$A\epsilon$ (\%)}&\textbf{$A\epsilon$ (\%)}&\textbf{$A\epsilon$ (\%)}\\
\hline
$M_N$ = 50 GeV   &$2.052$&$6.320\times10{-1}$&$5.581\times10{-1}$&$5.450\times10{-1}$&$4.791\times10{-1}$\\
$W^{\pm}Z/H/\gamma$ &$1.599\times10^{2}$&$7.696\times10^{-1}$&$5.737\times10^{-1}$&$5.652\times10^{-1}$&$5.430\times10^{-2}$\\
$ZZ/\gamma$ &$2.400\times10^{1}$&$7.990\times10^{-1}$&$7.066\times10^{-1}$&$6.967\times10^{-1}$&$9.740\times10^{-2}$\\
$\tau\tau+nj$ &$9.559\times10^{5}$&$3.680\times10^{-4}$&	$3.476\times10^{-4}$&$3.476\times10^{-4}$&$6.134\times10{-5}$\\
$t\bar{t}+nj$&$2.987\times10^4$&$2.164\times10^{-2}$&$4.353\times10^{-3}$&$5.804\times10^{-4}$&$1.878\times10^{-4}$\\
\hline\hline

\multirow{2}{*}{\textbf{Process}}&\textbf{$\sigma$}&\textbf{Preselection}&\textbf{$P_T^{miss}<$ 40 GeV}&\textbf{b veto}&\textbf{Invariant Mass Selection}\\
&(fb)&\textbf{$A\epsilon$ (\%)}&\textbf{$A\epsilon$ (\%)}&\textbf{$A\epsilon$ (\%)}&\textbf{$A\epsilon$ (\%)}\\
\hline
$M_N$ = 75 GeV   &$9.104\times10^{-2}$&$4.364\times10^{-1}$&$3.429\times10^{-1}$&$3.368\times10^{-1}$&$1.619\times10^{-1}$\\
$W^{\pm}Z/H/\gamma$ &$1.599\times10^{2}$&$7.696\times10^{-1}$&$5.737\times10^{-1}$&$5.652\times10^{-1}$&$5.860\times10^{-2}$\\
$ZZ/\gamma$ &$2.400\times10^{1}$&$7.990\times10^{-1}$&$7.066\times10^{-1}$&$6.967\times10^{-1}$&$1.010\times10^{-2}$\\
$\tau\tau+nj$ &$9.559\times10^{5}$&$3.680\times10^{-4}$&	$3.476\times10^{-4}$&$3.476\times10^{-4}$&$8.179\times10{-5}$\\
$t\bar{t}+nj$&$2.987\times10^4$&$2.164\times10^{-2}$&$4.353\times10^{-3}$&$5.804\times10^{-4}$&$2.561\times10^{-4}$\\
\hline\hline

\end{tabular}
\end{center}
\caption{The two opposite-sign $ \tau $s selection flow table for HNLs with benchmark points of $M_N$ = 25, 50 and 75 GeV with $U^2_{\tau N}=10^{-5}$. The preselection and invariant mass selection are written in the main text. The $A\epsilon$ for each selection is the total accepted efficiency in each step. }
\label{os_cut_soft}
\end{table}

\begin{table}[t!]
\tiny
\begin{center}
\begin{tabular}{c|cccccc }

\hline\hline
\multicolumn{7}{c}{\textbf{Two Opposite-Sign $\tau$s Selection Flow Table}}\\
\hline\hline
\multirow{2}{*}{\textbf{Process}}&\textbf{$\sigma$}&\textbf{Preselection}&\textbf{$P_T^{miss}<$ 85/2 GeV}&\textbf{b veto}&\textbf{$H_T < 200$ GeV}&\textbf{Invariant Mass Selection}\\
&(fb)&\textbf{$A\epsilon$ (\%)}&\textbf{$A\epsilon$ (\%)}&\textbf{$A\epsilon$ (\%)}&\textbf{$A\epsilon$ (\%)}&\textbf{$A\epsilon$ (\%)}\\
\hline
$M_N$ = 85 GeV   &$1.101\times10^{-2}$&$1.386$&$7.410\times10^{-1}$&$7.236\times10^{-1}$&$6.538\times10^{-1}$&$6.471\times10^{-1}$\\
$W^{\pm}Z/H/\gamma$ &$1.031\times10^{2}$&$4.402$&$2.004$&$1.971$&$	1.874$&$1.078$\\
$ZZ/\gamma$ &$2.082\times10^{1}$&5.275&2.616&2.572&2.490&$1.339$\\
$\tau\tau+nj$ &$9.561\times10^{5}$&$2.024\times10^{-4}$&	$1.518\times10^{-4}$&$1.417\times10^{-4}$&$1.214\times10^{-4}$&$1.012\times10^{-4}$\\
$t\bar{t}+nj$&$2.864\times10^4$&$2.712\times10^{-1}$&$4.881\times10^{-2}$&$7.395\times10^{-3}$&$20118\times10^{-3}$&$9.980\times10^{-4}$\\
\hline\hline

\multirow{2}{*}{\textbf{Process}}&\textbf{$\sigma$}&\textbf{Preselection}&\textbf{$P_T^{miss}<$ 100/2 GeV}&\textbf{b veto}&\textbf{$H_T < 200$ GeV}&\textbf{Invariant Mass Selection}\\
&(fb)&\textbf{$A\epsilon$ (\%)}&\textbf{$A\epsilon$ (\%)}&\textbf{$A\epsilon$ (\%)}&\textbf{$A\epsilon$ (\%)}&\textbf{$A\epsilon$ (\%)}\\
\hline
$M_N$ = 100 GeV   &$8.745\times10^{-3}$&$1.723$&$9.382\times10^{-1}$&$9.144\times10^{-1}$&$8.201\times10^{-1}$&$8.140\times10^{-1}$\\
$W^{\pm}Z/H/\gamma$ &$1.031\times10^{2}$&4.402&2.403&2.363&2.232&$1.516$\\
$ZZ/\gamma$ &$2.082\times10^{1}$&5.275&3.078&3.026&2.916&$1.938$\\
$\tau\tau+nj$ &$9.561\times10^{5}$&$2.024\times10^{-4}$&	$1.822\times10^{-4}$&$1.720\times10^{-4}$&$1.417\times10^{-4}$&$1.012\times10^{-4}$\\
$t\bar{t}+nj$&$2.864\times10^4$&$2.712\times10^{-1}$&$6.499\times10^{-2}$&$9.595\times10^{-3}$&$3.076\times10^{-3}$&$2.159\times10^{-3}$\\
\hline\hline

\multirow{2}{*}{\textbf{Process}}&\textbf{$\sigma$}&\textbf{Preselection}&\textbf{$P_T^{miss}<$ 125/2 GeV}&\textbf{b veto}&\textbf{$H_T < 200$ GeV}&\textbf{Invariant Mass Selection}\\
&(fb)&\textbf{$A\epsilon$ (\%)}&\textbf{$A\epsilon$ (\%)}&\textbf{$A\epsilon$ (\%)}&\textbf{$A\epsilon$ (\%)}&\textbf{$A\epsilon$ (\%)}\\
\hline
$M_N$ = 125 GeV   &$3.597\times10^{-3}$&$4.683$&$3.186$&$3.101$&$2.609$&$2.605$\\
$W^{\pm}Z/H/\gamma$ &$1.031\times10^{2}$&$4.402$&$2.927$&$2.875$&$	2.694$&$2.154$\\
$ZZ/\gamma$ &$2.082\times10^{1}$&5.275&3.694&3.630&3.472&$2.778$\\
$\tau\tau+nj$ &$9.561\times10^{5}$&$2.024\times10^{-4}$&	$1.923\times10^{-4}$&$1.822\times10^{-4}$&$1.518\times10^{-4}$&$1.316\times10^{-4}$\\
$t\bar{t}+nj$&$2.864\times10^4$&$2.712\times10^{-1}$&$9.252\times10^{-2}$&$1.328\times10^{-2}$&$4.196\times10^{-3}$&$3.544\times10^{-3}$\\
\hline\hline

\multirow{2}{*}{\textbf{Process}}&\textbf{$\sigma$}&\textbf{Preselection}&\textbf{$P_T^{miss}<$ 150/2 GeV}&\textbf{b veto}&\textbf{$H_T < 200$ GeV}&\textbf{Invariant Mass Selection}\\
&(fb)&\textbf{$A\epsilon$ (\%)}&\textbf{$A\epsilon$ (\%)}&\textbf{$A\epsilon$ (\%)}&\textbf{$A\epsilon$ (\%)}&\textbf{$A\epsilon$ (\%)}\\
\hline
$M_N$ = 150 GeV   &$1.800\times10^{-3}$&$6.746$&$4.808$&$4.682$&$3.630$&$3.626$\\
$W^{\pm}Z/H/\gamma$ &$1.031\times10^{2}$&$4.402$&$3.314$&$3.254$&$3.020$&$2.631$\\
$ZZ/\gamma$ &$2.082\times10^{1}$&5.275&4.125&4.051&3.844&$3.378$\\
$\tau\tau+nj$ &$9.561\times10^{5}$&$2.024\times10^{-4}$&	$1.923\times10^{-4}$&$1.822\times10^{-4}$&$1.518\times10^{-4}$&$1.518\times10^{-4}$\\
$t\bar{t}+nj$&$2.864\times10^4$&$2.712\times10^{-1}$&$1.209\times10^{-1}$&$1.719\times10^{-2}$&$5.214\times10^{-3}$&$4.643\times10^{-3}$\\
\hline\hline

\end{tabular}
\end{center}
\caption{The same as Tab.~\ref{os_cut_soft}, but for HNLs with benchmark points of $M_N$ = 85, 100, 125 and 150 GeV with $U^2_{\tau N}=10^{-5}$.}
\label{os_cut_hard}
\end{table}

\subsection{Opposite-sign tau leptons plus a muon}

Now we turn to our results for the channel of two opposite-sign
$\tau$s, $\mu$ and $mET$. 
The cut flow tables for $M_N < m_W$ ($M_N = 25, 50, 75$ GeV) 
and $ M_N > m_W $ ($M_N = 85, 100, 125, 150$ GeV) are shown in 
Tables \ref{os_cut_soft} 
and \ref{os_cut_hard}, respectively. 
Again, we set $U^2_{\tau N}=10^{-5}$ for all benchmark points. 
We list four major SM backgrounds in these two tables: 
$W^{\pm} Z/H/\gamma$, $ZZ/\gamma$, $\tau\tau + n j$ and $t\bar{t}+nj$. 
The $\tau \tau + n j$ is the dominant one among them. 
The notation of \textbf{Preselection} includes 
Eqs.~(\ref{eq:trigger-os1}), (\ref{eq:trigger-os2}) and (\ref{eq:veto-os-tau}) 
and \textbf{Invariant Mass Selection} includes Eqs.~(\ref{eq:mNcut}) and (\ref{eq:mWcut2}) (when $M_N < m_W$).

\begin{figure}[h]
\centering
	\begin{subfigure}{0.45\textwidth}
         \centering
        \includegraphics[width=0.8\linewidth]{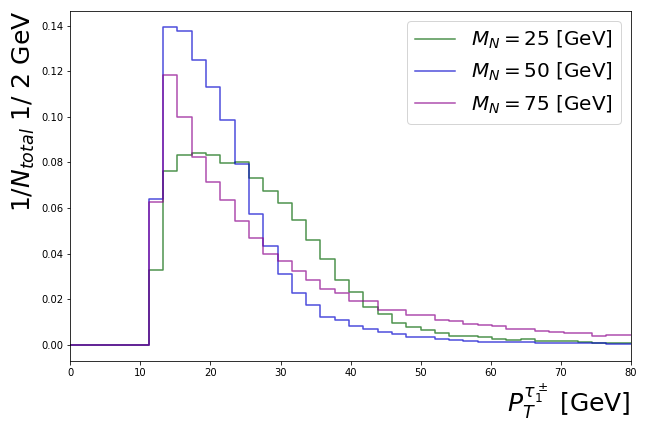}
         \caption{The $p_T$ distribution for leading $\tau^\pm$}
     \end{subfigure}
     \begin{subfigure}{0.45\textwidth}
         \centering
        \includegraphics[width=0.8\linewidth]{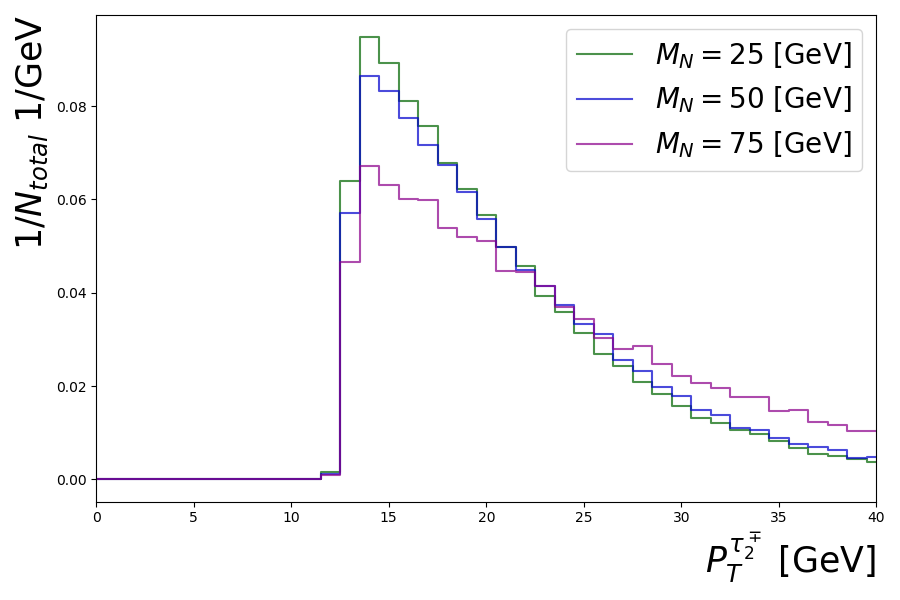}
         \caption{The $p_T$ distribution for subleading $\tau^\mp$}
     \end{subfigure}
     \begin{subfigure}{0.45\textwidth}
         \centering
        \includegraphics[width=0.8\linewidth]{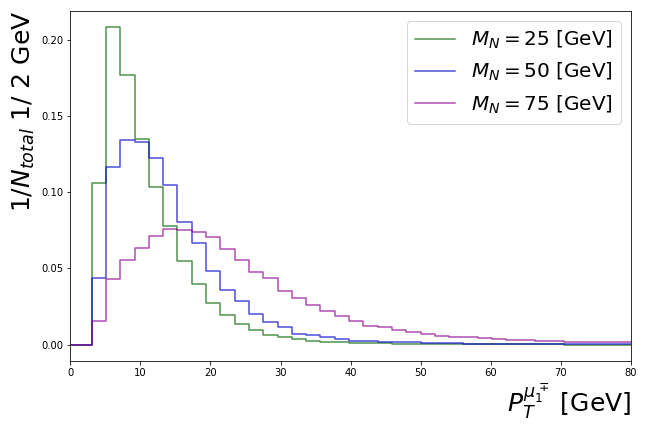}
         \caption{The $p_T$ distribution for leading $\mu^\mp$}
     \end{subfigure}
	 \begin{subfigure}{0.45\textwidth}
         \centering
        \includegraphics[width=0.8\linewidth]{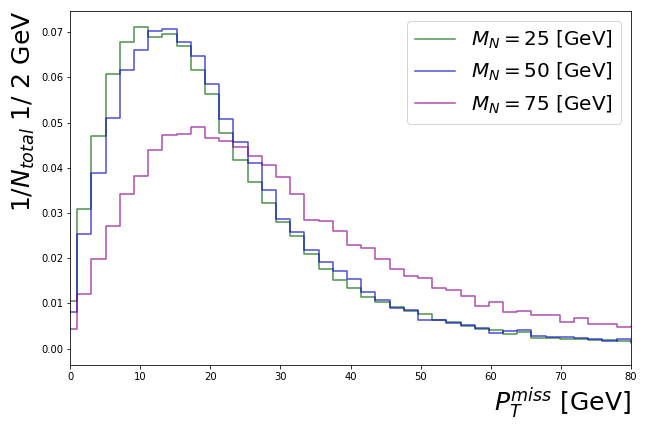}
         \caption{The $p_T$ distribution for missing energy}
     \end{subfigure}
     \begin{subfigure}{0.45\textwidth}
         \centering
        \includegraphics[width=0.8\linewidth]{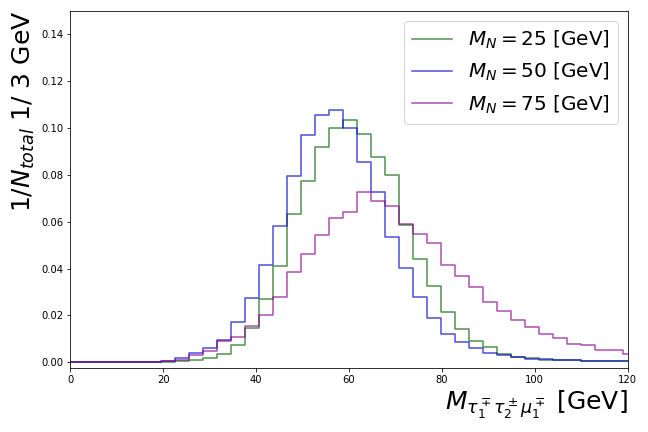}
         \caption{The distribution for $M_{\tau_1^\mp\tau_2^\pm\mu_1^\mp}$  }
     \end{subfigure}
	 \begin{subfigure}{0.45\textwidth}
         \centering
        \includegraphics[width=0.8\linewidth]{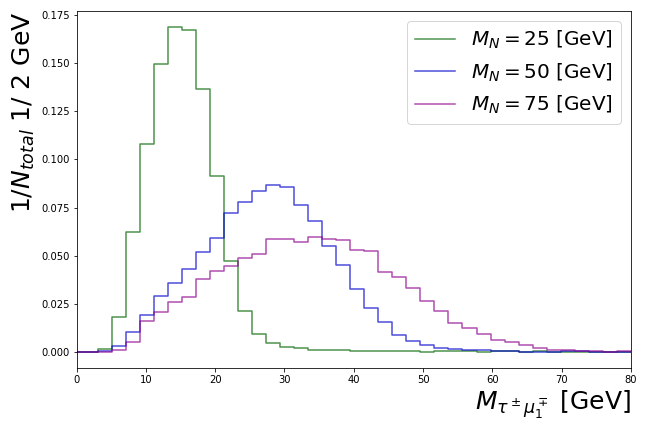}
         \caption{The distribution for $M_{\tau^\pm\mu_1^\mp}$  }
     \end{subfigure}
     \begin{subfigure}{0.45\textwidth}
         \centering
        \includegraphics[width=0.8\linewidth]{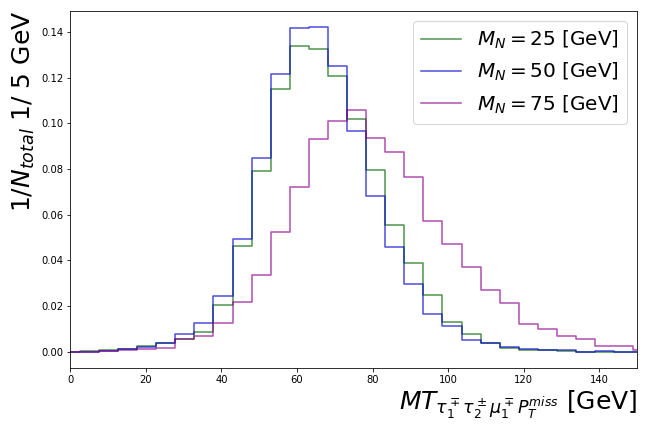}
         \caption{The distribution for $MT_{\tau_1^\mp\tau_2^\pm\mu_1^\mp P_T^{miss}}$  }
     \end{subfigure}
	 \begin{subfigure}{0.45\textwidth}
         \centering
        \includegraphics[width=0.8\linewidth]{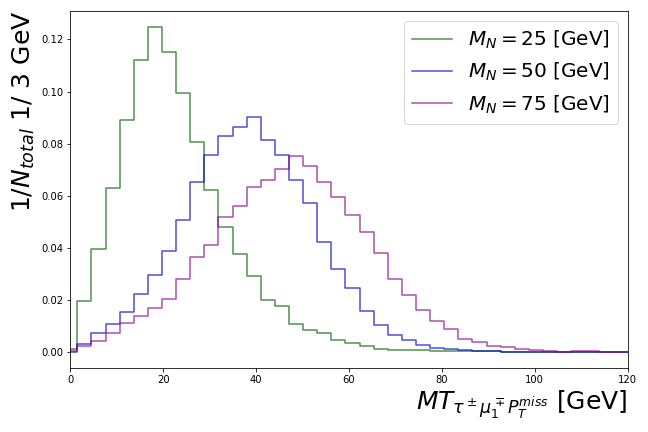}
         \caption{The distribution for $MT_{\tau^\pm\mu_1^\mp P_T^{miss}}$  }
     \end{subfigure}
\caption{
\textbf{Two Opposite-Sign $\tau$s:} 
various kinematical distributions for the signal with the
benchmark points of $M_N = 25$, $50$ and $75$ GeV. 
Notice the distributions in (e), (f), (g) and (h) passed
the preselection criteria.
}\label{fig:M_N_os_Soft}
\end{figure}

\begin{figure}[h]
\centering
	\begin{subfigure}{0.45\textwidth}
         \centering
        \includegraphics[width=0.8\linewidth]{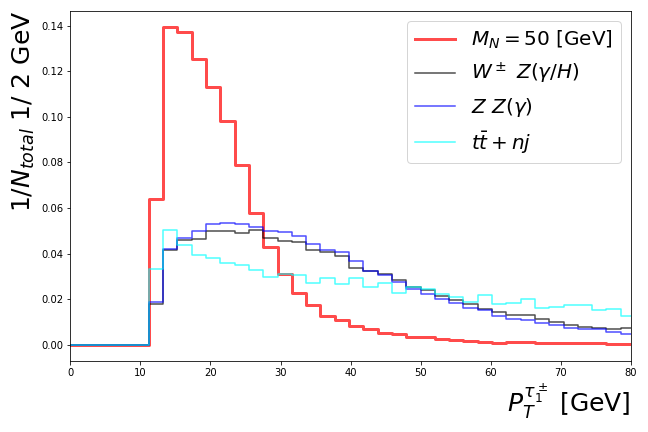}
         \caption{The $p_T$ distribution for leading $\tau^\pm$}
     \end{subfigure}
     \begin{subfigure}{0.45\textwidth}
         \centering
        \includegraphics[width=0.8\linewidth]{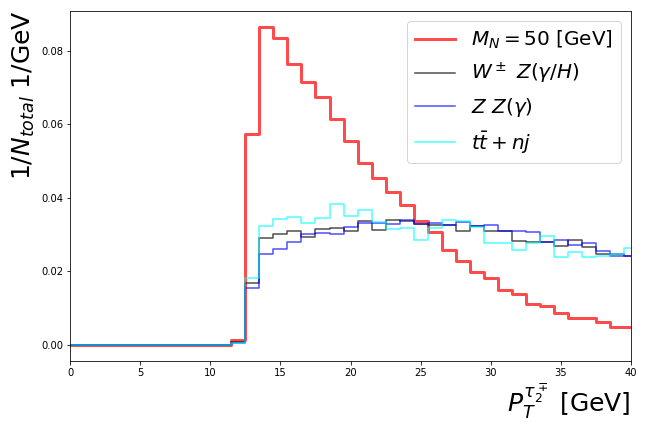}
         \caption{The $p_T$ distribution for subleading $\tau^\mp$}
     \end{subfigure}
     \begin{subfigure}{0.45\textwidth}
         \centering
        \includegraphics[width=0.8\linewidth]{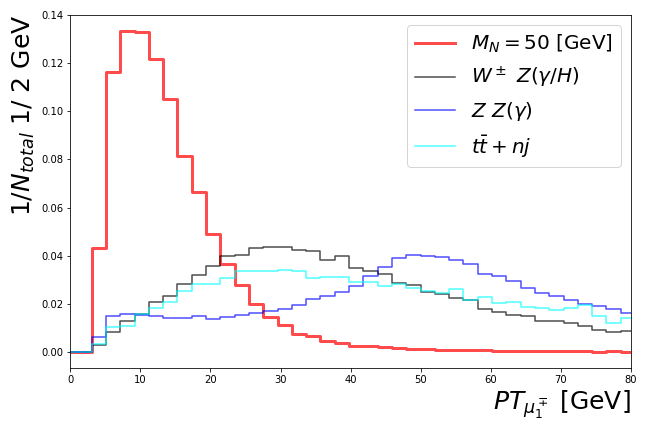}
         \caption{The $p_T$ distribution for leading $\mu^\mp$}
     \end{subfigure}
	 \begin{subfigure}{0.45\textwidth}
         \centering
        \includegraphics[width=0.8\linewidth]{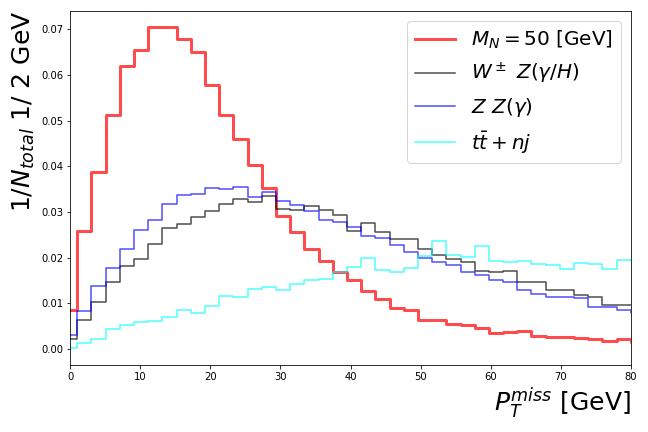}
         \caption{The $p_T$ distribution for missing energy}
     \end{subfigure}
     \begin{subfigure}{0.45\textwidth}
         \centering
        \includegraphics[width=0.8\linewidth]{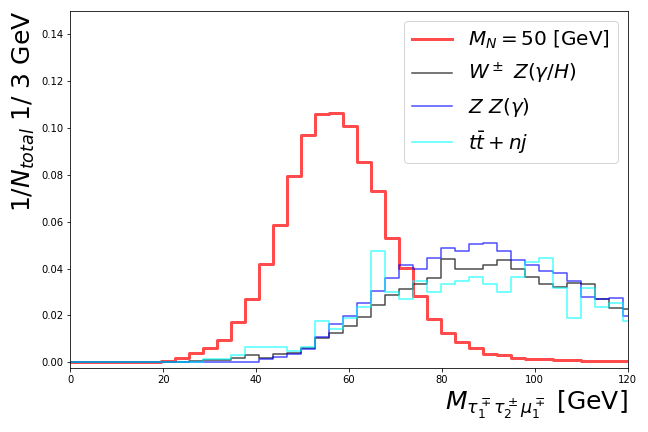}
         \caption{The distribution for $M_{\tau_1^\mp\tau_2^\pm\mu_1^\mp}$  }
     \end{subfigure}
	 \begin{subfigure}{0.45\textwidth}
         \centering
        \includegraphics[width=0.8\linewidth]{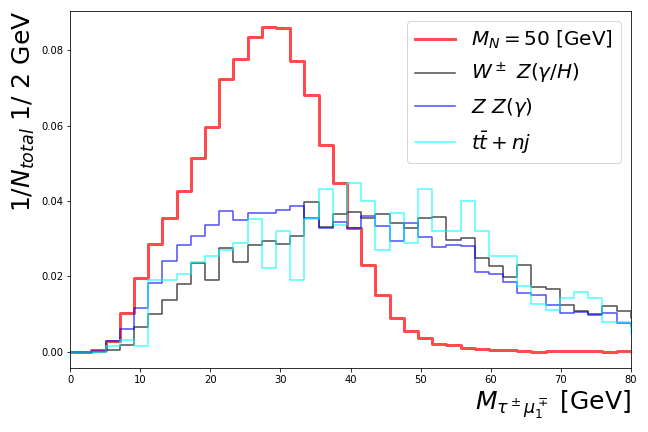}
         \caption{The distribution for $M_{\tau^\pm\mu_1^\mp}$  }
     \end{subfigure}
     \begin{subfigure}{0.45\textwidth}
         \centering
        \includegraphics[width=0.8\linewidth]{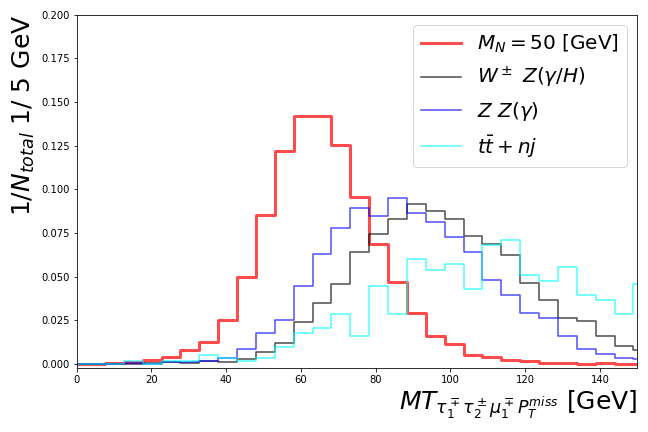}
         \caption{The distribution for $MT_{\tau_1^\mp\tau_2^\pm\mu_1^\mp P_T^{miss}}$  }
     \end{subfigure}
	 \begin{subfigure}{0.45\textwidth}
         \centering
        \includegraphics[width=0.8\linewidth]{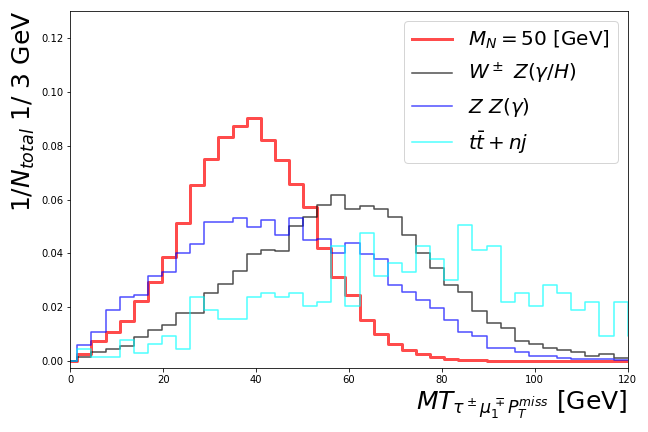}
         \caption{The distribution for $MT_{\tau^\pm\mu_1^\mp P_T^{miss}}$  }
     \end{subfigure}
\caption{
The same as Fig.~\ref{fig:M_N_os_Soft}, 
but for the signal with the benchmark point of $M_N = 50$ GeV and
major SM backgrounds.
}\label{fig:M_N_os_50}
\end{figure}

For $ M_N < m_W $, after passing all selection cuts, 
we can find the signal efficiencies around $0.16 \mathchar`- 0.61\%$, 
the efficiencies of $W^{\pm} Z/H/\gamma$ and $ZZ/\gamma$ are
less than $5.9\times 10^{-2}\%$ and $9.7\times 10^{-2}\%$, 
and that of $\tau \tau + n j$ and $t \bar{t} + n j$ are even smaller, 
less than $8.2\times 10^{-5}\%$ and $5.1\times 10^{-5}\% $, respectively.
\footnote{
  Again, the tiny efficiencies of $\tau \tau + n j$ and
  $t\bar{t} + n j$ also cause unavoidable large statistical
  fluctuations, even we already generated more than $5 \times 10^6$
  and $4 \times 10^6$ Monte Carlo events for them separately.}
Various kinematical distributions for the signal
with $M_N = 25$, $50$ and $75$ GeV are shown in Fig.~\ref{fig:M_N_os_Soft}. 
These distributions are similar to Fig.~\ref{fig:M_N_ss_Soft} except
for panels (a), (b) and (c) in Fig.~\ref{fig:M_N_os_Soft}.
This is due to the different helicity structures
between $N \rightarrow \tau^{+} l^{-}_\alpha \overline{\nu_\alpha}$ 
and $N \rightarrow \tau^{-} l^{+}_\alpha \nu_\alpha$
that involve the $W$ propagator with only the left-handed interaction, and 
causing the variation of $P^l_T$ and $P^{\tau}_T$ distributions.
In Fig~\ref{fig:M_N_os_50}, 
we also display these kinematical distributions for the signal $M_N = 50$
GeV and three major SM backgrounds. 
We do not show kinematical distributions for $\tau \tau + n j$ process 
because only very few events can pass the preselection criteria. 
As we expected, these selection criteria can also successfully
distinguish most parts of the signal from SM backgrounds.

\begin{figure}[h]
\centering
	\begin{subfigure}{0.45\textwidth}
         \centering
        \includegraphics[width=0.8\linewidth]{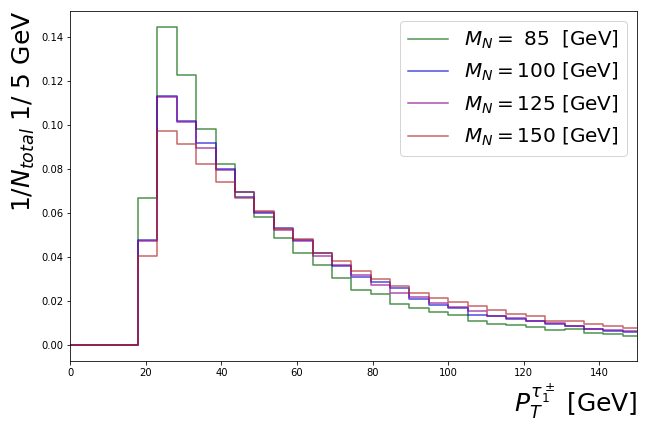}
         \caption{The $p_T$ distribution for leading $\tau^\pm$}
     \end{subfigure}
     \begin{subfigure}{0.45\textwidth}
         \centering
        \includegraphics[width=0.8\linewidth]{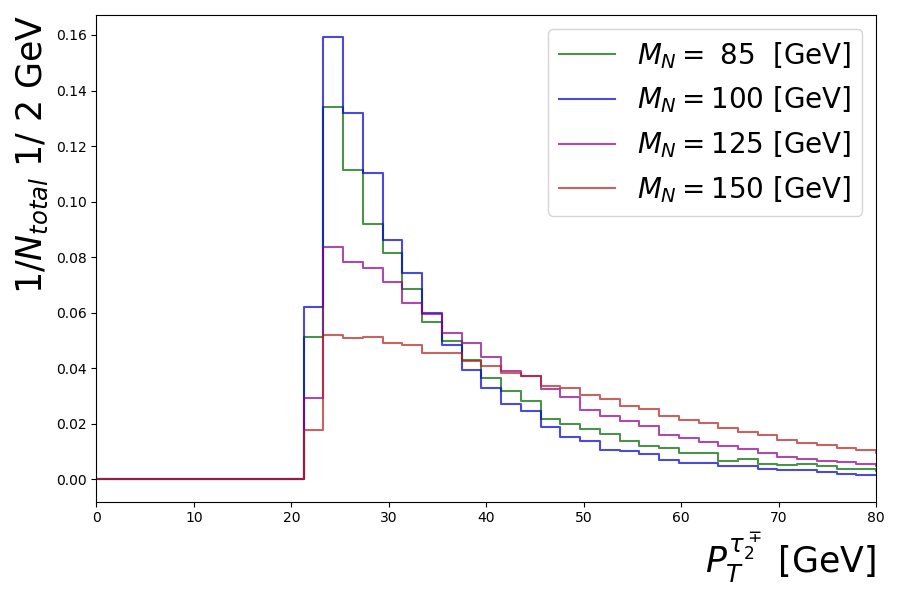}
         \caption{The $p_T$ distribution for subleading $\tau^\mp$}
     \end{subfigure}
     \begin{subfigure}{0.45\textwidth}
         \centering
        \includegraphics[width=0.8\linewidth]{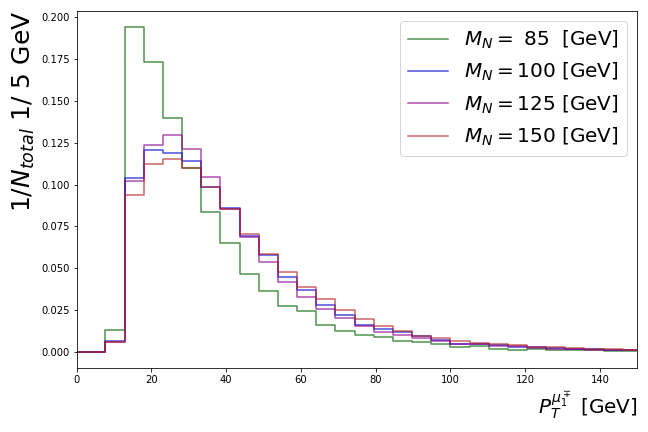}
         \caption{The $p_T$ distribution for leading $\mu^\mp$}
     \end{subfigure}
	 \begin{subfigure}{0.45\textwidth}
         \centering
        \includegraphics[width=0.8\linewidth]{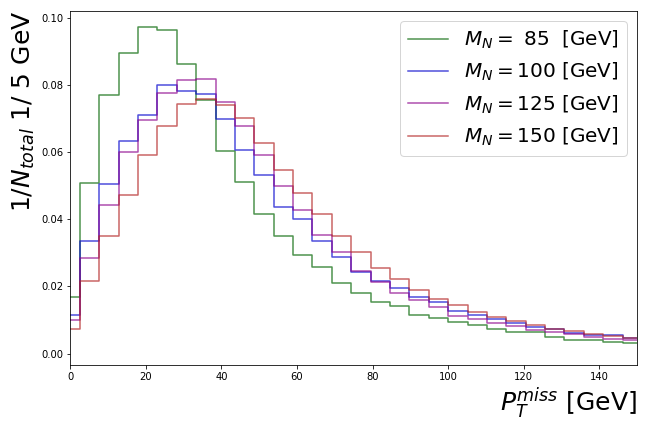}
         \caption{The $p_T$ distribution for missing energy}
     \end{subfigure}
	 \begin{subfigure}{0.45\textwidth}
         \centering
        \includegraphics[width=0.8\linewidth]{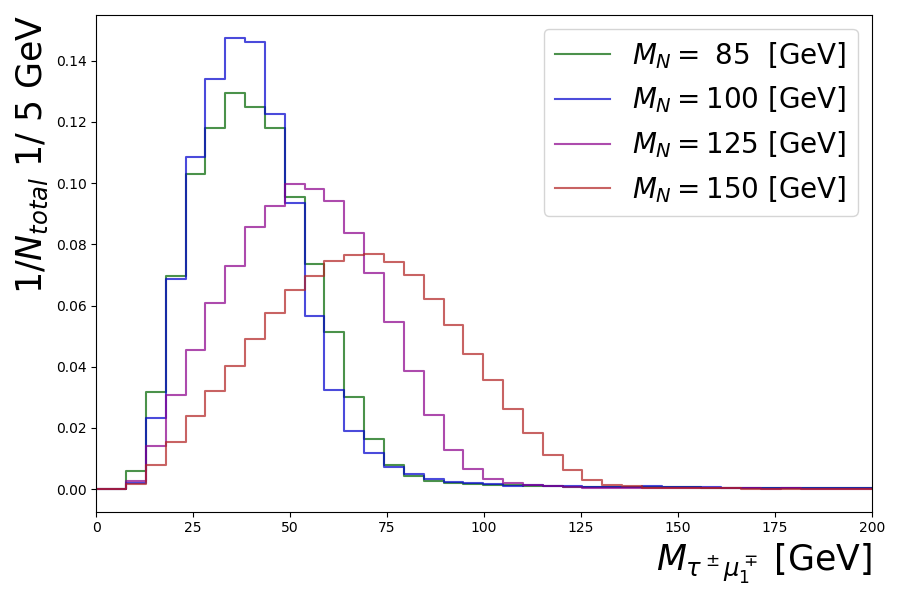}
         \caption{The distribution for $M_{\tau^\pm\mu_1^\mp}$  }
     \end{subfigure}
     \begin{subfigure}{0.45\textwidth}
         \centering
        \includegraphics[width=0.8\linewidth]{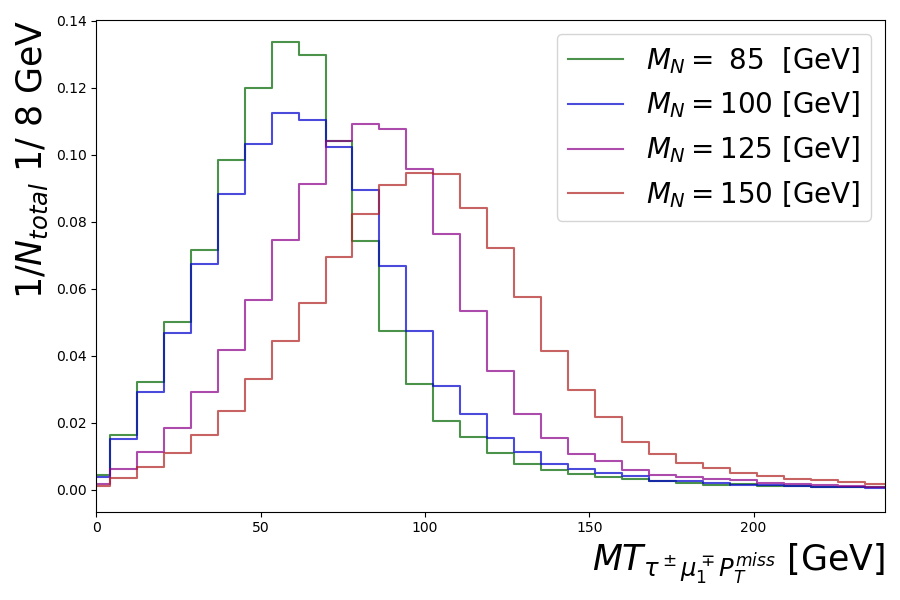}
         \caption{The distribution for $MT_{\tau^\pm\mu_1^\mp P_T^{miss}}$  }
     \end{subfigure}
\caption{
\textbf{Two Opposite-Sign $\tau$s:} 
various kinematical distributions for the signal with the
benchmark points of $M_N = 85$, $100$, $125$ and $150$ GeV. 
Notice the distributions in (e) and (f) passed the preselection criteria. 
}\label{fig:M_N_os_Hard}
\end{figure}

\begin{figure}[h]
\centering
	\begin{subfigure}{0.45\textwidth}
         \centering
        \includegraphics[width=0.8\linewidth]{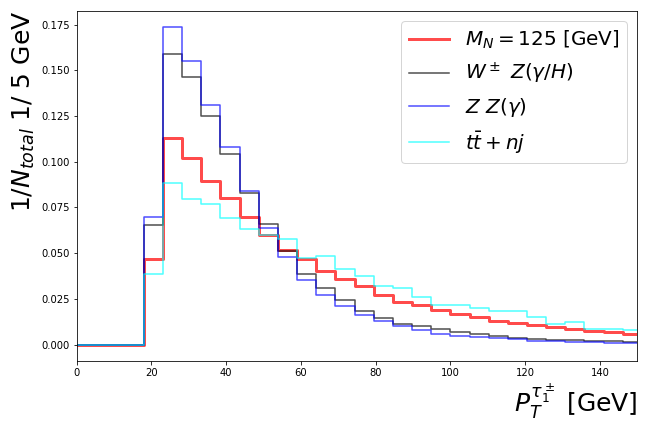}
         \caption{The $p_T$ distribution for leading $\tau^\pm$}
     \end{subfigure}
     \begin{subfigure}{0.45\textwidth}
         \centering
        \includegraphics[width=0.8\linewidth]{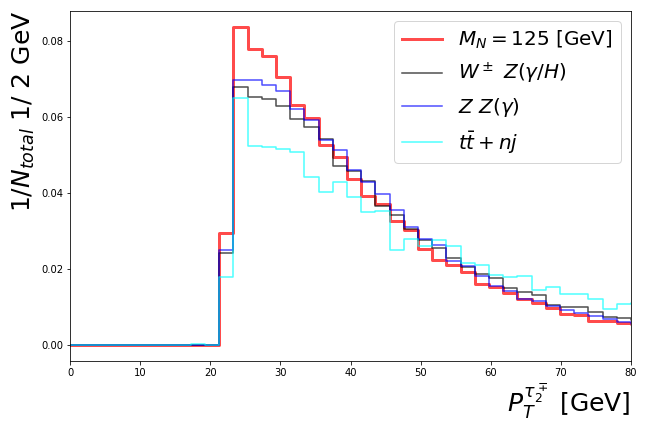}
         \caption{The $p_T$ distribution for subleading $\tau^\mp$}
     \end{subfigure}
     \begin{subfigure}{0.45\textwidth}
         \centering
        \includegraphics[width=0.8\linewidth]{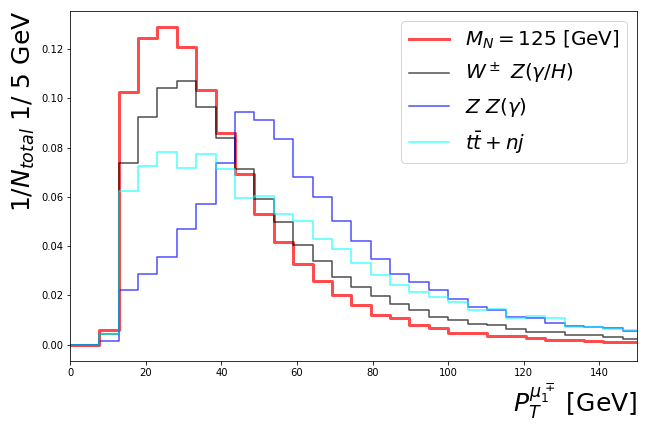}
         \caption{The $p_T$ distribution for leading $\mu^\mp$}
     \end{subfigure}
	 \begin{subfigure}{0.45\textwidth}
         \centering
        \includegraphics[width=0.8\linewidth]{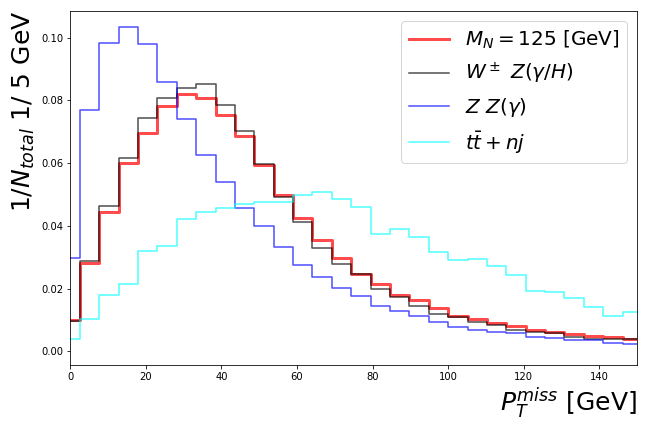}
         \caption{The $p_T$ distribution for missing energy}
     \end{subfigure}
	 \begin{subfigure}{0.45\textwidth}
         \centering
        \includegraphics[width=0.8\linewidth]{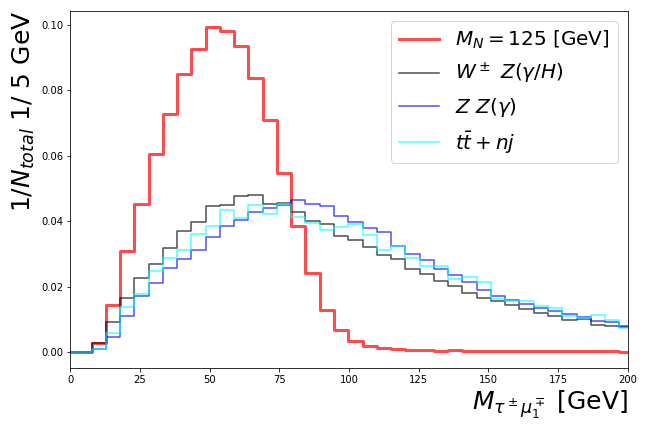}
         \caption{The distribution for $M_{\tau^\pm\mu_1^\mp}$  }
     \end{subfigure}
     \begin{subfigure}{0.45\textwidth}
         \centering
        \includegraphics[width=0.8\linewidth]{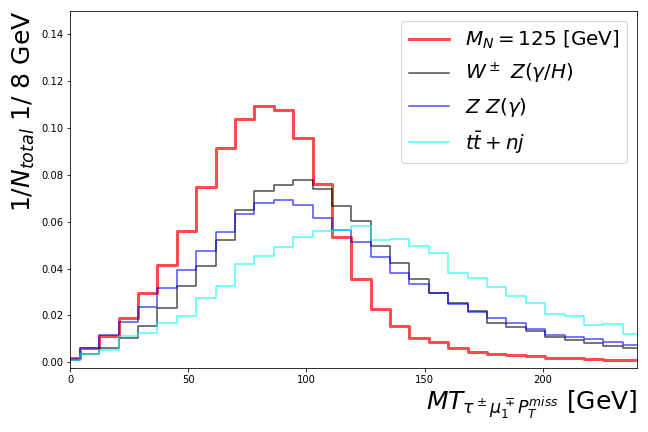}
         \caption{The distribution for $MT_{\tau^\pm\mu_1^\mp P_T^{miss}}$  }
     \end{subfigure}
\caption{
The same as Fig.~\ref{fig:M_N_os_Hard}
but for the signal with the benchmark point of $M_N = 125$ GeV and
major SM backgrounds.
}\label{fig:M_N_os_125}
\end{figure}

For $M_N > m_W$, 
after imposing all selection cuts, we can find the signal
efficiencies around $0.65 \mathchar`- 3.63\%$, 
the efficiencies of $W^{\pm} Z/H/\gamma$ and $ZZ/\gamma$ are
less than $2.63 \%$ and $3.38 \%$, 
and those of $\tau \tau + n j$ and $t \bar{t} + n j$ are even smaller, 
less than $1.52 \times 10^{-4} \%$ and $4.64 \times 10^{-3} \%$, respectively. 
Various kinematical distributions for the signal with
$M_N = 85$, $100$, $125$ and $150$ GeV 
are shown in Fig.~\ref{fig:M_N_os_Hard}. 
Again, these distributions are similar to Fig.~\ref{fig:M_N_ss_Hard}. 
In Fig.~\ref{fig:M_N_os_125}, 
we also display these kinematical distributions for the signal $M_N = 125$ GeV and three major SM backgrounds. 
Again, kinematical distributions for $\tau \tau + n j$ process 
are not shown in Fig.~\ref{fig:M_N_os_125} for the same reason.
It is clear that both $M_{\tau^\pm \mu_1^\mp}$ and $M_{T_{\tau^\pm \mu_1^\mp P_T^{miss}}}$
are useful variables to discriminate the HNL signal from the backgrounds.

\begin{figure}
\centering
\includegraphics[clip,width=0.49\linewidth]{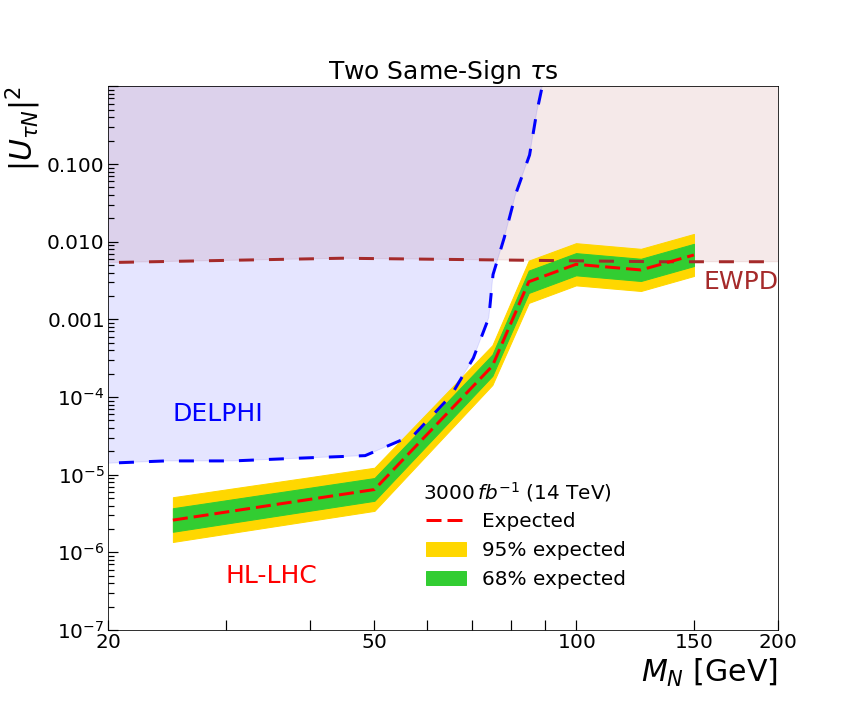}
\includegraphics[clip,width=0.49\linewidth]{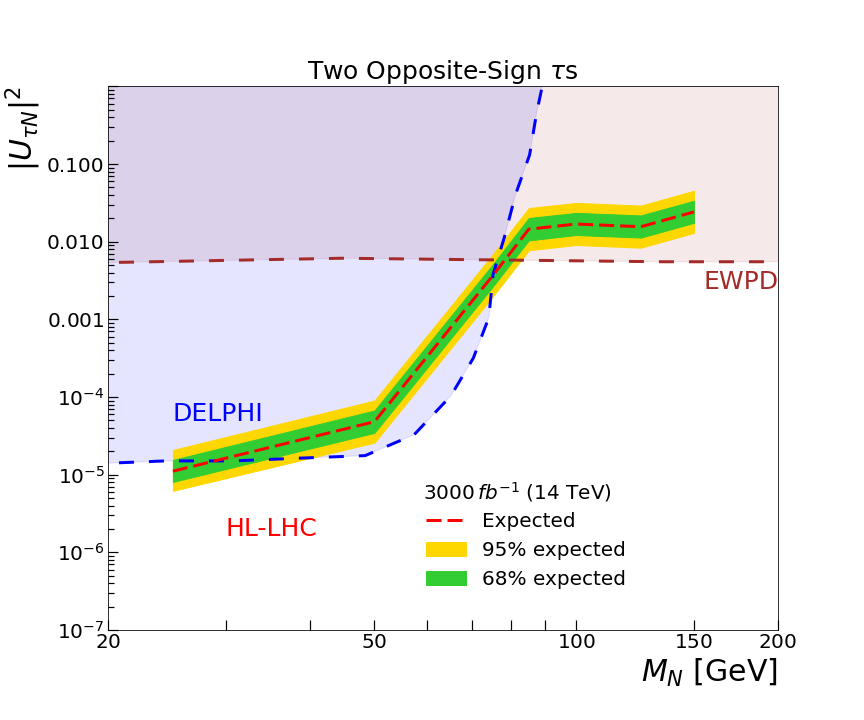}
\caption{
  The expected sensitivity reach of $|U_{\tau N}|^2$ as a function
 of the mass $M_N$ of the HNL for the
same-sign $\tau$ selection (left panel) and
opposite-sign $\tau$ selection (right panel) 
at $ \sqrt{s} = 14 $ TeV with an integrated luminosity
$\mathcal{L} = 3000 fb^{-1}$. 
The exclusion region at $95 \% \  (68 \%)$ CL in the $M_N$ vs.
$|U_{\tau N}|^2$ plane is shown in the yellow (green) band. 
The constraints from EWPD and DELPHI of Fig.~\ref{fig:tau_cons} are added for
comparison.
}\label{fig:CLs}
\end{figure}

Finally, the interpretation of our signal-background analysis results
at $\sqrt{s} = 14$ TeV with an integrated
luminosity $\mathcal{L} = 3000 fb^{-1}$ 
is presented in the left (right) panel of Fig.~\ref{fig:CLs} 
for two-same-sign $\tau$ selection (two-opposite-sign $\tau$ selection). 
The exclusion region at $95\%\,  (68 \%)$ CL in the $M_N$ vs.
$|U_{\tau N}|^2$ plane
is shown in the yellow (green) band.
Those SM backgrounds without MLM matching could have some level of theoretical uncertainties coming from higher order corrections as large as $+(50 \mathchar`- 100)\%$. Here we take into account these uncertainties by allowing a factor of 2 in the background calculation as a conservative estimation.
The constraints from EWPD and DELPHI of Fig.~\ref{fig:tau_cons} are
added for comparison.
We estimate the background uncertainties as $\sqrt{B}$
(we consider only the statistical one in this work) in the
CLs method~\cite{Read:2002hq} 
where $B$ is the total background event numbers.
Also, the background-only hypothesis is assumed and Gaussian
distributions are used for nuisance parameters. 
The \textbf{RooStats} package~\cite{Moneta:2010pm} is applied 
to estimate the confident interval with Asymptotic calculator and
one-sided Profile Likelihood. 
We observe that the sensitivity bounds from HL-LHC
can be stronger than LEP and EWPD constraints in some parameter space, 
especially for two-same-sign $\tau$ selection 
which can reach down to $ |U_{\tau N}|^2 \approx 5\times 10^{-6} $
for $ M_N \lesssim 50$ GeV.
These regions are close to the boundaries between the prompt
and long-lived decays of HNLs at the LHC scale. 
Hence, our study in this paper can serve as a complementary sensitivity
reach of Ref.~\cite{Cottin:2018nms} 
to make HNL searches in the channel
$pp \rightarrow W^{\pm(\ast)} + X \rightarrow \tau^{\pm} N + X $ more complete.

\section{Conclusions}\label{Sec:Conclusions}

The puzzle of tiny neutrino masses and the origin of the
matter-antimatter asymmetry of the Universe are two vital issues
beyond the standard model.  Electroweak scale Type-I seesaw mechanism
is one of the highly-motivated proposals to explain them simultaneously while
maintaining the detectability of the new particles. 
The model can be tested in present or near-future experiments
including the LHC to tell if one or more heavy neutral leptons
exist at the electroweak scale.
The discovery of heavy neutral leptons will become a concrete evidence of
new physics without any doubt.

Among numerous ways to search for heavy neutral leptons in various mass
ranges, the LHC can still serve as the most powerful machine to probe
$\mathcal{O} (10-100)$ GeV heavy neutral leptons in the present as
shown in Figs.~\ref{fig:e_cons} and \ref{fig:mu_cons}. Since there are fewer collider studies of the mixing between $ \nu_{\tau} $ and HNL in literature compared with those of $\nu_e$
and $\nu_{\mu}$ for the HNL of mass in the electroweak scale as shown in Fig.~\ref{fig:tau_cons}, we focus on the channel
$pp \rightarrow W^{\pm(\ast)} + X \rightarrow \tau^{\pm} N + X$ to
search for heavy neutral leptons at the LHC 14 TeV in this work.

The targeted signature in this study consists of three prompt charged
leptons which includes at least two tau leptons. 
We further classify our simulations and event selections according
to two same-sign $\tau$s {\it or} two
opposite-sign $\tau$s for revealing the Majorana nature of heavy
neutral leptons.  After the signal-background analysis, we can observe
these event selections can pick out most parts of the signal against
SM backgrounds, especially for the $M_N < m_W$ benchmark points as
shown in Tables~\ref{ss_cut_soft} and \ref{os_cut_soft} and
Figs.~\ref{fig:M_N_ss_N50},~\ref{fig:M_N_os_50}.  We summarize our
predictions for the testable bounds from HL-LHC in Fig.~\ref{fig:CLs}
which is stronger than the previous LEP constraint and Electroweak
Precision Data (EWPD).  It is obvious that the selection of two
same-sign $\tau$s is more powerful than two opposite-sign $\tau$s and
it can reach down to $ |U_{\tau N}|^2 \approx 5\times 10^{-6} $ for $M_N
\lesssim 50$ GeV.
We should emphasize even this work is based in the context of $\nu$MSM with Majorana neutrinos, our analysis can also be applied to models with Dirac-like/pseudo-Dirac heavy neutrinos with and without charged lepton flavor violation.

\section*{Acknowledgment} 
The work of H.I. was partially supported by JSPS KAKENHI Grant Number 18H03708.
The work of Y.-L.C. and K.C. was supported by the Taiwan MoST with the grant
number MOST-107-2112- M-007-029-MY3.

\newpage

\appendix

\section{Formulas for Heavy Neutral Lepton partial decay widths}\label{app:decay_width}

For the low mass region ($M_N \ll m_{W,Z}$), 
we follow the calculations in Ref.~\cite{Atre:2009rg,Helo:2010cw,Helo:2013prd} for the partial decay widths of $N$. 
Notice that we consider the inclusive approach, 
and take the parameter $\mu _0 \sim m_{\eta ^{\prime}} = 957.78 \pm 0.06$ MeV 
for the mass threshold from which we start taking into account hadronic contributions via $q \overline{q}$ production. 
\begin{enumerate}
\item For $ N \rightarrow l^-_\alpha l^+_\beta \nu_{\beta}, N 
\rightarrow 
l^+_\alpha l^-_\beta \overline{\nu_{l_\beta}} $ and $\alpha \neq \beta$
\begin{equation}
\begin{aligned}
\Gamma (N \rightarrow l^{-}_\alpha l^{+}_\beta \nu _{\beta}) 
&= 
\Gamma (N \rightarrow l^+_\alpha l^-_\beta \overline{\nu_{\beta}}) \\
&= 
|U_{\alpha N}|^2 \frac{G_F^2}{192 \pi ^3} M_N^5 I_1(y_{l_\alpha},y_{\nu _{\beta}},y_{l_\beta})\\
&\equiv 
|U_{\alpha N}|^2\Gamma ^{(l_\alpha l_\beta \nu_\beta)}\,.
\end{aligned}
\end{equation}
\item For $ N \rightarrow \nu_{\alpha} l^{-}_\beta l^{+}_\beta, 
N \rightarrow \overline{\nu_{\alpha}} l^{+}_\beta l^{-}_\beta$ 
\begin{equation}
\begin{aligned}
\Gamma (N \rightarrow \nu_{\alpha} l^{-}_\beta l^{+}_\beta) 
&= 
\Gamma (N \rightarrow \overline{\nu_{\alpha}} l^{+}_\beta l^{-}_\beta) \\
&= 
|U_{\alpha N}|^2 \frac{G_F^2}{96\pi ^3} M_N^5 
\Big[ \left( g^l_L g^l_R + \delta _{l_\alpha l_\beta} g^l_R \right) 
I_2 \left( y_{\nu _{\alpha}}, y_{l_\beta}, y_{l_\beta} \right) \\
&+ 
\left( \left( g^l_L \right)^2 + \left( g^l_R \right)^2 + \delta _{l_\alpha l_\beta} (1+2 g^l_L) \right) 
I_1 \left( y_{\nu _{l_\alpha}}, y_{l_\beta}, y_{l_\beta} \right) \Big] \\
&\equiv 
|U_{\alpha N}|^2\Gamma ^{(\nu_\alpha l_\beta l_\beta)}\,.
\end{aligned}
\end{equation}
\item For $ N \rightarrow \nu_{l_\alpha} \nu_{l_\beta} \overline{\nu_{l_\beta}}, 
N \rightarrow \overline{\nu_{\alpha}} \overline{\nu_{\beta}}\nu_{\beta} $
\begin{equation}
\begin{aligned}
\sum_{\beta=e,\mu,\tau} \Gamma (N \rightarrow \nu _{\alpha} \nu _{\beta} \overline{\nu _{l_\beta}}) 
&= 
\sum _{\beta=e,\mu,\tau} \Gamma (N \rightarrow \overline{\nu_{\alpha}} \overline{\nu_{\beta}}\nu_{\beta}) \\
&= 
|U_{\alpha N}|^2 \frac{G_F^2}{96 \pi ^3} M_N^5 \\
&\equiv |U_{\alpha N}|^2 \Gamma ^{(3\nu)}\,.
\end{aligned}
\end{equation}
\item For $N \rightarrow l^{-}_\alpha U \overline{D}, 
N\rightarrow l^{+}_\alpha \overline{U} D$
\begin{equation}
\begin{aligned}
\Gamma (N \rightarrow l^{-}_\alpha U \overline{D}) 
&= 
\Gamma (N \rightarrow l^{+}_\alpha \overline{U} D) \\
&= 
|U_{\alpha N}|^2 |V_{UD}|^2 \frac{G_F^2}{64\pi ^3} M_N^5 I_1 \left( y_{l_\alpha}, y_U, y_D \right) \\
&\equiv 
|U_{l_\alpha N}|^2 \Gamma ^{(lUD)}\,.
\end{aligned}
\end{equation}
\item For $ N\rightarrow \nu_{\alpha} q \overline{q}, 
N \rightarrow \overline{\nu_{\alpha}} \overline{q} q$
\begin{equation}
\begin{aligned}
\Gamma (N \rightarrow\nu_{\alpha} q \overline{q}) 
&= 
\Gamma ( N \rightarrow \overline{\nu_{\alpha}} \overline{q} q) \\
&= 
|U_{\alpha N}|^2 \frac{G_F^2}{32\pi ^3} M_N^5 
\Big[ g^q_L g^q_R I_2 \left( y_{\nu _{\alpha}}, y_q, y_q \right) 
+ \left( \left( g^q_L \right)^2 + \left( g^q_R \right)^2 \right) I_1 \left( y_{\nu _{\alpha}}, y_q, y_q \right) \big] \\
&\equiv 
|U_{\alpha N}|^2\Gamma ^{(\nu qq)}\,.
\end{aligned}
\end{equation}
\end{enumerate}
Here we denoted $y_i = m_i /M_N$ with $m_i = m_{l,q}$ and $U = u, c$, $D = d, s, b$ and $q = u, d, c, s, b$. 
For lepton and quark masses, 
we apply the values from PDG 2018~\cite{Tanabashi:2018oca}. 

The SM neutral current couplings of leptons and quarks are 
\begin{equation}
\begin{aligned}
g^l_L &= - \frac{1}{2} + \sin^2\theta_W\,, 
\quad 
q^U_L = \frac{1}{2} - \frac{2}{3} \sin^2\theta_W\,, 
\quad 
q^D_L = - \frac{1}{2} + \frac{1}{3} \sin^2\theta_W\,, \\
g^l_R &= \sin^2\theta_W\,, 
\quad 
q^U_R = - \frac{2}{3} \sin^2 \theta_W\,, 
\quad 
q^D_R = \frac{1}{3} \sin^2\theta_W\,.
\end{aligned}
\end{equation}

The kinematical functions used above are
\begin{align}
\lambda (x,y,z) 
&= 
x^2+y^2+z^2-2xy-2yz-2xz\,,\\
I_1(x,y,z) 
&= 
12 \int^{(1-z)^2}_{(x+y)^2} \frac{ds}{s}(s-x^2-y^2)(1+z^2-s) 
\lambda ^{1/2}(s,x^2,y^2) \lambda ^{1/2}(1,s,z^2)\,,\\
I_2(x,y,z) 
&= 
24 yz \int^{(1-x)^2}_{(y+z)^2} \frac{ds}{s}(1+x^2-s) 
\lambda ^{1/2}(s,y^2,z^2) \lambda ^{1/2}(1,s,x^2)\,.
\end{align}

For the medium mass region ($M_N \lesssim m_{t}$), 
we take into account the both effects of on-shell and off-shell $W$ and $Z$ bosons 
by including the width of these gauge bosons in the propagators. 
We follow the calculations in Ref.~\cite{Atre:2009rg,Liao:2017jiz} for the the partial decay widths of $N$. 
Notice all the SM fermion masses of the final states have been neglected to simplify our calculations. 
\begin{enumerate}
\item For $N \rightarrow l^-_\alpha l^+_\beta \nu_{\beta}, 
N \rightarrow l^+_\alpha l^-_\beta \overline{\nu_{\beta}}$ and $\alpha \neq \beta$
\begin{equation}
\begin{aligned}
\Gamma (N \rightarrow l^{-}_\alpha l^{+}_\beta \nu _{\beta}) 
&= 
\Gamma (N \rightarrow l^+_\alpha l^-_\beta \overline{\nu_{\beta}})\\
&= 
|U_{\alpha N}|^2 F_N \left( M_N, m_W, \Gamma_W \right) \\
&\equiv |U_{l_1N}|^2\Gamma ^{(l_1l_2\nu)}\,.
\end{aligned}
\end{equation}
\item For $ N \rightarrow \nu_{\alpha} l^{-}_\beta l^{+}_\beta, 
N \rightarrow \overline{\nu_{\alpha}} l^{+}_\beta l^{-}_\beta$
\begin{equation}
\begin{aligned}
\Gamma ( N \rightarrow \nu_{\alpha} l^{-}_\beta l^{+}_\beta) 
&= 
\Gamma ( N \rightarrow \overline{\nu_{\alpha}} l^{+}_\beta l^{-}_\beta)\\
&= 
|U_{\alpha N}|^2 \Big[ F_N \left( M_N, m_W, \Gamma_W \right) 
+ 3 \left( \left( g^l_L \right)^2 + \left( g^l_R \right)^2 \right) F_N \left( M_N, m_Z, \Gamma_Z \right)\\
&+ 2 g^l_L F_S \left( M_N, m_W, \Gamma_W, m_Z, \Gamma_Z \right) \Big] \\
&\equiv 
|U_{\alpha N}|^2\Gamma ^{(l_2l_2\nu)}\,.
\end{aligned}
\end{equation}
\item For $N \rightarrow \nu_{\alpha} \nu_{\beta} \overline{\nu_{\beta}}, 
N \rightarrow \overline{\nu_{\alpha}} \overline{\nu_{\beta}} \nu_{\beta}$
\begin{equation}
\begin{aligned}
\sum_{\beta=e,\mu,\tau} \Gamma ( N \rightarrow \nu_{\alpha} \nu_{\beta} \overline{\nu_{\beta}}) 
&= 
\sum_{\beta=e,\mu,\tau} \Gamma ( N \rightarrow \overline{\nu_{\alpha}} \overline{\nu_{\beta}} \nu_{\beta})\\ 
&= 
|U_{\alpha N}|^2 \frac{1}{4} (2+4) F_N \left( M_N, m_Z, \Gamma_Z \right) \\
&\equiv 
|U_{\alpha N}|^2 \Gamma^{(3\nu)}\,.
\end{aligned}
\end{equation}
\item For $N \rightarrow l^{-}_\alpha U \overline{D}, 
N \rightarrow l^{+}_\alpha \overline{U} D$
\begin{equation}
\begin{aligned}
\Gamma ( N \rightarrow l^{-}_\alpha U \overline{D}) 
&= 
\Gamma ( N \rightarrow l^{+}_\alpha \overline{U} D)\\
&= 
|U_{\alpha N}|^2 |V_{UD}|^2 N_c F_N \left( M_N, m_W, \Gamma_W \right)\\
&\equiv 
|U_{\alpha N}|^2\Gamma ^{(lUD)}\,.
\end{aligned}
\end{equation}
\item For $N \rightarrow \nu_{\alpha} q \overline{q}, 
N \rightarrow \overline{\nu_{l_\alpha}} \overline{q} q$
\begin{equation}
\begin{aligned}
\Gamma ( N \rightarrow \nu_{\alpha} q \overline{q}) 
&= 
\Gamma ( N \rightarrow \overline{\nu_{\alpha}} \overline{q} q)\\
&= 
|U_{\alpha N}|^2 N_c \left( \left( g^q_L \right)^2 
+ \left( g^q_R \right)^2 \right) F_N \left( M_N, m_Z, \Gamma_Z \right) \\
&\equiv 
|U_{\alpha N}|^2\Gamma^{(\nu qq)}\,.
\end{aligned}
\end{equation}
\end{enumerate}
where $N_c = 3$ is the number of color degrees of freedom for quarks. 

The functions $F_N$ is 
\begin{align}
F_N \left( M_N, m_W, \Gamma_W \right) 
= 
\frac{G^2_F M_N}{\pi^3} \int^{\frac{M_N}{2}}_0 dE_1 
\int^{\frac{M_N }{2}}_{\frac{M_N}{2}-E_1} \left( |P_W|^2 \frac{1}{2} \left( M_N -2E_2 \right) E_2 \right) dE_2\,, 
\end{align}
where $P_W$ comes from the propagator of the $W$ boson with the form, 
\begin{align}
P_W 
= 
\frac{m^2_W}{q^2-m^2_W + i \Gamma_W m_W}\,,
\end{align}
where $q^2 = M^2_N - 2M_N E_1$ and $\Gamma_W$ is the total decay width of $W$. 
We can simply obtain $F_N(M_N, m_Z, \Gamma_Z)$ by taking $(m_W, \Gamma_W) \rightarrow (m_Z, \Gamma_Z)$.

On the other hand, the function $F_S$ is given by 
\begin{align}
F_S 
= 
\frac{G^2_F M_N}{\pi^3} \int^{\frac{M_N}{2}}_0 dE_1 
\int^{\frac{M_N }{2}}_{\frac{M_N}{2}-E_1} \left( \left( P_W P^{\ast}_Z + P^{\ast}_W P_Z \right) 
\frac{1}{2} \left( M_N -2E_2 \right) E_2 \right) dE_2\,,
\end{align}
and $P_Z$ comes from the propagator of the $Z$ boson with the form, 
\begin{align}
P_Z 
= 
\frac{m^2_Z}{q^2_3-m^2_Z+i\Gamma_Z m_Z}\,,
\end{align}
where $ q^2_3 = M^2_N - 2M_N E_3 $ with $ E_3 = M_N - E_1 - E_2 $ considering the decay of $ N $ at rest.

Besides, we also take into account the $N$ partial decay width to the
Higgs boson 
and an active neutrino when $N$ is heavier than the Higgs boson, 
\begin{align}
\Gamma ( N \rightarrow \nu_\alpha H) 
&= 
\frac{g^2}{64\pi m^2_W} | U_{\alpha N} |^2 M^3_N  \left( 1-\frac{m^2_H}{M^2_N} \right)^2 \\
&\equiv 
|U_{\alpha N}|^2\Gamma ^{(\nu H)}\,.
\end{align}

Finally, we represent the total decay width of $N$ as 
\begin{align}
\Gamma _N 
&= 
\sum_{\alpha, \beta, \mathcal{H}} 
\Big[ 2 \times \Gamma ( N \rightarrow l^-_\alpha \mathcal{H}^+ ) 
+ 2 \times \Gamma ( N \rightarrow l^-_\alpha l^+_\beta \nu _{\beta}) 
+ \Gamma ( N \rightarrow \nu _{\beta} \mathcal{H}^0 ) \\
&+ 
\Gamma ( N \rightarrow l^-_\beta l^+_\beta \nu _{\alpha}) 
+ \Gamma ( N \rightarrow \nu _{\alpha} \nu _{\beta} \overline{\nu _{\beta}}) \Big] 
+ \Gamma ( N \rightarrow \nu_\alpha H)\,,
\end{align}
where we denoted the hadronic states $ \mathcal{H}^+ = \overline{d}u, \overline{s}u, \overline{d}c, \overline{s}c, \overline{b}u, \overline{b}c $ and $ \mathcal{H}^0 = \overline{q}q $.
Then we further simplify $ \Gamma _N $ as
\begin{align}
\Gamma _N 
= 
a_e(M_N) \cdot |U_{eN}|^2 + a_{\mu} (M_N) \cdot |U_{\mu N}|^2 + a_{\tau}(M_N)\cdot |U_{\tau N}|^2\,,
\end{align}
where
\begin{align}
a_\alpha (M_N) 
= 
2 \times \Gamma^{(l_\alpha \mathcal{H})} + \Gamma^{(\nu \mathcal{H})} + \Gamma ^{(3\nu)} 
+ \sum_{\beta} \left( \Gamma ^{(l_\beta l_\beta \nu)} + 2 \times \Gamma ^{(l_\alpha l_\beta \nu)} \right) 
+ \Gamma ^{(\nu H)}\,,
\end{align}
with $\alpha, \beta = e, \mu , \tau $.

\section{Extra cut flow tables and kinematical distributions}\label{app:tables_figures}

\begin{table}[h!]
\scriptsize
\begin{center}
\begin{tabular}{c|ccccc }
\hline\hline
\multicolumn{6}{c}{\textbf{Two Same-Sign $ \tau $s Selection Flow Table}}\\
\hline\hline
\multirow{2}{*}{\textbf{Process}}&\textbf{$\sigma$}&\textbf{Preselection}&\textbf{$P_T^{miss}<$ 40 GeV}&\textbf{b veto}&\textbf{Invariant Mass Selection}\\
&(fb)&\textbf{$A\epsilon$ (\%)}&\textbf{$A\epsilon$ (\%)}&\textbf{$A\epsilon$ (\%)}&\textbf{$A\epsilon$ (\%)}\\
\hline
$M_N$ = 25 GeV   &$2.851$&$2.535$&$2.348$&$	2.289$&$	1.640$\\
$W^{\pm}W^{\pm}W^{\mp}$ &$1.828\times10^{-1}$&$1.331$	&$7.078\times10^{-1}$&$6.935\times10^{-1}$	&$5.280\times10^{-2}$\\
$W^+W^-Z/H/\gamma$ &$1.065\times10^{-1}$ &$8.680\times10^{-1}$ &$4.174\times10^{-1}$ &$4.085\times10^{-1}$ &$3.785\times10^{-2}$\\
$t\bar{t}+nj$&$2.357\times10^4$ &$6.471\times10^{-2}$ &$1.287\times10^{-2}$ &$	1.759\times10^{-3}$ &$	9.637\times10^{-5}$\\
\hline\hline

\multirow{2}{*}{\textbf{Process}}&\textbf{$\sigma$}&\textbf{Preselection}&\textbf{$P_T^{miss}<$ 40 GeV}&\textbf{b veto}&\textbf{Invariant Mass Selection}\\
&(fb)&\textbf{$A\epsilon$ (\%)}&\textbf{$A\epsilon$ (\%)}&\textbf{$A\epsilon$ (\%)}&\textbf{$A\epsilon$ (\%)}\\
\hline
$M_N$ = 50 GeV   &$2.068$&$2.931$&$2.683$&$2.612$&$2.368$\\
$W^{\pm}W^{\pm}W^{\mp}$ &$1.828\times10^{-1}$&$1.331$	&$7.078\times10^{-1}$&$6.935\times10^{-1}$	&$1.628\times10^{-1}$\\
$W^+W^-Z/H/\gamma$ &$1.065\times10^{-1}$ &$8.680\times10^{-1}$ &$4.174\times10^{-1}$ &$4.085\times10^{-1}$ &$1.129\times10^{-1}$\\
$t\bar{t}+nj$&$2.357\times10^4$ &$6.471\times10^{-2}$ &$1.287\times10^{-2}$ &$	1.759\times10^{-3}$ &$	3.373\times10^{-4}$\\
\hline\hline

\multirow{2}{*}{\textbf{Process}}&\textbf{$\sigma$}&\textbf{Preselection}&\textbf{$P_T^{miss}<$ 40 GeV}&\textbf{b veto}&\textbf{Invariant Mass Selection}\\
&(fb)&\textbf{$A\epsilon$ (\%)}&\textbf{$A\epsilon$ (\%)}&\textbf{$A\epsilon$ (\%)}&\textbf{$A\epsilon$ (\%)}\\
\hline
$M_N$ = 75 GeV   &$8.935\times10^{-2}$&$1.588$&$1.302$&$1.270$&$8.049\times10^{-1}$\\
$W^{\pm}W^{\pm}W^{\mp}$ &$1.828\times10^{-1}$&$1.331$	&$7.078\times10^{-1}$&$6.935\times10^{-1}$	&$1.787\times10^{-1}$\\
$W^+W^-Z/H/\gamma$ &$1.065\times10^{-1}$ &$8.680\times10^{-1}$ &$4.174\times10^{-1}$ &$4.085\times10^{-1}$ &$1.231\times10^{-1}$\\
$t\bar{t}+nj$&$2.357\times10^4$ &$6.471\times10^{-2}$ &$1.287\times10^{-2}$ &$	1.759\times10^{-3}$ &$	4.337\times10^{-4}$\\
\hline\hline
\end{tabular}
\end{center}
\caption{
The same as Table~\ref{ss_cut_soft} except for the change of $ 15 < P^{\tau_1 (\tau_2)}_T < 50 (30)\; {\rm GeV} $ in \textbf{Preselection}.
}
\label{ss_cut_app}
\end{table}

\begin{table}[h!]
\scriptsize
\begin{center}
\begin{tabular}{c|cccccc }

\hline\hline
\multicolumn{6}{c}{\textbf{Two Opposite-Sign $\tau$s Selection Flow Table}}\\
\hline\hline
\multirow{2}{*}{\textbf{Process}}&\textbf{$\sigma$}&\textbf{Preselection}&\textbf{$P_T^{miss}<$ 40 GeV}&\textbf{b veto}&\textbf{Invariant Mass Selection}\\
&(fb)&\textbf{$A\epsilon$ (\%)}&\textbf{$A\epsilon$ (\%)}&\textbf{$A\epsilon$ (\%)}&\textbf{$A\epsilon$ (\%)}\\
\hline
$M_N$ = 25 GeV   &$2.291$&$1.808$&$1.627$&$1.591$&$1.432$\\
$W^{\pm}Z/H/\gamma$ &$1.599\times10^{2}$&$6.689\times10^{-1}$&$5.146\times10^{-1}$&$5.059\times10^{-1}$&$3.080\times10^{-2}$\\
$ZZ/\gamma$ &$2.400\times10^{1}$&$6.681\times10^{-1}$&$5.859\times10^{-1}$&$5.791\times10^{-1}$&$6.070\times10^{-2}$\\
$\tau\tau+nj$ &$9.559\times10^{5}$&$3.067\times10^{-4}$&	$3.067\times10^{-4}$&$3.067\times10^{-4}$&$2.045\times10^{-5}$\\
$t\bar{t}+nj$&$2.987\times10^4$&$2.560\times10^{-2}$&$4.994\times10^{-3}$&$5.660\times10^{-4}$&$4.043\times10^{-5}$\\
\hline\hline

\multirow{2}{*}{\textbf{Process}}&\textbf{$\sigma$}&\textbf{Preselection}&\textbf{$P_T^{miss}<$ 40 GeV}&\textbf{b veto}&\textbf{Invariant Mass Selection}\\
&(fb)&\textbf{$A\epsilon$ (\%)}&\textbf{$A\epsilon$ (\%)}&\textbf{$A\epsilon$ (\%)}&\textbf{$A\epsilon$ (\%)}\\
\hline
$M_N$ = 50 GeV   &$2.052$&$1.595$&$1.431$&$1.396$&$1.322$\\
$W^{\pm}Z/H/\gamma$ &$1.599\times10^{2}$&$6.689\times10^{-1}$&$5.146\times10^{-1}$&$5.059\times10^{-1}$&$1.029\times10^{-1}$\\
$ZZ/\gamma$ &$2.400\times10^{1}$&$6.681\times10^{-1}$&$5.859\times10^{-1}$&$5.791\times10^{-1}$&$1.439\times10^{-1}$\\
$\tau\tau+nj$ &$9.559\times10^{5}$&$3.067\times10^{-4}$&	$3.067\times10^{-4}$&$3.067\times10^{-4}$&$1.022\times10^{-4}$\\
$t\bar{t}+nj$&$2.987\times10^4$&$2.560\times10^{-2}$&$4.994\times10^{-3}$&$5.660\times10^{-4}$&$2.022\times10^{-4}$\\
\hline\hline

\multirow{2}{*}{\textbf{Process}}&\textbf{$\sigma$}&\textbf{Preselection}&\textbf{$P_T^{miss}<$ 40 GeV}&\textbf{b veto}&\textbf{Invariant Mass Selection}\\
&(fb)&\textbf{$A\epsilon$ (\%)}&\textbf{$A\epsilon$ (\%)}&\textbf{$A\epsilon$ (\%)}&\textbf{$A\epsilon$ (\%)}\\
\hline
$M_N$ = 75 GeV   &$9.104\times10^{-2}$&$7.987\times10^{-1}$&$6.439\times10^{-1}$&$6.279\times10^{-1}$&$4.506\times10^{-1}$\\
$W^{\pm}Z/H/\gamma$ &$1.599\times10^{2}$&$6.689\times10^{-1}$&$5.146\times10^{-1}$&$5.059\times10^{-1}$&$1.125\times10^{-1}$\\
$ZZ/\gamma$ &$2.400\times10^{1}$&$6.681\times10^{-1}$&$5.859\times10^{-1}$&$5.791\times10^{-1}$&$1.538\times10^{-1}$\\
$\tau\tau+nj$ &$9.559\times10^{5}$&$3.067\times10^{-4}$&	$3.067\times10^{-4}$&$3.067\times10^{-4}$&$1.431\times10^{-4}$\\
$t\bar{t}+nj$&$2.987\times10^4$&$2.560\times10^{-2}$&$4.994\times10^{-3}$&$5.660\times10^{-4}$&$2.426\times10^{-4}$\\
\hline\hline

\end{tabular}
\end{center}
\caption{
The same as Table~\ref{os_cut_soft} except for the change of $ 15 < P^{\tau_1 (\tau_2)}_T < 50 (30)\; {\rm GeV} $ in \textbf{Preselection}.
}
\label{os_cut_app}
\end{table}

\begin{figure}
\centering
\begin{minipage}{1.0\linewidth}
\includegraphics[clip,width=0.4\linewidth]{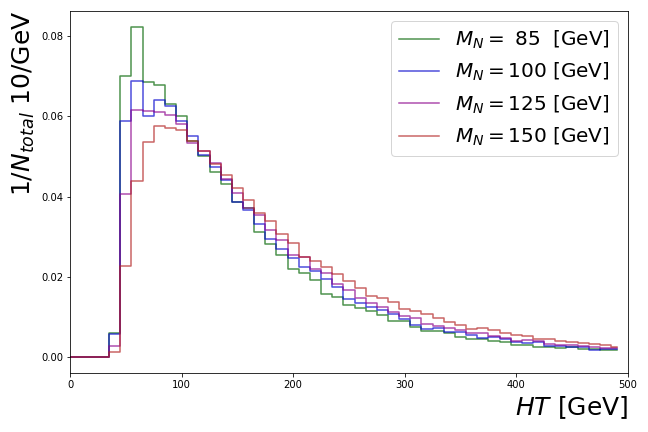}
\hspace{5mm}
\includegraphics[clip,width=0.4\linewidth]{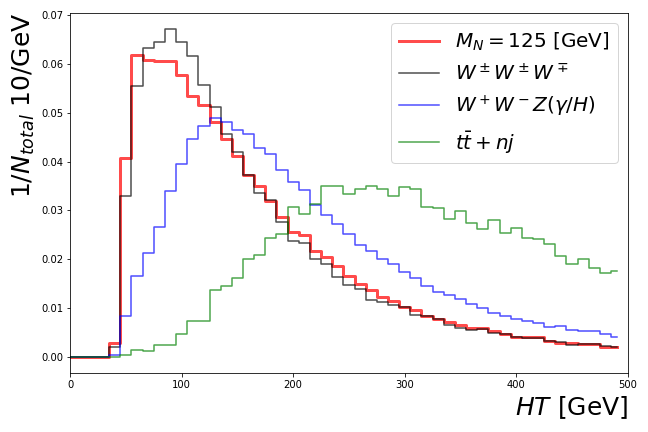}
\includegraphics[clip,width=0.4\linewidth]{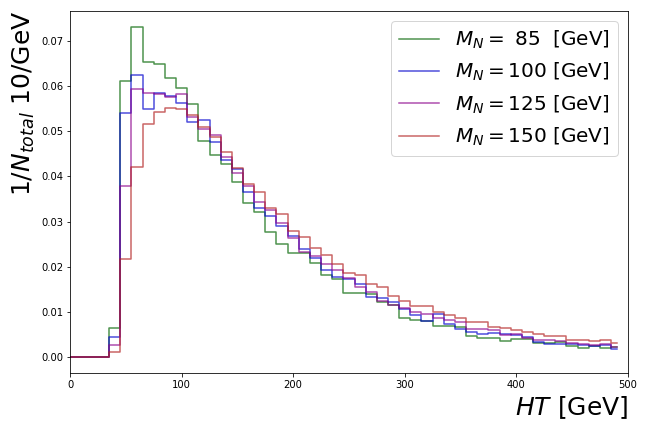}
\hspace{5mm}
\includegraphics[clip,width=0.4\linewidth]{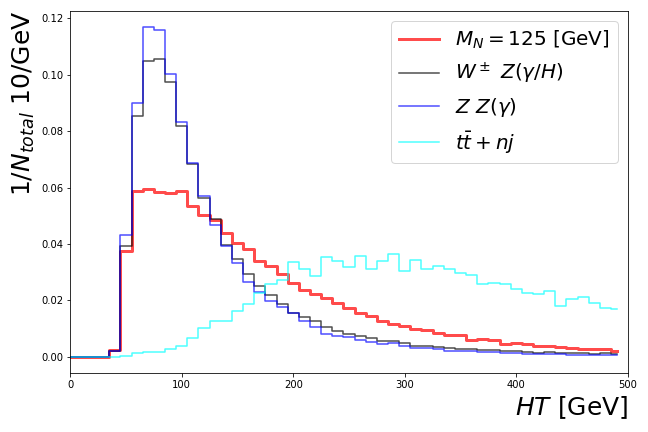}
\end{minipage}
\caption{
The inclusive $H_T$ distributons for the same-sign $\tau$ selection (upper panel) and opposite-sign $\tau$ selection (lower panel). Notice all distributions passed the preselection criteria.
}\label{fig:HT}
\end{figure}

In this appendix, we collect some extra cut flow tables and kinematical distributions which are not shown in the main text. First, inspired from the Ref.~\cite{Florez:2016lwi,Aboubrahim:2017aen}, the situation with $ 15 < P^{\tau_1 (\tau_2)}_T < 50 (30)\; {\rm GeV} $ may also be possible for $M_N < m_W$ and this selection can enhance the signal sensitivity reach. Therefore, we list this kind of event slection for the two same-sign $\tau$s selection flow table in Table~\ref{ss_cut_app} and {\color{blue}two} opposite-sign $\tau$s selection flow table in Table~\ref{os_cut_app} for readers as a reference. 
Second, in order to remove the extra hadronic activity from SM backgrounds for $M_N > m_W$, the inclusive scalar sum of jet $E_T$, $H_T$, which is defined in Eq.~(5.20) of Ref.~\cite{Pascoli:2018heg} is applied in our analysis. The inclusive $H_T$ distributons are shown in Fig.~\ref{fig:HT} for the same-sign $\tau$ selection (upper panel) and opposite-sign $\tau$ selection (lower panel). The selection $H_T < 200\; {\rm GeV}$ can effectively reduce the hadronic activity from $t\overline{t}$ associated processes.



\begin{thebibliography}{99}
\bibitem{Minkowski:1977sc} 
  P.~Minkowski,
  Phys.\ Lett.\  {\bf 67B}, 421 (1977).
  doi:10.1016/0370-2693(77)90435-X



\bibitem{Yanagida:1979as} 
  T.~Yanagida,
  Conf.\ Proc.\ C {\bf 7902131}, 95 (1979).



\bibitem{Yanagida:1980xy} 
  T.~Yanagida,
  Prog.\ Theor.\ Phys.\  {\bf 64}, 1103 (1980).
  doi:10.1143/PTP.64.1103



\bibitem{GellMann:1980vs} 
  M.~Gell-Mann, P.~Ramond and R.~Slansky,
  Conf.\ Proc.\ C {\bf 790927}, 315 (1979)
  [arXiv:1306.4669 [hep-th]].



\bibitem{Ramond:1979}
P.~Ramond, 
in {\em Talk given at the Sanibel Symposium}, 
Palm Coast, Fla., Feb.~25-Mar.~2, 1979, preprint CALT-68-709
(retroprinted as hep-ph/9809459).



\bibitem{Glashow:1979}
S.~L.~Glashow,
in {\em Proc. of the Carg\'ese  Summer Institute on Quarks and Leptons},
Carg\'ese, July 9-29, 1979, 
eds. M.~L\'evy et. al, , (Plenum, 1980, New York), p707.



\bibitem{Mohapatra:1979ia} 
  R.~N.~Mohapatra and G.~Senjanovic,
  Phys.\ Rev.\ Lett.\  {\bf 44}, 912 (1980).
  doi:10.1103/PhysRevLett.44.912



\bibitem{Fukugita:1986hr} 
  M.~Fukugita and T.~Yanagida,
  Phys.\ Lett.\ B {\bf 174}, 45 (1986).
  doi:10.1016/0370-2693(86)91126-3



\bibitem{Davidson:2002qv} 
  S.~Davidson and A.~Ibarra,
  Phys.\ Lett.\ B {\bf 535}, 25 (2002)
  doi:10.1016/S0370-2693(02)01735-5
  [hep-ph/0202239].



\bibitem{Ruchayskiy:2012si} 
  O.~Ruchayskiy and A.~Ivashko,
  JCAP {\bf 1210}, 014 (2012)
  doi:10.1088/1475-7516/2012/10/014
  [arXiv:1202.2841 [hep-ph]].



\bibitem{Asaka:2005an} 
  T.~Asaka, S.~Blanchet and M.~Shaposhnikov,
  Phys.\ Lett.\ B {\bf 631}, 151 (2005)
  doi:10.1016/j.physletb.2005.09.070
  [hep-ph/0503065].



\bibitem{Asaka:2005pn} 
  T.~Asaka and M.~Shaposhnikov,
  Phys.\ Lett.\ B {\bf 620}, 17 (2005)
  doi:10.1016/j.physletb.2005.06.020
  [hep-ph/0505013].



\bibitem{Akhmedov:1998qx} 
  E.~K.~Akhmedov, V.~A.~Rubakov and A.~Y.~Smirnov,
  Phys.\ Rev.\ Lett.\  {\bf 81}, 1359 (1998)
  doi:10.1103/PhysRevLett.81.1359
  [hep-ph/9803255].



\bibitem{Shrock:1980vy} 
  R.~E.~Shrock,
  Phys.\ Lett.\  {\bf 96B}, 159 (1980).
  doi:10.1016/0370-2693(80)90235-X



\bibitem{Shrock:1980ct} 
  R.~E.~Shrock,
  Phys.\ Rev.\ D {\bf 24}, 1232 (1981).
  doi:10.1103/PhysRevD.24.1232



\bibitem{Shrock:1981wq} 
  R.~E.~Shrock,
  Phys.\ Rev.\ D {\bf 24}, 1275 (1981).
  doi:10.1103/PhysRevD.24.1275



\bibitem{Atre:2009rg} 
  A.~Atre, T.~Han, S.~Pascoli and B.~Zhang,
  JHEP {\bf 0905}, 030 (2009)
  doi:10.1088/1126-6708/2009/05/030
  [arXiv:0901.3589 [hep-ph]].



\bibitem{Asaka:2011pb} 
  T.~Asaka, S.~Eijima and H.~Ishida,
  JHEP {\bf 1104}, 011 (2011)
  doi:10.1007/JHEP04(2011)011
  [arXiv:1101.1382 [hep-ph]].



\bibitem{Asaka:2016rwd} 
  T.~Asaka and H.~Ishida,
  Phys.\ Lett.\ B {\bf 763}, 393 (2016)
  doi:10.1016/j.physletb.2016.10.070
  [arXiv:1609.06113 [hep-ph]].



\bibitem{Abada:2019bac} 
  A.~Abada, C.~Hati, X.~Marcano and A.~M.~Teixeira,
  JHEP {\bf 1909}, 017 (2019)
  doi:10.1007/JHEP09(2019)017
  [arXiv:1904.05367 [hep-ph]].



\bibitem{Chun:2019nwi} 
  E.~J.~Chun, A.~Das, S.~Mandal, M.~Mitra and N.~Sinha,
  Phys.\ Rev.\ D {\bf 100}, no. 9, 095022 (2019)
  doi:10.1103/PhysRevD.100.095022
  [arXiv:1908.09562 [hep-ph]].



\bibitem{Bryman:2019bjg} 
  D.~A.~Bryman and R.~Shrock,
  Phys.\ Rev.\ D {\bf 100}, 073011 (2019)
  doi:10.1103/PhysRevD.100.073011
  [arXiv:1909.11198 [hep-ph]].



\bibitem{Kersten:2007vk} 
  J.~Kersten and A.~Y.~Smirnov,
  Phys.\ Rev.\ D {\bf 76}, 073005 (2007)
  doi:10.1103/PhysRevD.76.073005
  [arXiv:0705.3221 [hep-ph]].



\bibitem{Blondel:2014bra} 
  A.~Blondel {\it et al.} [FCC-ee study Team],
  Nucl.\ Part.\ Phys.\ Proc.\  {\bf 273-275}, 1883 (2016)
  doi:10.1016/j.nuclphysbps.2015.09.304
  [arXiv:1411.5230 [hep-ex]].



\bibitem{Deppisch:2015qwa} 
  F.~F.~Deppisch, P.~S.~Bhupal Dev and A.~Pilaftsis,
  New J.\ Phys.\  {\bf 17}, no. 7, 075019 (2015)
  doi:10.1088/1367-2630/17/7/075019
  [arXiv:1502.06541 [hep-ph]].



\bibitem{Drewes:2016jae} 
  M.~Drewes, B.~Garbrecht, D.~Gueter and J.~Klaric,
  JHEP {\bf 1708}, 018 (2017)
  doi:10.1007/JHEP08(2017)018
  [arXiv:1609.09069 [hep-ph]].



\bibitem{Cai:2017mow} 
  Y.~Cai, T.~Han, T.~Li and R.~Ruiz,
  Front.\ in Phys.\  {\bf 6}, 40 (2018)
  doi:10.3389/fphy.2018.00040
  [arXiv:1711.02180 [hep-ph]].



\bibitem{Helo:2018qej} 
  J.~C.~Helo, M.~Hirsch and Z.~S.~Wang,
  JHEP {\bf 1807}, 056 (2018)
  doi:10.1007/JHEP07(2018)056
  [arXiv:1803.02212 [hep-ph]].



\bibitem{Liu:2019qfa} 
  N.~Liu, Z.~G.~Si, L.~Wu, H.~Zhou and B.~Zhu,
  Phys.\ Rev.\ D {\bf 101}, no. 7, 071701 (2020)
  doi:10.1103/PhysRevD.101.071701
  [arXiv:1910.00749 [hep-ph]].

\bibitem{Bondarenko:2018ptm}
K.~Bondarenko, A.~Boyarsky, D.~Gorbunov and O.~Ruchayskiy,
JHEP \textbf{11} (2018), 032
doi:10.1007/JHEP11(2018)032
[arXiv:1805.08567 [hep-ph]].

\bibitem{Cvetic:2019shl} 
  G.~Cvetič and C.~S.~Kim,
  Phys.\ Rev.\ D {\bf 100}, no. 1, 015014 (2019)
  doi:10.1103/PhysRevD.100.015014
  [arXiv:1904.12858 [hep-ph]].



\bibitem{Abada:2018sfh} 
  A.~Abada, N.~Bernal, M.~Losada and X.~Marcano,
  JHEP {\bf 1901}, 093 (2019)
  doi:10.1007/JHEP01(2019)093
  [arXiv:1807.10024 [hep-ph]].



\bibitem{Cottin:2018nms} 
  G.~Cottin, J.~C.~Helo and M.~Hirsch,
  Phys.\ Rev.\ D {\bf 98}, no. 3, 035012 (2018)
  doi:10.1103/PhysRevD.98.035012
  [arXiv:1806.05191 [hep-ph]].

\bibitem{Hernandez:2018cgc}
P.~Hernández, J.~Jones-Pérez and O.~Suarez-Navarro,
Eur. Phys. J. C \textbf{79} (2019) no.3, 220
doi:10.1140/epjc/s10052-019-6728-1
[arXiv:1810.07210 [hep-ph]].



\bibitem{Drewes:2019vjy}
M.~Drewes, A.~Giammanco, J.~Hajer and M.~Lucente,
Phys. Rev. D \textbf{101} (2020) no.5, 055002
doi:10.1103/PhysRevD.101.055002
[arXiv:1905.09828 [hep-ph]].



\bibitem{Andres:2017daw} 
  A.~Flórez, K.~Gui, A.~Gurrola, C.~Patiño and D.~Restrepo,
  Phys.\ Lett.\ B {\bf 778}, 94 (2018)
  doi:10.1016/j.physletb.2018.01.009
  [arXiv:1708.03007 [hep-ph]].

\bibitem{Pascoli:2018rsg} 
  S.~Pascoli, R.~Ruiz and C.~Weiland,
  Phys.\ Lett.\ B {\bf 786}, 106 (2018)
  doi:10.1016/j.physletb.2018.08.060
  [arXiv:1805.09335 [hep-ph]].


\bibitem{Gronau:1984ct}
M.~Gronau, C.~N.~Leung and J.~L.~Rosner,
Phys. Rev. D \textbf{29} (1984), 2539
doi:10.1103/PhysRevD.29.2539

\bibitem{Perl:1984yp}
M.~L.~Perl, T.~Barklow, A.~Boyarski, M.~Breidenbach, P.~Burchat, D.~L.~Burke, J.~Dorfan, G.~J.~Feldman, L.~D.~Gladney, G.~Hanson, K.~G.~Hayes, R.~J.~Hollebeek, W.~R.~Innes, J.~Jaros, D.~Karlen, A.~J.~Lankford, R.~R.~Larsen, B.~LeClaire, N.~Lockyer, V.~Luth, C.~Matteuzzi, R.~A.~Ong, B.~Richter, K.~Riles, M.~C.~Ross, D.~Schlatter, J.~M.~Yelton, C.~Zaiser, G.~S.~Abrams, D.~Amidei, A.~R.~Baden, J.~Boyer, F.~Butler, G.~Gidal, M.~S.~Gold, G.~Goldhaber, L.~Golding, J.~Haggerty, D.~Herrup, I.~Juricic, J.~A.~Kadyk, M.~E.~Nelson, P.~C.~Rowson, H.~Schellman, W.~B.~Schmidke, P.~D.~Sheldon, C.~de la Vaissiere, D.~R.~Wood, M.~E.~Levi and T.~Schaad,
Phys. Rev. D \textbf{32} (1985), 2859
doi:10.1103/PhysRevD.32.2859

\bibitem{Gilman:1985tr}
F.~J.~Gilman and S.~H.~Rhie,
Phys. Rev. D \textbf{32} (1985), 324-326
doi:10.1103/PhysRevD.32.324

\bibitem{Gilman:1986mz}
F.~J.~Gilman,
Comments Nucl. Part. Phys. \textbf{16} (1986) no.5, 231-247
SLAC-PUB-3898.

\bibitem{Hagiwara:1987ub}
K.~Hagiwara and S.~Komamiya,
Adv. Ser. Direct. High Energy Phys. \textbf{1} (1988), 785-859
doi:10.1142/9789814415613\_0013

\bibitem{Dittmar:1989yg} 
  M.~Dittmar, A.~Santamaria, M.~C.~Gonzalez-Garcia and J.~W.~F.~Valle,
  Nucl.\ Phys.\ B {\bf 332}, 1 (1990).
  doi:10.1016/0550-3213(90)90028-C



\bibitem{Ma:1989jpa} 
  E.~Ma and J.~T.~Pantaleone,
  Phys.\ Rev.\ D {\bf 40}, 2172 (1989).
  doi:10.1103/PhysRevD.40.2172



\bibitem{Dicus:1991wj} 
  D.~A.~Dicus and P.~Roy,
  Phys.\ Rev.\ D {\bf 44}, 1593 (1991).
  doi:10.1103/PhysRevD.44.1593



\bibitem{Chun:2017spz} 
  E.~J.~Chun {\it et al.},
  Int.\ J.\ Mod.\ Phys.\ A {\bf 33}, no. 05n06, 1842005 (2018)
  doi:10.1142/S0217751X18420058
  [arXiv:1711.02865 [hep-ph]].



\bibitem{Alva:2014gxa} 
  D.~Alva, T.~Han and R.~Ruiz,
  JHEP {\bf 1502}, 072 (2015)
  doi:10.1007/JHEP02(2015)072
  [arXiv:1411.7305 [hep-ph]].



\bibitem{Pascoli:2018heg} 
  S.~Pascoli, R.~Ruiz and C.~Weiland,
  JHEP {\bf 1906}, 049 (2019)
  doi:10.1007/JHEP06(2019)049
  [arXiv:1812.08750 [hep-ph]].



\bibitem{Alekhin:2015byh} 
  S.~Alekhin {\it et al.},
  Rept.\ Prog.\ Phys.\  {\bf 79}, no. 12, 124201 (2016)
  doi:10.1088/0034-4885/79/12/124201
  [arXiv:1504.04855 [hep-ph]].



\bibitem{Kling:2018wct} 
  F.~Kling and S.~Trojanowski,
  Phys.\ Rev.\ D {\bf 97}, no. 9, 095016 (2018)
  doi:10.1103/PhysRevD.97.095016
  [arXiv:1801.08947 [hep-ph]].



\bibitem{Curtin:2018mvb} 
  D.~Curtin {\it et al.},
  Rept.\ Prog.\ Phys.\  {\bf 82}, no. 11, 116201 (2019)
  doi:10.1088/1361-6633/ab28d6
  [arXiv:1806.07396 [hep-ph]].



\bibitem{Lee:2018pag} 
  L.~Lee, C.~Ohm, A.~Soffer and T.~T.~Yu,
  Prog.\ Part.\ Nucl.\ Phys.\  {\bf 106}, 210 (2019)
  doi:10.1016/j.ppnp.2019.02.006
  [arXiv:1810.12602 [hep-ph]].



\bibitem{Dercks:2018wum} 
  D.~Dercks, H.~K.~Dreiner, M.~Hirsch and Z.~S.~Wang,
  Phys.\ Rev.\ D {\bf 99}, no. 5, 055020 (2019)
  doi:10.1103/PhysRevD.99.055020
  [arXiv:1811.01995 [hep-ph]].



\bibitem{Alimena:2019zri} 
  J.~Alimena {\it et al.},
  arXiv:1903.04497 [hep-ex].



\bibitem{Aielli:2019ivi} 
  G.~Aielli {\it et al.},
  arXiv:1911.00481 [hep-ex].



\bibitem{Hirsch:2020klk} 
  M.~Hirsch and Z.~S.~Wang,
  Phys.\ Rev.\ D {\bf 101}, no. 5, 055034 (2020)
  doi:10.1103/PhysRevD.101.055034
  [arXiv:2001.04750 [hep-ph]].



\bibitem{delAguila:2008pw} 
  F.~del Aguila, J.~de Blas and M.~Perez-Victoria,
  Phys.\ Rev.\ D {\bf 78}, 013010 (2008)
  doi:10.1103/PhysRevD.78.013010
  [arXiv:0803.4008 [hep-ph]].



\bibitem{Akhmedov:2013hec} 
  E.~Akhmedov, A.~Kartavtsev, M.~Lindner, L.~Michaels and J.~Smirnov,
  JHEP {\bf 1305}, 081 (2013)
  doi:10.1007/JHEP05(2013)081
  [arXiv:1302.1872 [hep-ph]].



\bibitem{Basso:2013jka} 
  L.~Basso, O.~Fischer and J.~J.~van der Bij,
  EPL {\bf 105}, no. 1, 11001 (2014)
  doi:10.1209/0295-5075/105/11001
  [arXiv:1310.2057 [hep-ph]].



\bibitem{deBlas:2013gla} 
  J.~de Blas,
  EPJ Web Conf.\  {\bf 60}, 19008 (2013)
  doi:10.1051/epjconf/20136019008
  [arXiv:1307.6173 [hep-ph]].



\bibitem{Antusch:2015mia} 
  S.~Antusch and O.~Fischer,
  JHEP {\bf 1505}, 053 (2015)
  doi:10.1007/JHEP05(2015)053
  [arXiv:1502.05915 [hep-ph]].



\bibitem{Adriani:1992pq} 
  O.~Adriani {\it et al.} [L3 Collaboration],
  Phys.\ Lett.\ B {\bf 295}, 371 (1992).
  doi:10.1016/0370-2693(92)91579-X



\bibitem{Acciarri:1999qj} 
  M.~Acciarri {\it et al.} [L3 Collaboration],
  Phys.\ Lett.\ B {\bf 461}, 397 (1999)
  doi:10.1016/S0370-2693(99)00852-7
  [hep-ex/9909006].



\bibitem{Achard:2001qv} 
  P.~Achard {\it et al.} [L3 Collaboration],
  Phys.\ Lett.\ B {\bf 517}, 67 (2001)
  doi:10.1016/S0370-2693(01)00993-5
  [hep-ex/0107014].



\bibitem{Abreu:1996pa} 
  P.~Abreu {\it et al.} [DELPHI Collaboration],
  Z.\ Phys.\ C {\bf 74}, 57 (1997)
  Erratum: [Z.\ Phys.\ C {\bf 75}, 580 (1997)].
  doi:10.1007/s002880050370



\bibitem{Sirunyan:2018mtv} 
  A.~M.~Sirunyan {\it et al.} [CMS Collaboration],
  Phys.\ Rev.\ Lett.\  {\bf 120}, no. 22, 221801 (2018)
  doi:10.1103/PhysRevLett.120.221801
  [arXiv:1802.02965 [hep-ex]].



\bibitem{Sirunyan:2018xiv} 
  A.~M.~Sirunyan {\it et al.} [CMS Collaboration],
  JHEP {\bf 1901}, 122 (2019)
  doi:10.1007/JHEP01(2019)122
  [arXiv:1806.10905 [hep-ex]].



\bibitem{Aad:2019kiz} 
  G.~Aad {\it et al.} [ATLAS Collaboration],
  JHEP {\bf 1910}, 265 (2019)
  doi:10.1007/JHEP10(2019)265
  [arXiv:1905.09787 [hep-ex]].



\bibitem{Tanabashi:2018oca} 
  M.~Tanabashi {\it et al.} [Particle Data Group],
  Phys.\ Rev.\ D {\bf 98}, no. 3, 030001 (2018).
  doi:10.1103/PhysRevD.98.030001



\bibitem{Degrande:2016aje} 
  C.~Degrande, O.~Mattelaer, R.~Ruiz and J.~Turner,
  Phys.\ Rev.\ D {\bf 94}, no. 5, 053002 (2016)
  doi:10.1103/PhysRevD.94.053002
  [arXiv:1602.06957 [hep-ph]].



\bibitem{Alloul:2013bka} 
  A.~Alloul, N.~D.~Christensen, C.~Degrande, C.~Duhr and B.~Fuks,
  Comput.\ Phys.\ Commun.\  {\bf 185}, 2250 (2014)
  doi:10.1016/j.cpc.2014.04.012
  [arXiv:1310.1921 [hep-ph]].



\bibitem{Alwall:2014hca} 
  J.~Alwall {\it et al.},
  JHEP {\bf 1407}, 079 (2014)
  doi:10.1007/JHEP07(2014)079
  [arXiv:1405.0301 [hep-ph]].



\bibitem{Frederix:2018nkq} 
  R.~Frederix, S.~Frixione, V.~Hirschi, D.~Pagani, H.-S.~Shao and M.~Zaro,
  JHEP {\bf 1807}, 185 (2018)
  doi:10.1007/JHEP07(2018)185
  [arXiv:1804.10017 [hep-ph]].



\bibitem{Keung:1983uu} 
  W.~Y.~Keung and G.~Senjanovic,
  Phys.\ Rev.\ Lett.\  {\bf 50}, 1427 (1983).
  doi:10.1103/PhysRevLett.50.1427



\bibitem{Das:2017gke} 
  A.~Das, P.~Konar and A.~Thalapillil,
  JHEP {\bf 1802}, 083 (2018)
  doi:10.1007/JHEP02(2018)083
  [arXiv:1709.09712 [hep-ph]].



\bibitem{Das:2018usr} 
  A.~Das, S.~Jana, S.~Mandal and S.~Nandi,
  Phys.\ Rev.\ D {\bf 99}, no. 5, 055030 (2019)
  doi:10.1103/PhysRevD.99.055030
  [arXiv:1811.04291 [hep-ph]].



\bibitem{Sjostrand:2007gs} 
  T.~Sjostrand, S.~Mrenna and P.~Z.~Skands,
  Comput.\ Phys.\ Commun.\  {\bf 178}, 852 (2008)
  doi:10.1016/j.cpc.2008.01.036
  [arXiv:0710.3820 [hep-ph]].



\bibitem{deFavereau:2013fsa} 
  J.~de Favereau {\it et al.} [DELPHES 3 Collaboration],
  JHEP {\bf 1402}, 057 (2014)
  doi:10.1007/JHEP02(2014)057
  [arXiv:1307.6346 [hep-ex]].



\bibitem{Mangano:2006rw} 
  M.~L.~Mangano, M.~Moretti, F.~Piccinini and M.~Treccani,
  JHEP {\bf 0701}, 013 (2007)
  doi:10.1088/1126-6708/2007/01/013
  [hep-ph/0611129].



\bibitem{Alwall:2007fs} 
  J.~Alwall {\it et al.},
  Eur.\ Phys.\ J.\ C {\bf 53}, 473 (2008)
  doi:10.1140/epjc/s10052-007-0490-5
  [arXiv:0706.2569 [hep-ph]].



\bibitem{Cacciari:2008gp} 
  M.~Cacciari, G.~P.~Salam and G.~Soyez,
  JHEP {\bf 0804}, 063 (2008)
  doi:10.1088/1126-6708/2008/04/063
  [arXiv:0802.1189 [hep-ph]].



\bibitem{Cacciari:2011ma} 
  M.~Cacciari, G.~P.~Salam and G.~Soyez,
  Eur.\ Phys.\ J.\ C {\bf 72}, 1896 (2012)
  doi:10.1140/epjc/s10052-012-1896-2
  [arXiv:1111.6097 [hep-ph]].



\bibitem{ATLAS:2019uhp} 
  The ATLAS collaboration [ATLAS Collaboration],
  ATL-PHYS-PUB-2019-033.



\bibitem{Aad:2019qnd} 
  G.~Aad {\it et al.} [ATLAS Collaboration],
  Phys.\ Rev.\ D {\bf 101}, no. 5, 052005 (2020)
  doi:10.1103/PhysRevD.101.052005
  [arXiv:1911.12606 [hep-ex]].



\bibitem{Florez:2016lwi} 
  A.~Flórez, L.~Bravo, A.~Gurrola, C.~Ávila, M.~Segura, P.~Sheldon and W.~Johns,
  Phys.\ Rev.\ D {\bf 94}, no. 7, 073007 (2016)
  doi:10.1103/PhysRevD.94.073007
  [arXiv:1606.08878 [hep-ph]].



\bibitem{Aboubrahim:2017aen} 
  A.~Aboubrahim, P.~Nath and A.~B.~Spisak,
  Phys.\ Rev.\ D {\bf 95}, no. 11, 115030 (2017)
  doi:10.1103/PhysRevD.95.115030
  [arXiv:1704.04669 [hep-ph]].



\bibitem{Sirunyan:2018iwl} 
  A.~M.~Sirunyan {\it et al.} [CMS Collaboration],
  Phys.\ Lett.\ B {\bf 782}, 440 (2018)
  doi:10.1016/j.physletb.2018.05.062
  [arXiv:1801.01846 [hep-ex]].



\bibitem{Sirunyan:2019mlu} 
  A.~M.~Sirunyan {\it et al.} [CMS Collaboration],
  Phys.\ Rev.\ Lett.\  {\bf 124}, no. 4, 041803 (2020)
  doi:10.1103/PhysRevLett.124.041803
  [arXiv:1910.01185 [hep-ex]].



\bibitem{Liu:2019ayx} 
  J.~Liu, Z.~Liu, L.~T.~Wang and X.~P.~Wang,
  JHEP {\bf 1907}, 159 (2019)
  doi:10.1007/JHEP07(2019)159
  [arXiv:1904.01020 [hep-ph]].



\bibitem{menu}
https://twiki.cern.ch/twiki/pub/AtlasPublic/TriggerOperationPublicResults/menuTable.png



\bibitem{Read:2002hq} 
  A.~L.~Read,
  J.\ Phys.\ G {\bf 28}, 2693 (2002).
  doi:10.1088/0954-3899/28/10/313



\bibitem{Moneta:2010pm} 
  L.~Moneta {\it et al.},
  PoS ACAT {\bf 2010}, 057 (2010)
  doi:10.22323/1.093.0057
  [arXiv:1009.1003 [physics.data-an]].



\bibitem{Helo:2010cw} 
  J.~C.~Helo, S.~Kovalenko and I.~Schmidt,
  Nucl.\ Phys.\ B {\bf 853}, 80 (2011)
  doi:10.1016/j.nuclphysb.2011.07.020
  [arXiv:1005.1607 [hep-ph]].



\bibitem{Helo:2013prd} 
  J.~C.~Helo and S.~Kovalenko,
  Phys.\ Rev.\ D {\bf 89}, 073005 (2014)
  doi:10.1103/PhysRevD.89.073005
  [arXiv:1312.2900v1  [hep-ph]].
  
  

\bibitem{Liao:2017jiz} 
  W.~Liao and X.~H.~Wu,
  Phys.\ Rev.\ D {\bf 97}, no. 5, 055005 (2018)
  doi:10.1103/PhysRevD.97.055005
  [arXiv:1710.09266 [hep-ph]].



 
\end{thebibliography}
\end{document}